\documentclass[manuscript]{aastex}
\usepackage{enumerate}
\usepackage{multicol}
\usepackage{caption}

\accepted{11th February 2014}

\slugcomment{Submitted to {\it The Astrophysical Journal}}

\begin{document}

%Predefined words with units
\def\kms{\rm{kms$^{-1}$}}
\def\deg{$^\circ$}
\def\solar{_\odot}

%molecules
\def\CO{\mbox{$^{12}$CO(1-0) }}
\def\CCO{\mbox{$^{13}$CO(1-0) }}
\def\CCCO{\mbox{C$^{18}$O(1-0) }}
\def\CS{\mbox{CS(2-1) }}

%abbreviations
\def\COH{$W($CO$)$-to-$N(H_{2})$ }
\def\LCO{$L_{CO}$ }
\def\HH{H$_{2}$ }
\def\HHCO{H$_{2}$CO }
\def\WCO{$W_{CO}$ }
\def\NHH{$N(H_{2})$ }
\def\MHH{$M(H_{2})$ }
\def\MVIR{$M_{virial}$ }
\def\back{\emph{background} }
\def\IV{IVQ }
\def\I{IQ }
\def\IRAS{IRAS/CS }
\def\dv{$\Delta v(FWHM)$ }
\def\nh{$n(H_{2})$ }

%separacion de palabras
\hyphenation{mi-ssing}

\title{GIANT MOLECULAR CLOUDS AND MASSIVE STAR FORMATION IN THE SOUTHERN MILKY WAY}

\author{P. Garc\'\i a}

\affil{Departamento de Astronom\'\i a, Universidad de Chile,
 Casilla 36--D, Santiago, Chile}

\affil{ I. Physikalisches Institut, Universit$\mathrm{\ddot{a}}$t zu K$\mathrm{\ddot{o}}$ln, 
D-50937, Cologne, Germany}

\email{pablo@ph1.uni-koeln.de}

\author{L. Bronfman}

\affil{Departamento de Astronom\'\i a, Universidad de Chile,
 Casilla 36--D, Santiago, Chile}

\author{Lars-{\AA}ke Nyman}
\affil{Joint ALMA Observatory (JAO), Alonso de Cordova 3107, Vitacura, Santiago, Chile} 
\affil{European Southern Observatory (ESO), Alonso de Cordova 3107, Vitacura, Santiago, Chile}

\and

\author{T. M. Dame}
\affil{Harvard-Smithsonian Center for Astrophysics, 60 Garden Street, Cambridge MA 02138}

\begin{abstract}
The Columbia - U. de Chile CO Survey of the Southern Milky Way is used for separating the CO(1-0) emission of the fourth Galactic quadrant within the solar circle into its dominant components, giant molecular clouds (GMCs). After the subtraction of an axisymmetric model of the CO \back emission in the inner Southern Galaxy, 92 GMCs are identified, and for 87 of them the two-fold distance ambiguity is solved. Their total molecular mass is \MHH $=$ 1.14 $\pm$ 0.05 $\times$ 10$^{8}$ M$\solar$ accounting for around 40\% of the molecular mass estimated from an axisymmetric analysis of the \HH volume density in the Galactic disk \citep{bron88} \MHH$_{disk}$ $=$ 3.03 $\times$ 10$^{8}$ M$\solar$. The large scale spiral structure in the Southern Galaxy, within the solar circle, is traced by the GMCs in our catalog; 3 spiral arm segments: the \emph{Centaurus}, \emph{Norma}, and \emph{3-kpc expanding} arm are analyzed. After fitting a logarithmic spiral arm model to the arms, tangent directions at 310\deg, 330\deg, and 338\deg, respectively, are found, consistent with previous values from the literature. A complete CS(2-1) survey toward IRAS point-like sources with FIR colors characteristic of UC HII regions is used to estimate the massive star formation rate per unit \HH mass (MSFR), and the massive star formation efficiency ($\epsilon$) for GMCs. The average MSFR for GMCs is 0.41 $\pm$ 0.06 L$\solar$/M$\solar$, and for the most massive clouds in the \emph{Norma} arm it is 0.58 $\pm$ 0.09 L$\solar$/M$\solar$. Massive star formation efficiencies of GMCs are on average 3\% of their available molecular mass.\\

\end{abstract}


\clearpage
\newpage

\section{INTRODUCTION}

Within the Galactic disk, at 100 parsecs scales, the molecular gas is contained mostly in the form of giant molecular clouds (GMCs) \citep{dame86,bron88,williams97}. The large scale clumpy structure of the CO emission in longitude-velocity diagrams is clear evidence of the organization of the gas into these large objects \citep{bron88b}. The principal characteristic of GMCs is the role they play as tracers of the large scale structure in the Galaxy and as birthplaces of most of the massive stars in the Galactic disk. \\

As for the origin of GMCs, it is accepted that, in the disk of the Milky Way, they are formed as the molecular gas enters in the spiral wave pattern of the gravitational potential energy of the Galactic disk \citep{elmegreen94}. Tidal shear forces among GMCs are weaker within the spiral arms than for the interarm regions making the gravitational collapse more feasible \citep{luna06}. Other mechanisms, such as the growth by collisions are, apparently, too inefficient to reproduce the observed power law mass distribution of GMCs \citep{elmegreen93b,tan2012}.\\

GMCs are excellent tracers of large scale Galactic structure. The best example is found in the \emph{Carina} spiral arm, traced over 20 kpc in the outer Galaxy by more than 40 GMCs, between $l =$ 270\deg\textrm{ }and $l =$ 330\deg\textrm{ } in Galactic longitude \citep{grabelsky87,grabelsky88}. \citet{dame86} reconstructed a 3 spiral arm model for the first Galactic quadrant based on their catalog defined in CO and suggested that the containment of the largest molecular clouds in the arms demonstrates that CO emission is enhanced in the arms not only because the clouds are hotter as suggested by \citet{sanders85}, but mainly because they are larger and contain more mass.\\

GMCs are the known places of massive star formation. Massive stars ($M > 8$ M$\solar$) originate mainly inside dense, compact clumps (UC HII regions)  in giant molecular clouds \citep{evans99,maclow03,krum05b,tan05,luna06,mckee07,zinnecker07,schuller2010}. At large scales, \citet{bron00} showed that the radial distribution of massive star forming regions follows, on average, that of the molecular gas in the inner Galactic disk for all galactocentric radii and that the massive star formation is highest at the peak of the southern ``molecular ring'' (understood as an azimuthally averaged molecular gas distribution of the Galactic disk), between 0.5 $\leq$ R/R$\solar \leq $ 0.6 in galactocentric radius. For the fourth Galactic quadrant (\IV), \citet{luna06} showed that massive star formation occurs in regions with high molecular gas density, roughly coincident with the line of sight tangent to spiral arms. At smaller scales, giant molecular clouds harbor most of the massive star formation in the Galaxy \citep{luna06,maclow03,mckee07,zinnecker07}. An example of this is the extremely high velocity molecular outflow, signature of massive star formation \citep[and references therein]{krum05}, in the G331 region \citep{bron08,merello2013}.\\ 

While extensive work has been carried out to identify and find the physical characteristics and spatial distributions of GMCs in the first Galactic quadrant (\I) within the solar circle \citep{dame86,scoville87,solomon87}, no equivalent catalog of GMCs in the \IV within the solar circle has been published yet. For the \I, different methods have been used to define the GMCs: topologically closed surfaces in the three dimensional LBV (longitude, latitude, velocity) CO data phase space  \citep{scoville87,solomon87}; \emph{``clipping''} the CO data below a certain temperature threshold \citep{myers86}; and subtraction from the observed CO emission of a synthetic dataset generated from an axisymmetric model of the Galactic molecular gas distribution (background subtraction; \citep{dame86}), to ``extract'' the GMCs from the complex CO \back emission in which they are immersed. In this work, we use  the CO dataset from the Columbia - U. de Chile Southern CO Survey of the Milky Way \citep{bron88b} (hereinafter referred to as CO Survey) to define GMCs in the 3-dimensional phase space through the subtraction, from the CO data, of an axisymmetric model (ASM; \citep{bron88}) of the complex CO \back emission following \citet{dame86}. \\

We estimate the massive star formation rate and massive star formation efficiency of GMCs in our catalog by taking advantage of a complete \CS survey toward IRAS point-like sources with FIR colors characteristic of UC HII regions \citep{bron96} complemented by a new unpublished \CS survey (L. Bronfman et al., in preparation). Hereinafter we refer to the whole \CS dataset as \CS survey and to the IRAS point-like UC HII regions as \IRAS sources. We select massive star forming regions from the IRAS point source catalog, because most of the embedded massive star luminosity is emitted around $100 \mu m$ in the Mid- and Far-Infrared (see Figure 2 in \citet{faundez2004}). The IRAS point sources are selected by their FIR colors, using the \citet{wood89b} criterion for UC HII regions, which trace well the population of young embedded massive stars \citep{faundez2004} emitting also in the millimeter continuum. This continuum survey shows that massive cold cores, the very early stages of MSF, are associated to the central UC HII region. The \CS emission is a great tracer in that it is only excited in regions of densities above  $10^{4}$ - $10^{5}$ cm$^{-3}$. We detected the line emission for about 75\% of the candidates. This is a  complete sample of the most massive and luminous regions of massive star formation in the Galaxy. There are more recent Galactic surveys in the Mid-Infrared and Sub-Millimeter regions of the spectrum (GLIMPSE, MIPSGAL, and ATLASGAL), but (a) there is no complete line surveys to determine their kinematic information and (b) only a small fraction of the luminosity from massive star forming regions comes in the NIR  and sub-mm wavelengths \citep{faundez2004,tanti2012}.\\    

In Section \ref{catalog}  we present for the first  time a complete catalog of molecular clouds in the IV quadrant from $l = $ 300\deg\textrm{ }to $l = $ 348\deg, within the solar circle. Distances, masses, and other physical properties of these objects are determined, and their statistical distributions are discussed. In Section \ref{spiral}, the large scale spiral structure traced by GMCs within the Galactic disk is analyzed. In Section \ref{msfr} we study the massive star formation rate per unit \HH mass (MSFR) and massive star formation efficiency ($\epsilon$) for GMCs. In Section \ref{concl} the main conclusions of the present work are summarized. In Appendix \ref{app1} the effect of subtracting a model from the CO data is described, and in Appendix \ref{app2}, the two-fold distance ambiguity resolution for GMCs in our catalog is explained.\\

%%%%%%%%%%%%%%%%%%%%%%%%%%%%%%%%%%%%%%%%%%%%%%%%%%%%%%%%%%%%%%%%%%%%%
%%%%%%%%%%%%%%%%%%%%%%%%%%%%%%%%%%%%%%%%%%%%%%%%%%%%%%%%%%%%%%%%%%%%%
%%%%%%%%%%%%%%%%%%%%%%%%%%%%%%%%%%%%%%%%%%%%%%%%%%%%%%%%%%%%%%%%%%%%%
%%%%%%%%%%%%%%%%%%%%%%%%%%%%%%%%%%%%%%%%%%%%%%%%%%%%%%%%%%%%%%%%%%%%%

\section{GMCs IN THE FOURTH GALACTIC QUADRANT \label{catalog}}

\subsection{GMCs Identification\label{def}}

Giant molecular clouds, within the solar circle, are generally surrounded by and superimposed on an extended \back of CO emission \citep{dame83} while, outside the solar circle, GMCs appear to be isolated and well defined \citep{may97}. Therefore, one of the main difficulties we confront in describing GMCs is their definition in phase space (longitude-latitude-velocity). Figure \ref{fig1} shows a longitude-velocity diagram of the CO emission detected in the Columbia - U. de Chile CO Survey made by integrating the emission over the latitude range $b = \pm$ 2\deg. Superimposed on the CO data, 284 \IRAS sources, used in Section  \ref{msfr} to determine the massive star formation rate and efficiency of GMCs, are plotted as filled circles in a reddish color scale, representing the Far-Infrared flux of the sources. The principal characteristics of the CO survey are summarized in Table \ref{tbl-1}. The complexity of the emission is evident in the figure. The most accepted interpretation of such a \back is related to the presence of numerous smaller and more diffuse clouds than GMCs \citep{dame83,dame86,solomon87,bron88} distributed across the inner Galaxy. Since the \back is seen only in the longitude-velocity diagram, it is also plausible that such smaller and more diffuse clouds are largely confined to spiral arms, along with the GMCs. In addition to this problem, and for Galactic longitudes $l \ge $ 328\deg, the presence of more than one spiral arm feature along the line of sight \citep{russeil03} also contributes to the complex structure of the CO emission.\\ 

In order to identify the largest molecular clouds in the longitude-velocity diagram, similar to the analysis done by \citet{dame86} for the first Galactic quadrant, a synthetic dataset was generated using the ASM of the CO emission by \citet{bron88} (see the insert on the left lower corner of the Figure \ref{fig2}), and subtracted from the observed CO dataset, creating a LBV  background subtracted data cube. In this data cube, individual boxes were defined containing the main emission features. As a consistency check, we used the spatial and radial velocity distribution of the \IRAS sources to trace the extension of such structures in phase space. The physical information of GMCs was then derived from their spatial maps and spectra. A detailed explanation of the method utilized to define the clouds is presented in Appendix \ref{app1}. After the subtraction of the CO \back emission, the largest molecular clouds appear as isolated structures, allowing us to assign a $(v,b,l)$ box in phase space to each one of the 92 clouds presented in our catalog. GMCs in the \IV are presented in Figure \ref{fig2}, a longitude-velocity diagram of the model subtracted CO emission from the CO Survey in units of antenna temperature. \\ 

The observational information of each cloud, contained in its phase space, is utilized in deriving its physical parameters (radius, mass, etc). Integrating the phase space box over its angular extension, the composite spectrum of the cloud $T_{A}(v) = \sum T_{A}(v,l,b)  \Delta b  \Delta  l$, with $ \Delta b  = \Delta  l = $ 0\deg.125, is generated and a Gaussian profile fitted to obtain the radial velocity center ($V_{lsr}$), the velocity width (\dv), and intensity ($I_{CO} = \int T(v) dv$). In some cases because of the large velocity dispersion of the clouds, larger than the cloud separation, there is a partial blend of the CO profiles along the line of sight between two clouds. In those cases, and in order to properly estimate the CO intensity for each individual cloud, two Gaussian were fitted simultaneously. An example of a two Gaussian profile fit to recover the kinematic information and CO emission of the cloud GMC G333.625$-$0.125 is shown in Figure \ref{fig3_tvspect}. \\

The physical parameters such as position and angular size are derived from the spatial map for each GMC. An example of the spatial map for a GMC in our catalog is presented in Figure \ref{fig4_3} corresponding to GMC G331.500$-$0.125 (number 43 in Table \ref{tbl-2}). The limits in phase space were identified after the subtraction of the \back model emission. The bluish color scale represents the CO intensity of the cloud ($I(l,b) = \sum T_{A}(v,l,b)  \Delta v$, with  $\Delta v  = 1.3$ \kms) and the white square the CO peak intensity. Superimposed on the contours are 10 \IRAS sources associated to this cloud, identified as orange filled circles in the figure (the beam size for the CS sources is 50" from the SEST Telescope at 100 GHz). The correlation of the point sources with the gas results evident, proving that massive star formation occurs primarily in GMCs. Similar spatial maps for all the clouds in our catalog are available on request.\\

%%%%%%%%%%%%%%%%%%%%%%%%%%%%%%%%%%%%%%%%%%%%%%%%%%%%%%%%%%%%%%%%%%%%%%%%%%
\subsection{Distance Determination\label{dist}}

The determinations of heliocentric distances to GMCs is crucial to obtain many of their physical parameters. Currently, a huge effort is being made to obtain distances to massive star forming regions (such as water and methanol masers) in the northern hemisphere via trigonometric parallaxes \citep[The BeSSeL Survey]{sanna2013}, and it is planned to extend this work to southern hemisphere sources \citep{reid2009paralaxes}. Since this is an ongoing work, distances for southern massive star forming regions are not available yet. On the other hand, due to optical extinction in the Galactic disk, kinematic distances are usually the only ones available for objects beyond $\sim$ 3 kpc. Using the radial velocity of the source and a rotation curve (under the assumption of pure circular motion of the gas around the Galactic centre), a kinematic distance to the source can be determined. In the outer Galaxy, given the radial velocity of the source, a unique kinematic distance can be assigned to it. This is not the case for the inner Galaxy where, due to geometric effects, the radial velocity of the source can come from two different places along the same line of sight. This geometric effect is known as the two-fold distance ambiguity of the inner Galaxy. After the determination of the two possible kinematic distances for a source within the solar circle, the distance ambiguity must be removed in order to calculate distance dependent physical parameters. In this simple way, distances can be obtained for GMCs in our catalog.\\ 

In order to estimate the kinematic distances to the GMCs in the present work, we adopt the rotation curve derived by \citet{alvarez90} from the same dataset analyzed here, assuming an heliocentric distance to the Galactic center of R$\solar =$ 8.5 kpc and a circular velocity of the local standard of rest, $\Theta \solar =$ 220 \kms:

\begin{equation}\label{eq1}
\Theta (R) = 209.2 + 10.5\frac{R_{GAL}}{R\solar} \qquad [kms^{-1}].
\end{equation}

Effects such as blending along the line of sight and deviations from the common assumption of pure circular motion are among the main sources of uncertainty in kinematic distances. Large scale streaming motions have been observed in a number of regions \citep{burton88,brandandblitz93}. They produce deviations from pure circular motion and may introduce uncertainties of up to 5\% in the estimation of galactocentric radii with corresponding uncertainties in the estimated heliocentric distances that may go from 0.6 kpc to 1.7 kpc, if the streaming motion is along the line of sight \citep{luna06}. There are also several good examples in the fourth Galactic quadrant of large velocity perturbations caused by the action of energetic events such as, for instance, the large hole in the longitude-velocity diagram at 325\deg, $-$40 \kms\textrm{ }attributed by \citet{nyman87} to a single event causing multiple supernova explosions and stellar winds.\\

Based on measurements of trigonometric parallaxes and proper motions of high-mass star forming regions in the Galactic plane, \citet{reid2009} investigated deviations from a flat rotation curve by fitting a rotation curve of the form $\Theta (R) = \Theta\solar + (d\Theta/dR)(R - R\solar)$, with $\Theta \solar =$ 254 \kms\textrm{ }and R$\solar =$ 8.4 kpc. Two values for the derivative $d\Theta/dR$ are obtained: 1.9 and 2.3. We tested both models and found no significant differences with the kinematic distances derived from the rotation curve in equation \ref{eq1}. For both values of $d\Theta/dR$, kinematic distances for clouds closer than 7.5 kpc (68 GMCs) are systematically underestimated. For $\sim$ 80\% of these clouds, differences are less than 15\%, on the order of the estimated errors. For the remaining 20\% of these clouds, the largest difference is less than 30\%. For clouds with heliocentric distances larger than 7.5 kpc (19 GMCs), the distance estimations are essentially the same (less than 5\% difference).\\

The well known two-fold distance ambiguity is an important source of uncertainty in kinematic distances within the solar circle. In the present work, the ambiguity between far and near distance has been removed by several methods: (a) spatial association with optical objects and the existence of visual counterparts; (b) absorption measurements against HII regions or \IRAS sources associated to the cloud for which the distance ambiguity has been solved. Other criteria such as: the latitude criterion (distance off the Galactic plane),  the size-to-linewidth relationship also known as the \emph{``first Larson's Law"} \citep{larson81}, and continuity of spiral arms are complementary to the former criteria. The proximity of the CO radial velocity to the terminal velocity ($|v| < 10$ \kms) is used to assign the tangent distance to the cloud. From the 92 GMCs previously defined, we were able to solve the two-fold distance for 87 of them, which are the ones used in the coming sections to analyze the properties of GMCs in our catalog. A detailed analysis of the methods utilized in removing the two-fold distance ambiguity is presented in Appendix \ref{app2}.\\ 

After the distance ambiguity was removed for most of the GMCs in our catalog, we have compared our distance ambiguity resolution criteria with that used for the 6.7 GHz methanol (CH$_{3}$OH) masers by \citet[and references therein]{green_mcclure2011} as a consistency check. In most cases there is a good agreement between the distance assigned to the methanol masers and the distance determined in the present work for their parent GMCs. A detailed discussion of this consistency check can be found in appendix \ref{app2}.\\ 

%%%%%%%%%%%%%%%%%%%%%%%%%%%%%%%%%%%%%%%%%%%%%%%
\subsection{Physical Properties\label{param}}
 
The physical parameters characterizing the GMCs identified here are summarized in Table \ref{tbl-2}. The first column represents the identification number of each GMC, from smaller to larger Galactic longitudes, and from lower to higher Galactic latitudes. The second and third columns represent the Galactic longitude $l$ and Galactic latitude $b$ of the CO peak intensity in the spatial map of each GMC (as described in Figure \ref{fig4_3}). The fourth column represents $V_{lsr}$ of each cloud, estimated from the Gaussian fit to its composite spectrum. Radial velocities with a $\ast$ symbol were corrected by $+$12.2 \kms\textrm{ }to take into account the anomaly in terminal radial velocities between $l =$ 300\deg\textrm{ }- 312\deg\textrm{ }found by \citet{alvarez90}. The fifth column is the linewidth \dv of each GMC composite spectrum. Column 6 gives the heliocentric kinematic distance $D$ to each cloud. Column 7 gives the adopted distance (N = near, F = far, and T = tangent) and, as upper indexes, the criteria used in removing the distance ambiguity (see footnotes at the bottom of the table). Column 8 contains the radius for each GMC, defined as $R = R_{ang} \times D$, with $R_{ang}$ the effective angular radius defined as $\pi R_{ang}^{2} = A_{ang}$, and $A_{ang}$ the angular area. Since this criterion depends on the intensity threshold in the velocity integrated spatial map, above which the observed position belong to the cloud, \citet{ungerechts2000} proposed the estimation of the effective physical radius weighting it by the CO intensity of the cloud. We keep our estimation of the effective physical radius mainly to be consistent with the previous northern catalog of \citet{dame86}. Column 9 gives the Virial mass ($M_{virial}$) of GMCs in our catalog. Under the assumption of virial equilibrium, we estimate virial masses as follows:

\begin{equation}\label{eq4}
M_{virial} = \frac{5\Delta v(FWHM)^{2}R}{8\ln (2)G},
\end{equation}
where $G$ is the gravitational constant. The last column contains the molecular mass \MHH for each cloud estimated as follows :\\

\begin{equation}\label{eq2}
M(H_{2}) = m_{H_{2}}L_{CO} \chi.
\end{equation}
A mean molecular weight per \HH molecule corrected for helium abundance of $m_{H_{2}} =$ 2.72$m_{H}$ \citep{allen73} is used. The $\chi$ factor is the Galactic average \COH conversion factor $\chi = 1.56 \pm 0.05 \times 10^{20}$ [K \kms]$^{-1}$ cm$^{-2}$ \citep{hunter97} between the integrated CO main beam temperature $W_{CO}$ ($W_{CO} = \int T dv$), and the molecular column density $N(H_{2})$. The factor $L_{CO}$ represents the CO luminosity of the cloud estimated as:\\
 
\begin{equation}\label{eq3}
L_{CO} = \frac{I_{CO} D^{2}}{\eta},
\end{equation}
where $\eta = 0.82$ is the main beam efficiency correction of the antenna temperatures \citep{bron88b} and $I_{CO}$ in units of K \kms\textrm{ }sr$^{2}$ is the integrated CO intensity of the cloud. To obtain $I_{CO}$ for each GMC we perform the integral of  $T(v)$ over the Gaussian model fit for each cloud ($I_{CO} = \int T(v) dv$). However, the peak temperature of $T(v)$ depends strongly on the level of the background emission model used to define the clouds (see section  \ref{def}). To remove such uncertainty, before computing $I_{CO}$, the Gaussian fit is performed again over the original observed dataset, but fixing $V_{lsr}$ and \dv to the values obtained previously in the cloud definition process, and leaving the peak temperature as the only free parameter. The value of $I_{CO}$ is then calculated from this new Gaussian fit, and is independent of the level of the background model used to define the clouds. The procedure is further explained in Appendix \ref{app1}.\\

%%%%%%%%%%%%%%%%%%%%%%%%%%%%%%%%%%%%%%%%%%%%%%%

\subsection{Statistical Properties}

\subsubsection{Size-to-Linewidth Relationship}

The theoretical interpretation of the apparently universal size-to-linewidth relationship of molecular clouds has been a matter of debate for many years, and often different theoretical interpretations have been suggested. One of the first attempts was done by \citet{larson81}. They noticed that the power index ($ \sim 1/3$) is similar to the Kolmogorov law of incompressible turbulence and argued that the  observed non thermal linewidths originate from a common hierarchy of interstellar turbulent motions and the structures in the clouds can not have formed by simple gravitational collapse. On the other hand,  \citet{solomon87} argued that the Kolmogorov turbulent spectrum is ruled out by the ``new" data with a power index of $ \sim 1/2$. They suggest that the size-to-linewidth relation arises from the virial equilibrium of the clouds  since their mass determined dynamically agrees with other independent measurements and they are not in pressure equilibrium with warm/hot ISM. Presently, MHD-models have shown that supersonic turbulence is sufficient to explain the observed slope of the size-to-linewidth relation and that self-gravity may be important on large and small scales  \citep{evans99,kritsuk2007}.\\ 

The  size-to-linewidth relationship \dv $= AR^{\alpha}$ for clouds in Table  \ref{tbl-2} is presented in Figure \ref{fig5_dv}.  A least-squares fit yields:

\begin{equation}\label{eq5}
  \Delta v(FWHM) = ( 1. 26\pm 0.35 ) \textrm{ } R^{0.50 \pm 0.07}.
\end{equation}
The mean value of the full width and half maximum for clouds in our sample is \dv$ = $ 9.3 \kms\textrm{ }which implies a mean velocity dispersion $\sim$ 4.0 \kms. The mean physical radius of the clouds in Table \ref{tbl-2} is $R = $ 60 pc. The values $A  = 1. 26\pm 0.35 $ and $\alpha = 0.50 \pm 0.07$ are in agreement with the values in \citet{dame86} $A  = 1. 20\pm 0.22 $ and $\alpha = 0.50 \pm 0.05$ (since their estimations for the galactocentric radii do not scale linearly with $R\solar$, we have explicitly adopted their radii estimates in Figure \ref{fig5_dv} and \ref{fig6_density}). \citet{solomon87} found a power law relationship of the form $\sigma_{v} = 1.0 \pm 0.1 S^{0.5 \pm 0.05}$ , where $\sigma_{v}$ is the velocity dispersion of the cloud, and $S$ is its physical radius. The difference in the proportionality factor $A$ is due to the different definition of the physical radius of the clouds (geometrical average in their case) and the use of the velocity dispersion instead of the \dv for the fit. The slope of the relationship results surprisingly similar to the value obtained in the present work, specially if the different methods employed in defining GMCs are considered. The same occurs in \citet{scoville87}. They reported a relationship of the form $\sigma_{v} = 0.5 \pm 0.1 S^{0.55 \pm 0.05}$, close to our values, within the error.\\

\subsubsection{\HH Density-to-Size Relationship}

The  \HH volume density \nh can be expressed as a function of the cloud's mass and radius: 

\begin{equation}\label{eq6}
n(H_{2}) = \frac{M(H_{2})} {2m_{H}4\pi R^{3}/3}.
\end{equation}
The existence of a power law relationship between the \nh and $R$ for molecular clouds (\nh $ = BR^{-\beta}$) was first reported by \citet{larson81}. This relationship is shown in Figure  \ref{fig6_density} for the clouds in our catalog. A least-squares fit to the physical quantities of GMCs in Table \ref{tbl-2} gives:

\begin{equation}\label{eq7}
 n(H_{2}) = ( 8.91 \pm 4.31 ) \times 10^{2}\textrm{ } R^{-0.89 \pm 0.12}.
\end{equation}
The dispersion dispersion of the data points is large. This is mainly due to the radius dependence in equation \ref{eq6} and the corresponding difficulties in defining such a radius. Nonetheless, a trend in Figure \ref{fig6_density} is recognizable. The fit results are close to those values obtained by \citet{grabelsky87} for GMCs in the Carina spiral arm ($B = $ 2.67 $\pm$ 1.75 $\times$ 10$^{2}$, $\beta = $ 0.94 $\pm$ 0.16 ), in particular the slope (within error uncertainties). Since most clouds there do not suffer from the distance ambiguity and are very well defined, we believe that the determination of such an observational relationship should be more accurate for clouds in the outer Galaxy. On the other hand, the fit parameters are quite different from the values found by \citet{dame86} ($B = $ 3.6 $\pm$ 1.2 $\times$ 10$^{3}$, $\beta = $ 1.3 $\pm$ 0.1 ), even if GMCs with densities above 100 cm$^{-3}$ in their catalog are excluded and a new fit to their data is made, and there seems to be no easy way to conciliate both estimates. We believe this might be due to the different rotation curve utilized in their work.\\

\subsubsection{Virial Mass-to-CO Luminosity Relationship and the \COH Conversion Factor}

A relationship between virial mass and CO luminosity of the form  \MVIR $ = C$\LCO$^{\delta}$ is presented in Figure \ref{fig7_test_back_dvfix}. A power law fit represented by the dotted line yields:\\

\begin{equation}\label{eq9}
  M_{virial} = ( 13 \pm 8 ) \textrm{ } L_{CO}^{0.90 \pm 0.05}.
\end{equation}
The correlation between virial masses and CO luminosities results evident form the figure. The fit parameters $C =  13 \pm 8 $ and   $\delta = 0.90 \pm 0.05$ are similar to previous estimates in \citet{solomon87}, $C =  39 \pm 12 $ and $\delta = 0.81 \pm 0.03$, although the slope in our relationship is closer to unity than in their work. The difference in the proportionality factor $C$ is mainly due to the recovery of CO flux in our catalog. The well behaved correlation puts the CO luminosity as a good tracer of mass in the inner Galaxy.\\ 

A linear relationship between the virial mass and the CO luminosity might be used in estimating a \COH conversion factor for GMCs close to virial equilibrium. In such case, the slopes found in equations \ref{eq5} and \ref{eq7} are expected to follow the relationship $\alpha + \beta/2 = 1$ \citep{dame86,bertoldi1992}. For our sample, this relationship yields $\alpha + \beta/2 = 0.95$ supporting the idea that GMCs in our catalog are close to virial equilibrium.  If we consider that the virial mass is close to the real mass of the clouds, the $\chi$ factor can be evaluated directly from the proportionality between the virial and molecular masses. The solid straight line in Figure \ref{fig7_test_back_dvfix} represents a  least-squares fit procedure yielding $M_{virial} = ( 3.7 \pm 0.2 )\textrm{ }L_{CO} $. The corresponding \COH average conversion factor, if GMCs are in virial equilibrium, is: 

\begin{equation}\label{eq10}
    \chi_{GMCs} = 1.71 \pm 0.09 \times 10^{20} \textrm{ } \textrm{[(K \kms)$^{-1}$ cm$^{-2}$]}.
\end{equation}
This value is consistent within 10\% with the $\chi$  conversion factor of $1.56 \pm 0.05\times10^{20}$ [K \kms]$^{-1}$ cm$^{-2}$ from \citet{hunter97} utilized here (see Figure \ref{fig7_test_back_dvfix}).\\

From a theoretical point of view, \citet{shetty10} discussed possible variations in gas simulations of the $\chi$ factor for typical conditions in the Milky Way disk. They reported that the $\chi$ factor is not expected to be constant within individual molecular clouds, although in most cases it is similar to the Galactic value. From simulations, their best fit model resembling typical conditions of the disk, predicts an average value of $\chi = 2 \times 10^{20}$ [K \kms]$^{-1}$ cm$^{-2}$ which is somewhat  larger than our observational estimated value.  \citet{glover2011} showed that  ``there is a sharp cut-off in CO abundance at mean visual extinctions $A_{v} \leq 3$," where  photodissociation becomes important. This is not the case for \HH which is argued to be controlled principally by the product of density and the metallicity, and insensitive to photodissociation. This could imply a different conversion factor for clouds and inter-cloud medium. \citet{ungerechts2000} studied GMCs in the Perseus spiral arm traced by the $^{12}$CO and $^{13}$CO lines. They adopt  a factor $\chi = 1.9 \times 10^{20}$ [K \kms]$^{-1}$ cm$^{-2}$ which is larger than  the value of \citet{hunter97}, and suggest that the ``relatively large virial masses or equivalently, the low CO luminosities in relation to the linewidths show that $\chi$ could be even higher than the adopted value."\\

%%%%%%%%%%%%%%%%%%%%%%%%%%%%%%%%%%%%%%
\subsection{Molecular Mass Spectrum and Comparison with the Total Molecular Mass of the Galactic Disk\label{mms}}

The molecular mass spectrum of clouds in Table \ref{tbl-2} is shown in Figure \ref{fig8_sm_and_dvfix}. Following \citet{williams97} we adopted a logarithmic mass bin interval of $\Delta_{log} = 0.30$ in constructing the mass spectrum of the clouds. Since the shape of the mass spectrum depends strongly on the selected mass bin \citep{williams97,roso05}, the adopted mass bin interval allows us to make a direct comparison between our results and the results presented in \citet{williams97} in analyzing the catalogs of \citet{solomon87,scoville87}. The vertical dotted lines in Figure \ref{fig8_sm_and_dvfix} indicates our completeness limit (around $1.3 \times 10^{6}$ M$\solar$), where about 75\% molecular mass is concentrated toward the high mass end of the spectrum, meaning that we are in fact detecting most of the molecular mass in the catalog. Error bars are estimated as $\sqrt{N}$, where $N$ is the number of clouds that fall into each mass bin. Triangles represent the central mass of each mass bin. A least-squares fit to the data is performed in the range indicated by the solid line. The dashed line is an extrapolation of the fit to the lower mass end of the distribution. The least-squares fit yields:\\ 

\begin{equation}\label{eq11}
\frac{dN}{dM} \propto  M^{-1.50 \pm 0.40}.
\end{equation}

The mass distribution of GMCs determines how the molecular mass is distributed among GMCs. For the inner Galaxy, observational evidence has shown a slope of the mass spectrum for GMCs $\gamma < 2$ \citep{dame83,casoli84,sanders85,solomon87,solomon89,may97,williams97,blitz04,roso05,blitz07}. The value for the index of the mass distribution of GMCs in the inner Galaxy $\gamma = 1.50 \pm 0.40$ in the present work, $\gamma = 1.81 \pm 0.14$ for the \citet{solomon87} GMCs catalog, and of $\gamma = 1.67 \pm 0.25$ for the clouds in \citet{scoville87}, strongly suggests that most of the molecular mass is concentrated toward the largest molecular clouds in the Galactic disk.\\

The total molecular mass in form of GMCs in our catalog is \MHH $=$ 1.14 $\pm$ 0.05 $\times$ 10$^{8}$ M$\solar$. The most massive cloud in the catalog is GMC G337.750$+$0.000 with a molecular mass of \MHH $=$ 8.7 $\times$ 10$^{6}$ M$\solar$. This last result shows that that molecular mass upper limit M$_{max} = 6 \times 10^{6}$ M$\solar$, established in other GMCs catalogs, may depend on the way GMCs are defined. From stability arguments it is clear that a mass upper limit must exist, but here we find it is higher than the previous value in the literature.\\

\subsubsection{Comparison with Axisymmetric Model Mass}

The total molecular mass obtained for all the clouds in our catalog is of 1.14 $\pm$ 0.05 $\times$ 10$^{8}$ M$\solar$. The sum of the virial masses for all cloud is very similar, 1.21 $\pm$ 0.03 $\times$ 10$^{8}$ M$\solar$. Errors are estimated from the standard deviations of the physical parameters from the Gaussian fits. In comparison,  the molecular mass  derived, for the same region sampled here, from the axisymmetric analysis of \citet{bron88} is 3.03 $\times$ 10$^{8}$ M$\solar$.  Therefore, we account for only about 40\% of the axisymmetric model mass in our cloud decomposition analysis. Such kind of difference had been detected already for the first Galactic quadrant by \citet{williams97}, which estimated about 80\% of  axisymmetric model mass not accounted for in the catalogs of  \citet{solomon87} and \citet{scoville87}.\\

A possible explanation for this result may arise from the  $\chi$  conversion factor used here to determine the GMCs mass, which is the same conversion factor, averaged over the whole Galactic plane, used to calculate the axisymmetric model mass. It is possible to postulate different conversion factors for the GMCs and the inter-cloud medium (ICM). For instance, using the same conversion factor, half of the total mass is in GMCs and the other half in the ICM, but using a conversion factor for GMCs twice that for the ICM, one would obtain 2/3 of the mass in GMCs and 1/3 in the ICM. Such a difference between  $\chi_{GMCs}$  and $\chi_{ICM}$ is consistent with results obtained from comparing LTE column densities measured with $^{13}$CO with those obtained from CO observations of GMCs and ICM regions (A. Luna et al. 2013, in preparation).\\         

In terms of the distance, the molecular mass in our catalog is distributed as follows: 59\% is contained in 64 clouds at the near distance and 39\% in 20 clouds at the far distance (the remaining 2\% is contained in the 3 clouds for which the tangent distance was assigned) reflecting the fact that the model of the CO \back emission is less sensitive to the mass at the far distance. The beam dilution causes GMCs at the near distance to be more easily detected than the clouds at the far distance, but this difference is not too large as shown by the percentage of molecular mass detected at the near and far distance.\\         
 
%%%%%%%%%%%%%%%%%%%%%%%%%%%%%%%%%%%%%%%%%%%%%%%%%%%%%%%%%%%%%%%%%%%%%
%%%%%%%%%%%%%%%%%%%%%%%%%%%%%%%%%%%%%%%%%%%%%%%%%%%%%%%%%%%%%%%%%%%%%
%%%%%%%%%%%%%%%%%%%%%%%%%%%%%%%%%%%%%%%%%%%%%%%%%%%%%%%%%%%%%%%%%%%%%
%%%%%%%%%%%%%%%%%%%%%%%%%%%%%%%%%%%%%%%%%%%%%%%%%%%%%%%%%%%%%%%%%%%%%

\section{GMCs AS TRACERS OF THE LARGE SCALE STRUCTURE IN THE FOURTH GALACTIC QUADRANT: SPIRAL ARMS FEATURES\label{spiral}}

In this section we examine the large scale structure traced by the GMCs in our catalog. In longitude-velocity space, spiral arms features appear as opening loops as a consequence of the rotation of the Galaxy  \citep[Figure 5]{bron00}. Since we have already removed the distance ambiguity for the GMCs in our catalog, we attempt to reconstruct the spiral structure of the Southern Galaxy, within the solar circle, by following the position of the clouds in both longitude-velocity space and in the Galactic plane. \\

\subsection{Identification of Spiral Arms Features}

The large scale spiral structure can be inferred in the longitude-velocity diagram of the \IV. Using the model subtracted dataset and location of giant molecular clouds in the longitude-velocity diagram (only clouds with their two-fold distance ambiguity removed), we reconstruct a tentative picture of the spiral structure in the IV quadrant (Figure \ref{fig9}). Clouds with different colors are used for each spiral arm. We distinguish three main large scale features across the Galactic plane: the \emph{Centaurus} spiral arm (near and far sides traced traced by clouds in red and orange boxes respectively), the \emph{Norma} spiral arm (blue boxes for the near an far side of the arm), and the \emph{3-kpc expanding} arm (black boxes for the near and far side of the arm). We assign the cloud in the yellow box to the near side of the \emph{Carina} spiral arm identified by \citet{grabelsky87}. The clouds at positive velocities are tracing the far side of the \emph{Carina} arm \citep{bron86,grabelsky88} and, as they are beyond the solar circle, are out of the scope of the present work.\\ 

The most prominent spiral feature in Figure \ref{fig9} is by far the \emph{Centaurus} spiral arm, very clearly traced by 50 GMCs in our catalog, over 40\deg\textrm{ }in its near side (red boxes), and over 15\deg\textrm{ }in its far side (orange boxes). The near side of the arm extends roughly from 305\deg\textrm{ }to 348\deg, and from $-$70 \kms\textrm{ }to $-$30 \kms, while its far side extends roughly from 35\deg\textrm{ }to 321\deg, and from $-$20 \kms\textrm{ }to 0 \kms. Close to the far side velocity of the \emph{Centaurus} arm is the well known feature, the Coalsack at 303\deg\textrm{ }and 0 \kms.\\ 

The \emph{Norma} spiral arm is shown as blue boxes in Figure \ref{fig9}, extending roughly from 327\deg\textrm{ }to 348\deg, and from $-$110 \kms\textrm{ }to $-$50 \kms. The detection of the far side of the arm is particularly difficult in these case since, at some spots far emission overlaps with near emission. Such is the case for the GMC G334.125$+$0.500 (number 48 in Table \ref{tbl-2}). In the latitude-velocity maps of \citet{bron88b}, the near emission appears as wide lanes along Galactic latitude while the far emission has a small angular extension but it is very extended along the velocity axis. The far side of the \emph{Norma} arm also harbors the most massive clouds in our catalog, namely GMC G336.875$+$0.125 and GMC G337.750$+$0.000. This statement would change if GMC G342.750$+$0.000 (number 76 in Table \ref{tbl-2}) had the wrong distance assignment. We have adopted the near distance to this cloud after the evidence we have collected from the literature but its distance ambiguity resolution is still not as solid as in other cases. If the cloud was indeed at the far distance, it would be associated to the far side of the \emph{3-kpc expanding} and it would have the same molecular mass as the most massive cloud in our catalog: GMC G337.750$+$0.000, located at the far side of the \emph{Norma} spiral arm. For further discussion on cloud 76 see Appendix \ref{app2}.\\ 

The \emph{3-kpc expanding} arm, the closest to the center of the Galaxy, is traced by clouds on its near side, between 335\deg\textrm{ }and 348\deg, and between $-$150 \kms\textrm{ }and $-$100 \kms, and on its far side, between 345\deg\textrm{ }and 348\deg, and  between $-$100 \kms\textrm{ }and $-$60 \kms. The near side of the \emph{3-kpc expanding} arm has long been recognized in CO and 21 cm longitude-velocity diagrams as a nearly linearly feature in the range $l = $ 348\deg\textrm{ }to 12\deg, with an expanding motion of $-$53 \kms\textrm{ }toward $l = $ 0\deg. Recently, \citet{dame2008} identified the far side of the arm over the same longitude range as a similar parallel feature displaced $\sim$ 100 \kms\textrm{ }to positive velocities. At longitudes further from the Galactic center, the loci of the near and far arms are difficult to trace, and theoretical predictions vary widely (e.g. \citet{cohen1976,romero2011b}). We find 13 clouds at $l < $ 338\deg\textrm{ }with velocities and kinematic distances that suggest they trace the near side of the arm, and 4 more that may trace the far side (GMC 339.125$+$0.000, GMC 345.125$−$0.250, GMC 346.000$+$0.000, and GMC 347.250$+$0.000). Although the velocities of these clouds, $-$100 \kms\textrm{ }to $-$60 \kms, are far below the velocity of the far arm at $l > $ 348\deg\textrm{ }as traced by \citet{dame2008}, just such a sharp drop in the far arm velocities at lower longitudes is predicted by the recent modeling of \citet{romero2011b}.  Furthermore, the derived distances of these clouds are consistent with that of the far arm as estimated by \citet{dame2008}.\\

In the following, we define limits in Galactic longitude and CO VLSR velocity to produce spatial maps of each one of the arms. In Figure \ref{fig9}, the insert on the left lower corner contains the radial velocity and Galactic longitude limits (orange lines) for each one of the spiral arms over plotted on the CO data of the Columbia Survey. We set such limits following the distribution of the GMCs in the spiral arms. The limits in radial velocity and Galactic longitude of the boxes defined for the clouds belonging to the same spiral segment are use to establish the extension of the arm in longitude-velocity space. The view in Galactic longitude and latitude of the spiral arms is presented in Figure \ref{fig10}. In the figure, the CO Columbia Survey (without the subtraction of the axisymmetric model of the \back emission) is used. The color scale of the gas represents the CO intensity ($I(l,b) =  \sum T_{A}(v,l,b) \times \Delta v$) of the arms. The integration limits along the velocity axis were taken from the insert in Figure \ref{fig9}. For the \emph{3-kpc expanding} arm and the \emph{Norma} arm, their near and far sides are plotted on the same map. Over plotted on the figure are the 284 \IRAS sources utilized in the present work (section \ref{msfr}) to estimate the massive star formation rate per unit \HH mass and massive star formation efficiency. The sources are presented as filled circles in a reddish color scale representing the flux of each \IRAS source (see Figure \ref{fig1}). The correlation of the CO intensity and the distribution of \IRAS sources results evident, being the latter well concentrated toward the plane within $b = \pm$ 1\deg\textrm{ }in all three arms. Since most parts of the arms are traced by GMCs, the role that they play as places of most of the massive star formation in the Galactic disk is evident in the figure.\\

\subsection{Empirical Model of the Spiral Arms in the Southern Galaxy, Within the Solar Circle}  
   
A face-on view of molecular clouds in the Galaxy, including our results and previous ones, is presented in Figure \ref{fig11}. In the fourth Galactic quadrant, GMCs in our catalog are plotted as filled circles with the corresponding color to its parent spiral arm  (as defined in Figure \ref{fig9}) over the physical area covered in the present work (area filled with gray color), between Galactic longitudes $l = $ 300\deg\textrm{ }and $l =$ 348\deg. The size of the circles is proportional to the molecular mass of GMCs (last column in Table \ref{tbl-2}). Also plotted are giant molecular clouds (black filled circles) from the catalog of \citet{grabelsky87}, tracing the \emph{Carina} spiral arm outside the solar circle. For the first Galactic quadrant, GMCs from the catalog of \citet{dame86} are plotted as black filled circles between Galactic longitudes $l = $ 12\deg\textrm{ }and $l = $ 60\deg, tracing the \emph{Sagittarius}, \emph{Scutum}, and \emph{4-kpc} spiral arms, as identified by the author. For the catalogs of \citet{grabelsky87} and \citet{dame86} their heliocentric distances were corrected by a factor 0.85 to account for the different R$\solar$ adopted, and molecular masses are those given by the authors, for simplicity. The dotted circle between the Sun and the Galactic Center represents the ``tangent distances'' in the inner Galaxy, i.e., the distance at the CO terminal velocity. At the position of the Galactic Center, the \emph{``molecular bar"} is represented as a dashed-dotted line. The parameters for the molecular bar were taken from \citet{englmaier99} (for consistency with the work of \citet{russeil03}), with a radius of 3.5 kpc and an orientation angle of 22\deg.5 measured from the Galactic Center and in clockwise direction. Currently, there is a relative broad consensus that our Galaxy is a moderate barred Galaxy, and that the Galactic bar has two main components: a triaxial bulge (also referred as the ``thick bar'') inclined in an angle with values found in the literature between 15\deg\textrm{ }and 30\deg\textrm{ }and with a semimajor axis between 3.1 kpc and 3.5 kpc long, and a long ``thin bar'' inclined $\sim$ 45\deg\textrm{ }(though more recent works suggest angles between 25\deg\textrm{ }and 35\deg) with a semimajor axis of 4 kpc. The former is mainly traced by old star population while the later is traced by current star formation such as methanol masers \citep[and references therein]{green2011,romero2011a}. Whether the angular separation between the thick and thin bars is real or just a projection artifact is still a matter of debate \citep{romero2011a}. In terms of star formation activity, an increase in star formation is expected where the thin bar and the \emph{3-kpc expanding} arm meet. Between Galactic longitude 345\deg\textrm{ }and 351\deg, and radial velocities $-$30 \kms\textrm{ }and $+$10 \kms\textrm{ }such an increase in star formation activity is observed in the methanol maser distribution of \citet{green2011}. It is interesting to notice that, if one assumes a 45\deg\textrm{ }inclination angle and semimajor axis of 3.4 kpc for the thin long bar, the three GMCs at the far side of the \emph{3-kpc expanding} arm would match its southern end almost exactly.\\

We aim to quantify empirically, the parameters describing the principal spiral features presented here. Following \citet{russeil03} we fit a logarithmic spiral arm model to the positions of the clouds tracing each of the three spiral features seen in Figure \ref{fig9} as:\\

\begin{equation}\label{eq15}
R(\phi) = r_{\circ} e^{-p\phi}.
\end{equation}

In the logarithmic spiral arm model, the origin of the reference frame to measure the angle $\phi$ (in radians) is set to the position of the Galactic center, and $\phi$ is measured clockwise, with $\phi = $ 0\deg\textrm{ }at Galactic longitude $l =$ 0\deg. The relationship in equation \ref{eq15} is defined by two parameters: $r_{\circ}$ (kpc) meaning the initial radius of the spiral arm, and the pitch angle $p$, defined as the angle between the tangent to the corresponding galactocentric radius at a certain point in the spiral arm, and the tangent direction to the arm at the same position. It is important to notice that the pitch angle is measured clockwise in the present work yielding, by definition, only positive values, meaning that the $-p\phi$ expression in equation \ref{eq15} results positive and, by consequence, the galactocentric radius of the arm increases as the angle $\phi$ moves toward negative values, i.e., into the fourth Galactic quadrant. The model in Equation \ref{eq15} is not intended to account for possible variations of the pitch angle along the spiral arm \citep{russeil03}. In the fitting procedure, we weighted all clouds by the error in galactocentric radius. Other weights such as molecular mass, yield the same results within error uncertainties. In Table \ref{tbl-4} we summarized the results of the fit procedures for the spiral arms identified in our sample. The last column in Table \ref{tbl-4} accounts for the tangent direction to each spiral arm seen from the Sun, i.e., as measured in Galactic longitude. The spiral arm models for the clouds in our catalog are plotted as thick color lines in Figure \ref{fig11}. Tangent directions to the model spiral arms are plotted as color straight lines.\\

In the following, we discuss the face-on view of GMCs in the Galaxy. The most prominent feature, as expected from Figure \ref{fig9}, is the \emph{Centaurus} arm, traced over 10 kpc in the inner Galaxy. The tangent direction to the arm model in Table \ref{tbl-4}  around 310\deg\textrm{ }is consistent with previous estimates \citep{alvarez90,bron92,englmaier99}. The arm is quite open and spatially closer to the Carina spiral arm than to the other spiral arm segments in the IV quadrant. The pitch angle of this arm $p = $ 13\deg.4\textrm{ }is consistent with the values obtained by \citet{russeil03} ($\sim $ 11\deg) and \citet{dame2011} ($\sim $ 14\deg). The far side of the arm is better traced by GMCs than the near side mainly because, at far distances, the CO \back is much less prominent due to beam dilution effects. The \emph{Norma} spiral arm is the second major spiral feature in our catalog and contains the most massive clouds in our sample. The tangent direction to the model arm $\sim$ 330\deg\textrm{ }is consistent with the values found by \citet{alvarez90} and \citet{bron92} of 328\deg. The pitch angle of the arm $p =$ 6\deg.6 in Table \ref{tbl-4} is consistent with previous estimates \citep[and references therein]{russeil03}.\\ 

We put special attention to the \emph{3-kpc expanding} arm. The far side of the \emph{3-kpc expanding} arm appears well traced by four clouds, but the spiral structure disappears at the near distances. We extended the fit of the logarithmic model for this arm to the position of the Galactic bar. The correlation between the clouds at the far distance and the position of the bar results evident, and is also consistent with the distance determined by \citet{dame2008} for the far side of the arm (11.8 kpc). The tangent direction for the arm $\sim$ 338\deg\textrm{ }is also very consistent with the value found by \citet{alvarez90} and \citet{bron92} of 337\deg. A dense ridge of masers near $l =$ 338\deg\textrm{ }is identified as the tangent point of the \emph{3-kpc expanding} arm in the work of \citet{green2011}. Using the ATLASGAL 870 $\mu m$ Survey, \citet{beuther2012} identified an increase in the sub-millimeter clumps distribution at $l =$ 338\deg\textrm{ }also attributed to the tangent point of the \emph{3-kpc expanding} arm. These results are fully consistent with the tangent direction we find from the logarithmic spiral model of the \emph{3-kpc expanding} arm. The GMCs associated with this spiral arm might be responding to the presence of a Galactic bar, deviating their radial velocities from the assumption of pure circular motion. No previous estimation of the \emph{3-kpc expanding} arm pitch angle $p \sim$ 5\deg.6 in Table \ref{tbl-4} is found in the literature. On the observational side, the methanol masers distribution follows an oval structure in the longitude-velocity diagram that could be physically related to an elliptical structure in the face-on disk \citep{green2011}. Such structure accounts for the parallel lanes of the maser distribution at the far and near distances seen in the longitude-velocity diagram toward the Galactic Center and, to some degree, also accounts for the tangent point at $l =$ 338\deg. \citet{beuther2012} shows the \emph{3-kpc expanding} arm as a continuous elliptical structure in the Galactic disk following the work of \citet{reid2009}. On the modeling side, \citet{romero2011a} applied for the first time the ``Invariant Manifolds'' theory to model spiral arms features in the Galactic disk. \citet{romero2011b} approximately reproduces the observed longitude-velocity CO emission of the near and, to a lesser extent, the far side of the \emph{3-kpc expanding} arm. Their \emph{PMM04-2 bar} model naturally forms a continuos elliptical structure surrounding the composite Galactic bar (thick and thin bars). A continuos elliptical structure of the \emph{3-kpc expanding} arm seems to be also supported by the spatial distribution of GMCs associated to this arm in our catalog.\\

Concerning the region enclosed between the \emph{Norma} and the \emph{3-kpc expanding} arm in Figure \ref{fig11}, \citet{green2011} suggest that part of the \emph{Perseus} arm could harbor some of the methanol masers found toward the tangent direction of the \emph{3-kpc expanding} arm, and in the radial velocity range from $-$60 \kms\textrm{ }and $-$85 \kms. They suggest that these sources could be attributed to the origin of the \emph{Perseus} arm. In our catalog, only the GMC G339.125$+$0.250 associated to the far side of the \emph{Norma} arm falls within this range. In Figure \ref{fig11} we do not see any clear indication of the starting point of the arm in the region between the \emph{Norma} and the \emph{3-kpc expanding} arms, place where the starting point of the \emph{Perseus} arm would be found according to the spiral arms models of \citet{russeil03}. Since we are not sensitive to low mass clouds at the far distance, if only low mass GMCs were tracing the starting point of the \emph{Perseus} arm, we would not be able detect them.\\

The transformation from longitude-velocity phase space to geometrical space in the inner Galaxy has to be taken cautiously. Perturbations in the velocity field caused by density waves \citep{burton71} and energetic events, like supernovae, as well as by the cloud-cloud velocity dispersion, introduce large uncertainties in the derived kinematic distances (typically between 10\% and 20\%). Such effects will inevitably wash out much of our description of the spiral structure in the derived distribution of clouds, even if the clouds are confined to a well defined spiral pattern \citep{combes91}. This is particularly true for the clouds tracing the \emph{3-kpc expanding} arm, since its expanding velocity, around 53 \kms\textrm{ }introduces large uncertainties in to the position of the arm in phase and geometrical space. Another example of deviations of the pure circular motion is the hole at 329\deg, $-$60 \kms\textrm{ }in the longitude-velocity diagram surrounded by molecular clouds with velocity differences of up to 30 \kms\textrm{ }along the line of sight that, according to formaldehyde (H$_{2}$CO) absorption measurements, are probably at the near side of the \emph{Centaurus} arm. Usually, large variations in model parameters such as pitch angle, initial radius, and tangent direction are found depending on the tracer (21 cm emission, HII regions, stellar population, etc.) used to identify the large scale structure in the Galactic disk \citep{englmaier99,russeil03}.\\ 

Despite all difficulties involved in transforming CO radial velocities into heliocentric distances, the spiral structure in the southern Milky Way stands out clearly when traced by giant molecular clouds. In the present work, and with a simple logarithmic spiral arm model we reproduce some characteristics found by other authors, such as the tangent directions to the spiral arms in the southern Galaxy. More sophisticated models are necessary to account for the details of the parameters in each spiral arm. The purpose here is to test the consistency with previous work  of  the large scale spiral structure traced by GMCs in the fourth Galactic quadrant.\\

%%%%%%%%%%%%%%%%%%%%%%%%%%%%%%%%%%%%%%%%%%%%%%%%%%%%%%%%%%%%%%%%%%%%%
%%%%%%%%%%%%%%%%%%%%%%%%%%%%%%%%%%%%%%%%%%%%%%%%%%%%%%%%%%%%%%%%%%%%%
%%%%%%%%%%%%%%%%%%%%%%%%%%%%%%%%%%%%%%%%%%%%%%%%%%%%%%%%%%%%%%%%%%%%%
%%%%%%%%%%%%%%%%%%%%%%%%%%%%%%%%%%%%%%%%%%%%%%%%%%%%%%%%%%%%%%%%%%%%%

\section{MASSIVE STAR FORMATION RATE PER UNIT \HH MASS AND THE STAR FORMATION EFFICIENCY DERIVED FOR GMCs\label{msfr}}

There is close relationship between GMCs and massive stars in our study. While our CO Survey is the best available tracer of GMCs, the \CS survey is the best tracer of UC HII regions. We find a close relationship between GMCs and massive stars in our study. Since most of the UV photons emitted by massive stars are absorbed and re-emitted by the surrounding dust in the Far-Infrared part of the electromagnetic spectrum \citep{kennicutt98a}, the Far-Infrared (FIR) emission of UC HII regions is a good tracer of the massive star formation \citep{luna06}. On the other hand, the \CS emission requires high molecular gas densities, $10^{4}$ - $10^{5}$ cm$^{-3}$, to become excited, resulting also in a good tracer of massive star formation regions. In order to estimate the massive star formation rate per unit \HH mass (MSFR) and the star formation efficiency $\epsilon$ of giant molecular clouds in our catalog, we use the catalog of IRAS point-like sources. The FIR flux of the UC HII regions is obtained directly from the four bands ($12 \mu m$, $25 \mu m$, $60 \mu m$, and $100 \mu m$) of the IRAS point-like sources catalog. Since the IRAS catalog does not provide velocity information of the sources necessary to locate them along the radial velocity axis, we use the \CS line.  We obtain the kinematic information of the UC HII regions from the \CS survey of \citet{bron96} of IRAS point-like sources with FIR colors characteristic of UC HII regions in the whole Galaxy complemented by a new survey to be published elsewhere. The new data amounts to 19\% of the detections .\\ 

\begin{figure}
\epsscale{1.0}
\begin{center}
\includegraphics[angle=0,width=0.95\textwidth]{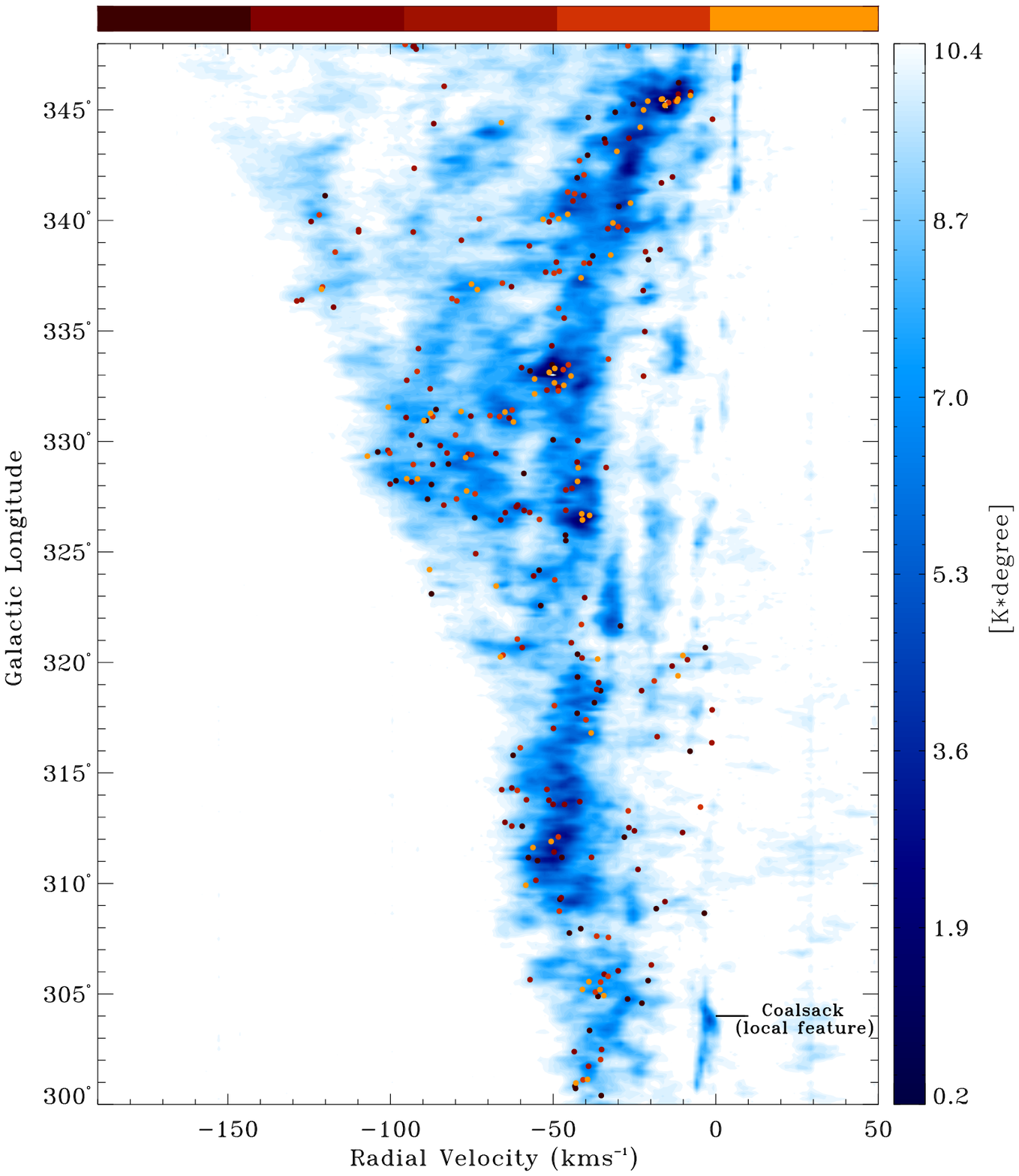}
\caption{\scriptsize{Longitude-velocity diagram of the Columbia Southern CO Survey and IRAS point-like sources with FIR colors of UC HII regions \citep{wood89b} (IRAS/CS sources) in the fourth Galactic quadrant. The Longitude-velocity diagram of the CO emission was obtained by integrating the data set over $b = \pm$ 2.0\deg. The bluish color scale denotes values of antenna temperature as $\int T_{A} db$. A brief summary of the survey parameters is presented in Table \ref{tbl-1}. For the IRAS/CS sources, the reddish color scale represents the FIR flux (94 - 704 Jy brown; 728 - 1590 Jy dark red; 1629 - 3561 Jy red; 3598 - 8140 Jy orange; and 8160 - 62864 Jy yellow). The kinematic information of the sources was obtained from the most complete current available \CS survey of IRAS point-like sources with FIR colors characteristic of UC HII regions \citep{bron96}, complemented by a new \CS unpublished survey (L. Bronfman et al., in preparation).}\label{fig1}} 
\end{center}
\end{figure}

\begin{figure}
\epsscale{1.0}
\begin{center}
\includegraphics[angle=0,width=0.95\textwidth]{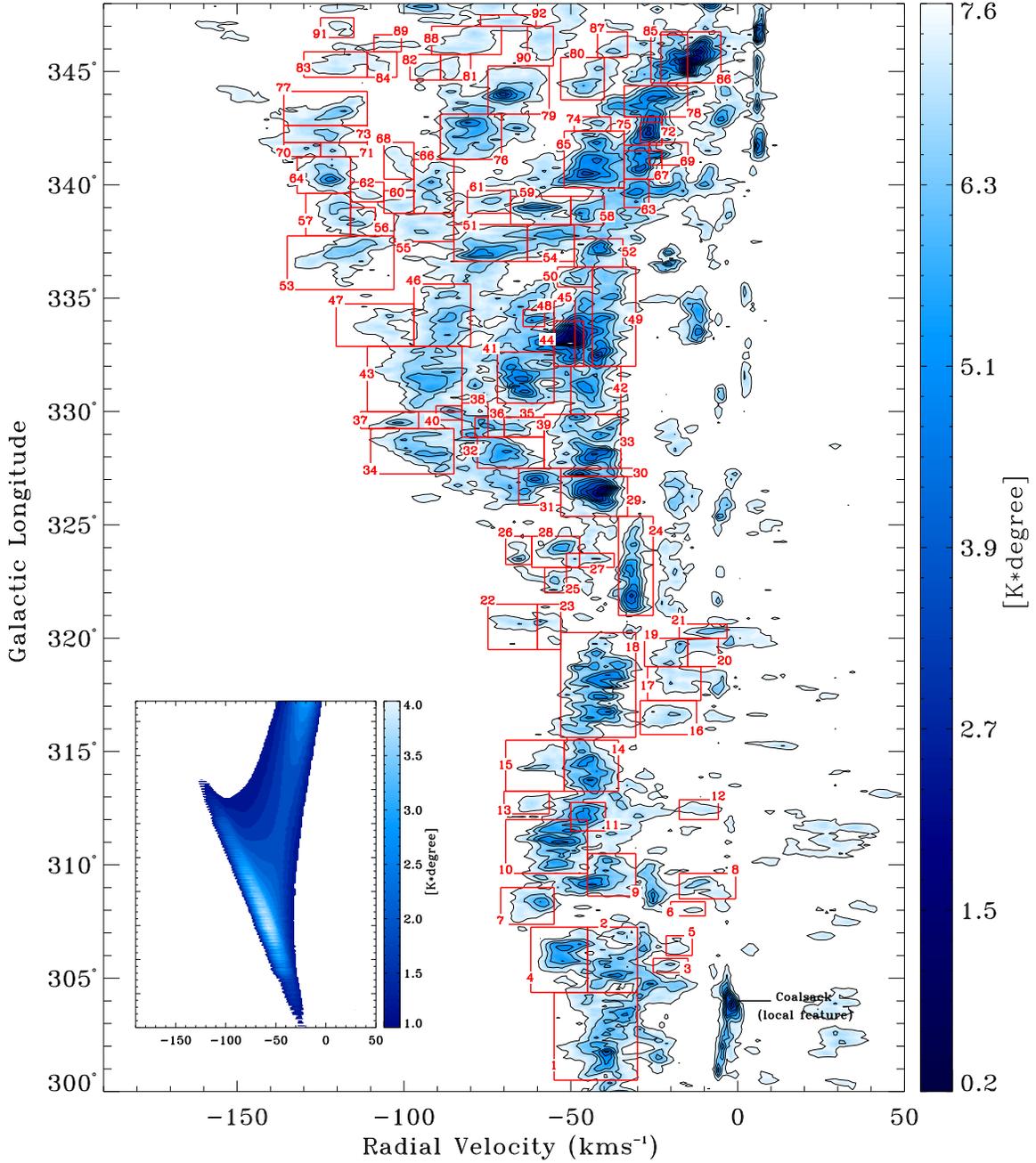}
\caption{\scriptsize{Giant Molecular Clouds in the fourth Galactic quadrant. The longitude-velocity diagram was obtained by integrating the model subtracted CO dataset across the Galactic plane over $b = \pm$ 2.0\deg. The blue color scale denotes values of $\int T_{A} db$. The first contour is located at 0.25 K degree. The axisymmetric model of the background emission is presented in the insert on the left lower corner of the figure. The model was subtracted from the data in Figure \ref{fig1} in order to isolate GMCs from their surrounding \back. The result is shown in the present figure.}\label{fig2}}
\end{center}
\end{figure}

\begin{figure}
\begin{center}
\includegraphics[angle=90,width=\textwidth]{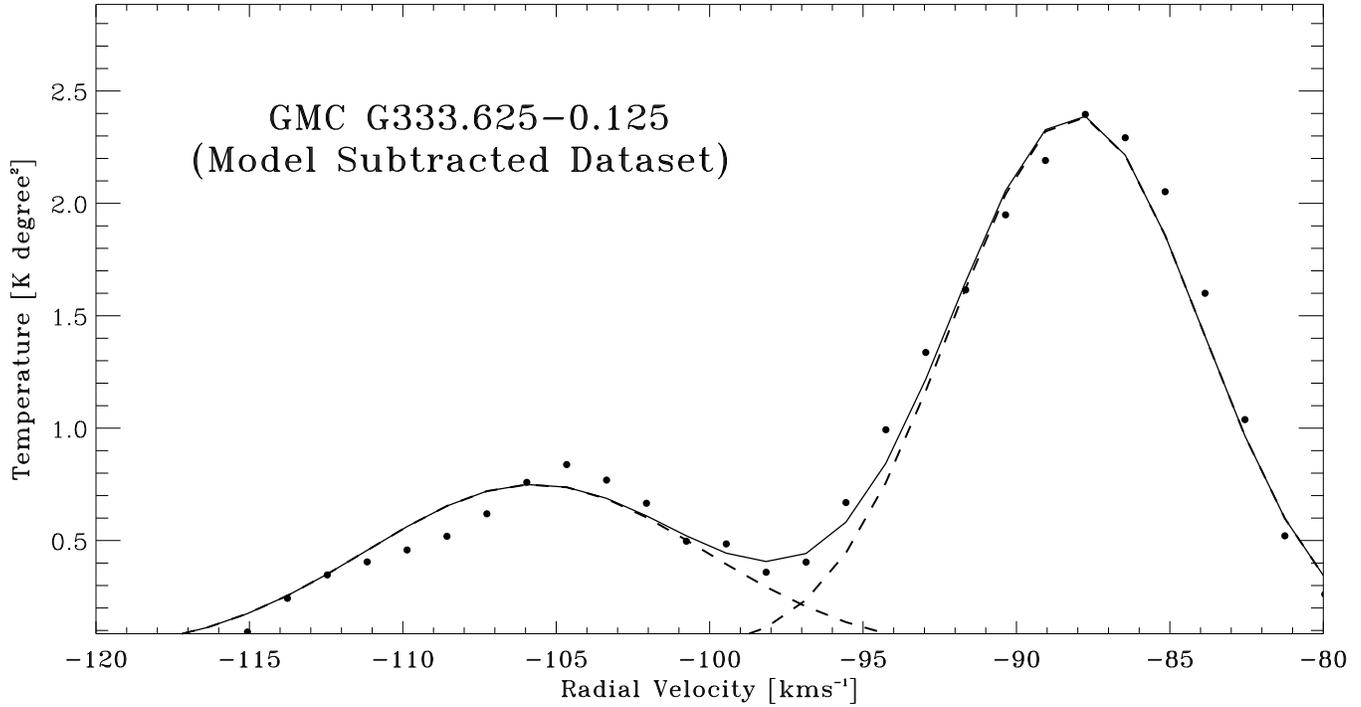}
\caption{Composite spectrum for GMC G333.625$-$0.125. Filled circles represent the antenna temperatures in the composite spectrum, after subtraction of the axisymmetric model (ASM) of the CO \back emission from the original dataset. The solid line shows the results of the two-component Gaussian fit to the composite spectrum. The dashed lines show each individual Gaussian obtained in the fit procedure.\label{fig3_tvspect}}  
\end{center}
\end{figure}

\begin{figure}
\begin{center}
\includegraphics[angle=90,width=\textwidth]{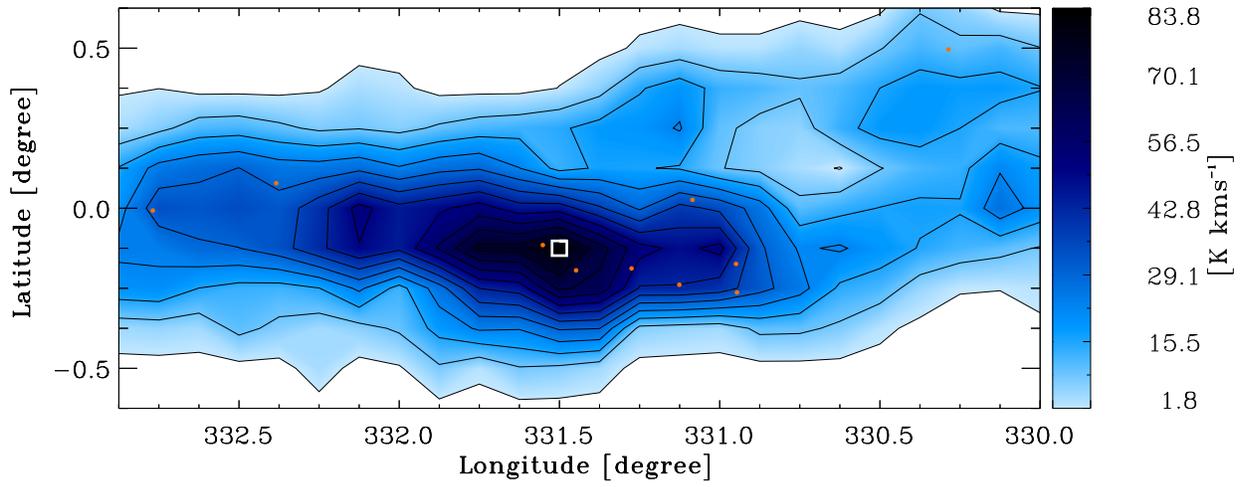}
\caption{\footnotesize{Spatial map for GMC G331.500$-$0.125. The bluish scale denotes values of CO intensity $\int T_{A} dv$. Orange dots denote the location of the \IRAS sources associated to this cloud in the present work. The white square shows the center of the cloud based on the CO peak intensity. All the intrinsic physical parameters for each cloud are derived from the information contained within the spatial map and composite spectrum of the cloud.\label{fig4_3}}} 
\end{center}
\end{figure}

\begin{figure}
\begin{center}
\includegraphics[angle=90,width=\textwidth]{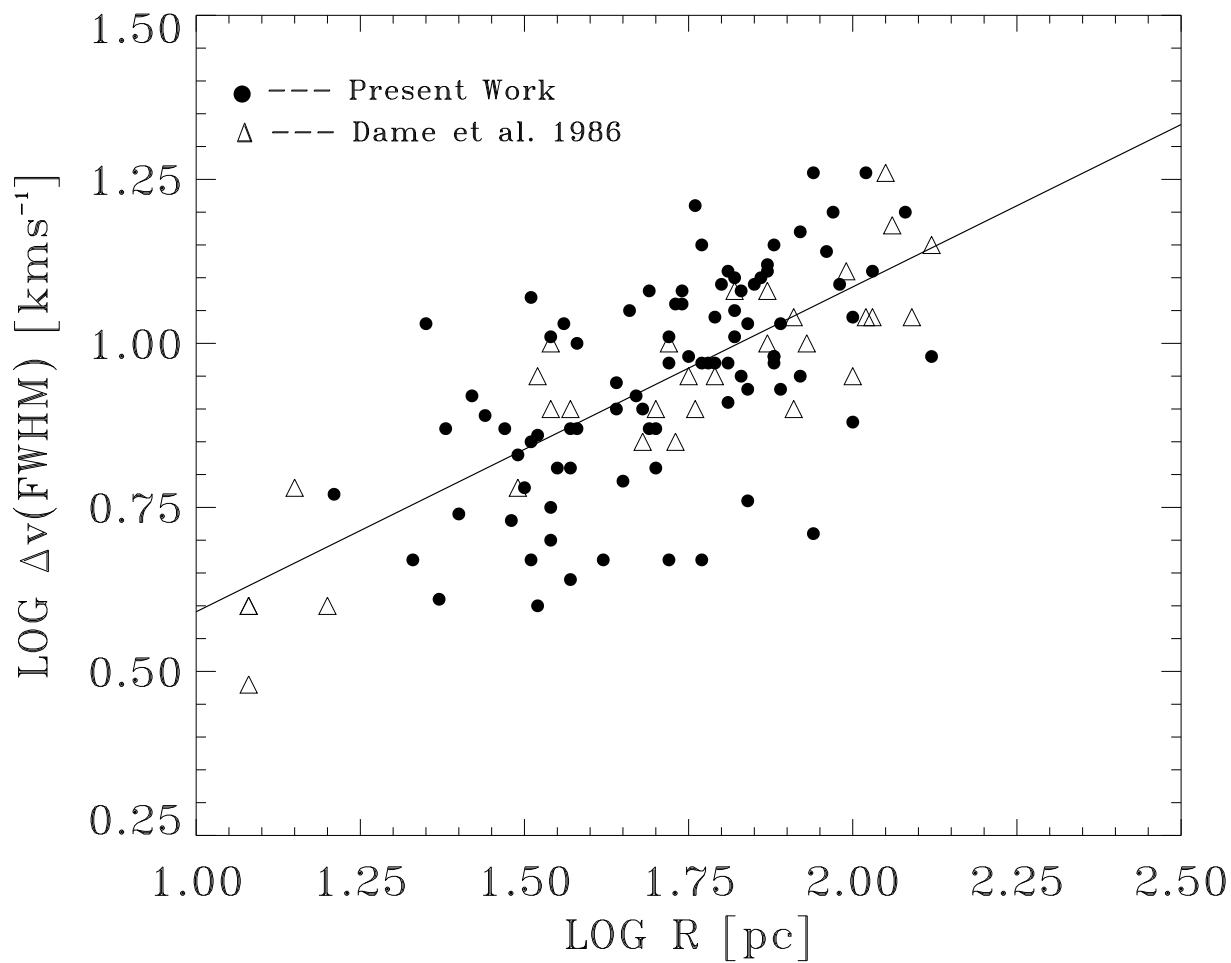}
\caption{Logarithm of the observed linewidth \dv vs. the logarithm of the effective physical radius $R$ of GMCs in Table \ref{tbl-2} (filled circles) and for GMCs in the catalog of \citet{dame86} (open triangles). The straight line is a least-squares fit to the clouds in our catalog given by the equation $\log$ \dv = 0.10 + 0.50$\log R$.\label{fig5_dv}} 
\end{center}
\end{figure}

\begin{figure}
\begin{center}
\includegraphics[angle=90,width=0.9\textwidth]{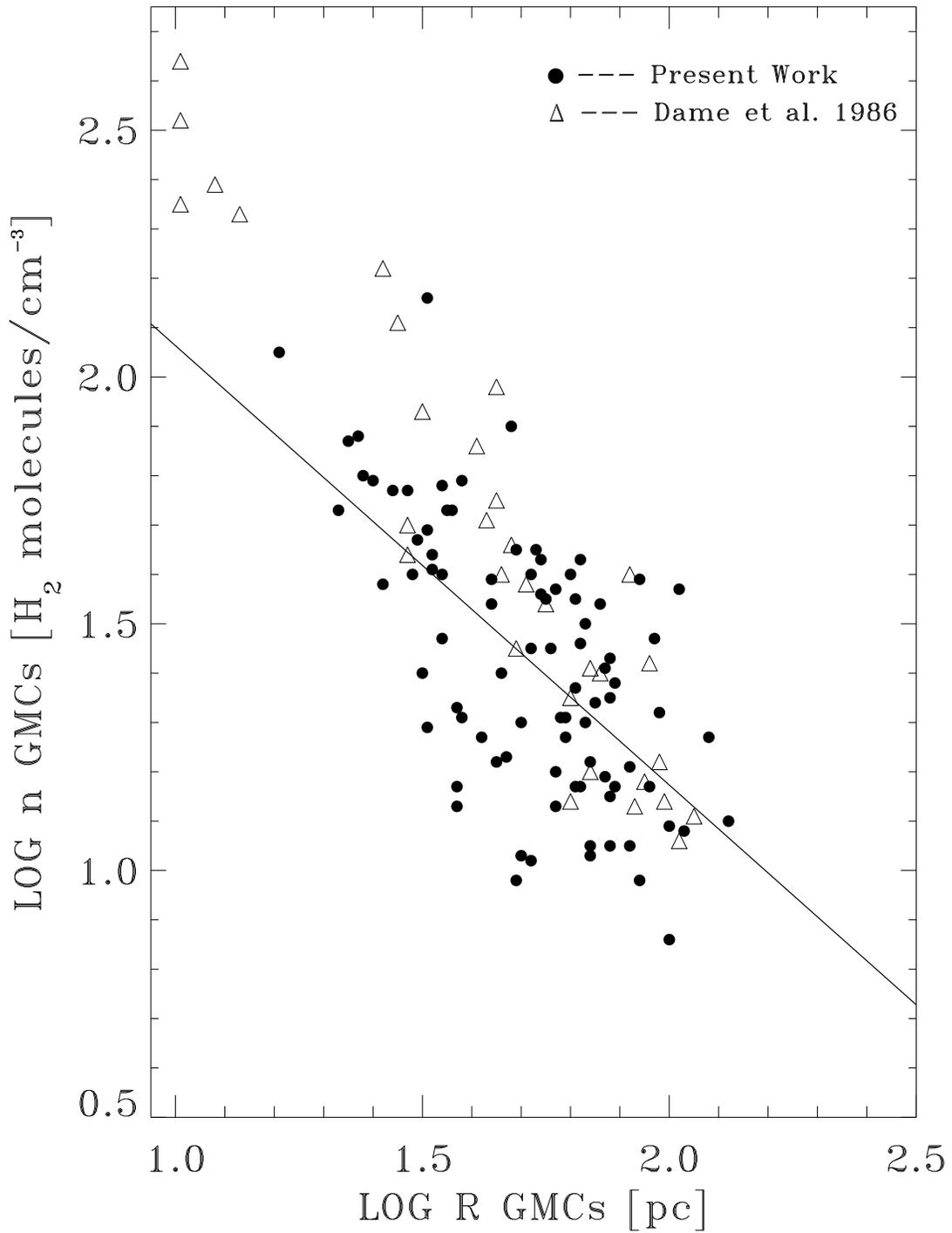}
\caption{\footnotesize{Logarithm of the \HH volume density corrected for helium \nh vs. the logarithm of the effective physical radius $R$ for GMCs in Table \ref{tbl-2} (filled circles) and for GMCs in the catalog of \citet{dame86} (open triangles). The straight line is a least-squares fit given by the equation $\log$ \nh = 2.95 $-$ 0.89$\log R$.\label{fig6_density}}} 
\end{center}
\end{figure}

\begin{figure}
\begin{center}
\includegraphics[angle=90,width=\textwidth]{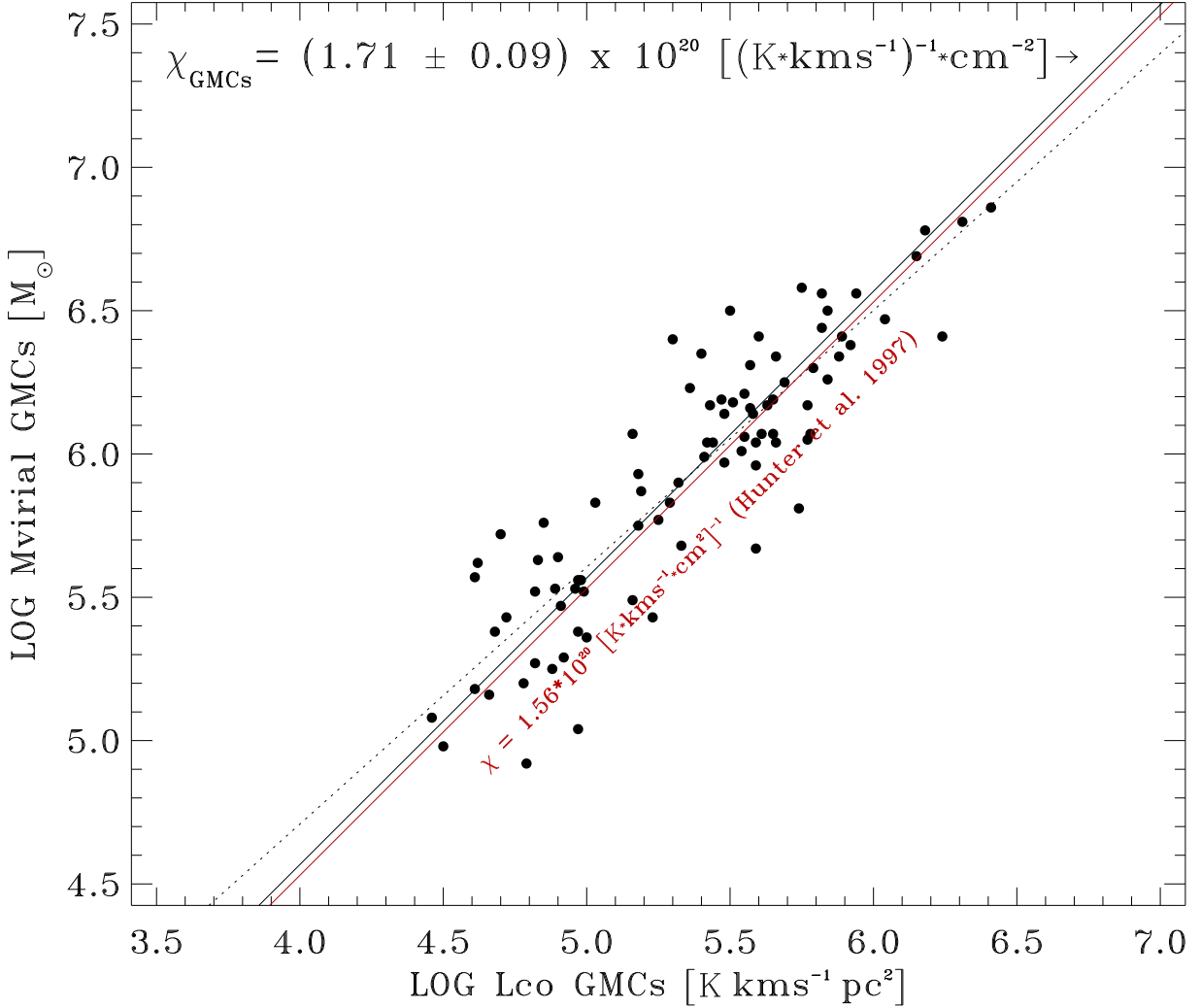}
\caption{Logarithm of the virial mass vs. the logarithm of the CO luminosity for GMCs in Table \ref{tbl-2}. The dotted straight line is a least-squares fit given by the equation $\log M_{virial}$ = 1.11 + 0.90$\log L_{CO}$. The solid black straight line is a least-squares fit given by the equation $\log M_{virial}$ = 0.57 $+$ $\log L_{CO}$. The red straight line represents the \COH factor utilized in the present work \citep{hunter97}.\label{fig7_test_back_dvfix}}  
\end{center}
\end{figure}

\begin{figure}
\epsscale{1.0}
\begin{center}
\includegraphics[angle=90,width=\textwidth]{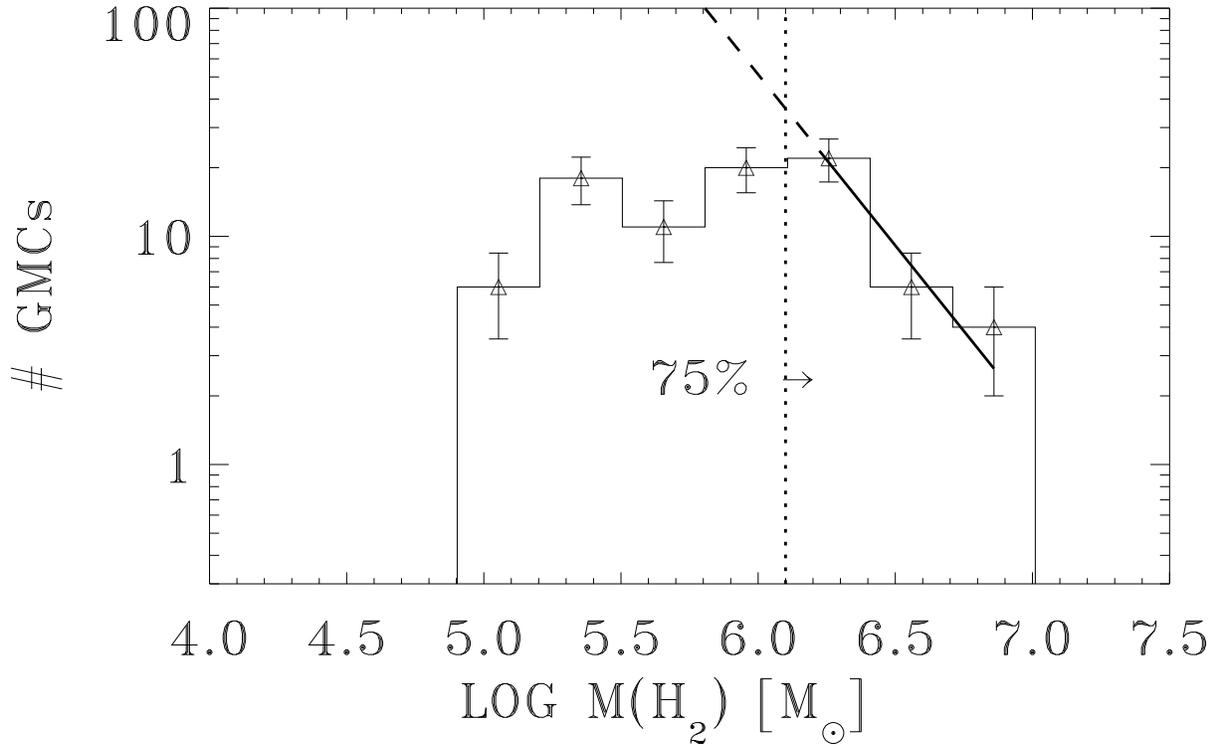}
\caption{Mass spectrum for GMCs in Table \ref{tbl-2}. A least-squares fit has been made over the range indicated by the solid line, being the dashed line an extrapolation to the lower mass end of the spectra. The logarithmic mass bin is $\Delta_{log}$ = 0.3. Small triangles represent the central mass in each mass bin and the dotted line represents the 75\% of the total molecular mass contained to the high mass end of the distribution. The slope of the distribution is $\gamma$ = 1.50 $\pm$ 0.40.\label{fig8_sm_and_dvfix}} 
\end{center}
\end{figure}

\begin{figure}
\epsscale{1.0}
\begin{center}
\includegraphics[angle=0,width=0.95\textwidth]{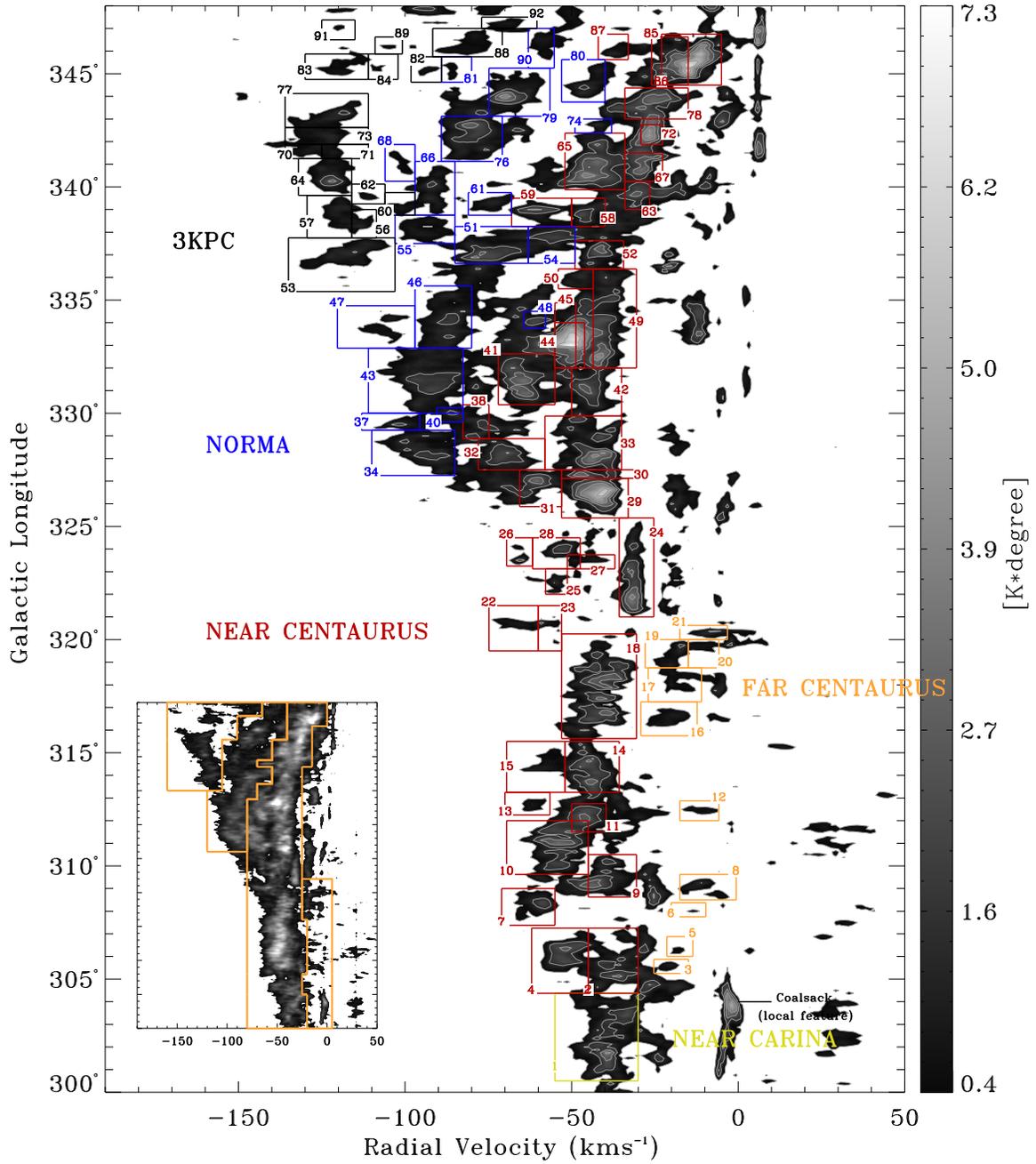}
\caption{\scriptsize{Giant Molecular Clouds in the fourth Galactic quadrant as tracers of the large scale spiral structure in the southern Galaxy. From the plot we identify three spiral arms: \emph{Centaurus} (red clouds tracing the near side and orange clouds tracing the far side);  \emph{Norma} (blue clouds tracing the near and far sides of the arm); and \emph{3-kpc expanding} arm (black box clouds tracing near and far sides of the arm). Based on spatial coincidence, we also identify one cloud (yellow box) belonging to the well known \emph{Carina} spiral arm \citep{grabelsky87}. Based on the distribution of GMCs in the longitude-velocity diagram, a tentative picture of the limits in CO radial velocity and Galactic longitude of the spiral features is presented in the insert on the left lower corner of the figure. The limits are plotted over the CO data of the Columbia Survey.}\label{fig9}} 
\end{center}
\end{figure}

\begin{figure}
\epsscale{1.0}
\begin{center}
\includegraphics[angle=0,width=0.95\textwidth]{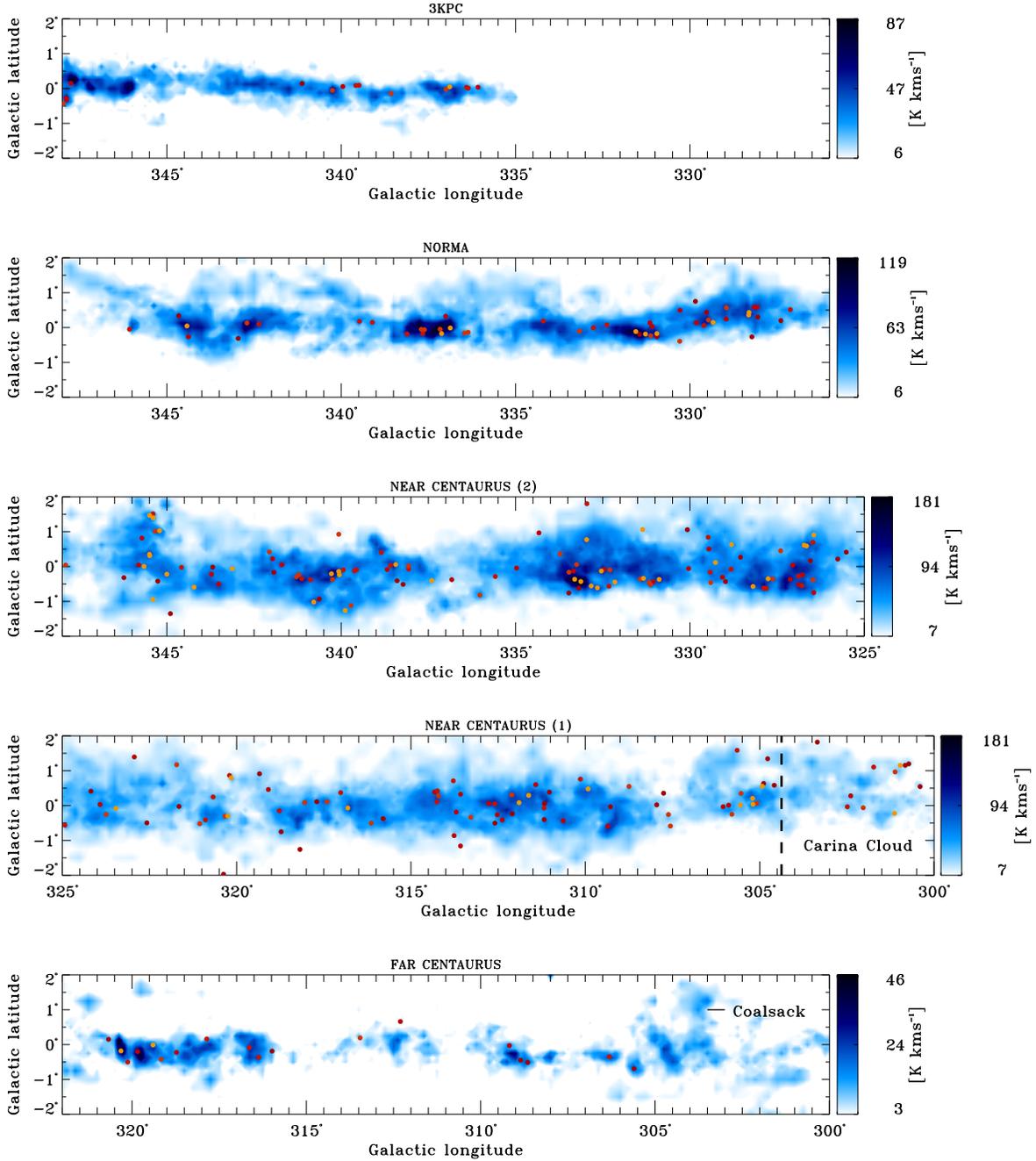}
\caption{\footnotesize{Spatial edge-on maps of the Galactic spiral arms obtained by integrating the CO data of the Columbia Survey across the corresponding velocity range presented in Figure \ref{fig9} for each arm. Contours denote values of CO intensity $I(l,b) = \int T_{A} dv$. Each map has its own intensity color scale (except for the two maps showing the near side of the \emph{Centaurus} arm, which share a common color scale), being the lowest intensity at 7$\sigma$ of the corresponding map, where $\sigma$ is the characteristic intensity noise of the map. The Spatial distribution of 284 IRAS/CS sources utilized in the present work along the spiral arms is also shown as superimposed filled circles.}\label{fig10}}
\end{center}
\end{figure}

\begin{figure}
\begin{center}
\includegraphics[angle=90,width=\textwidth]{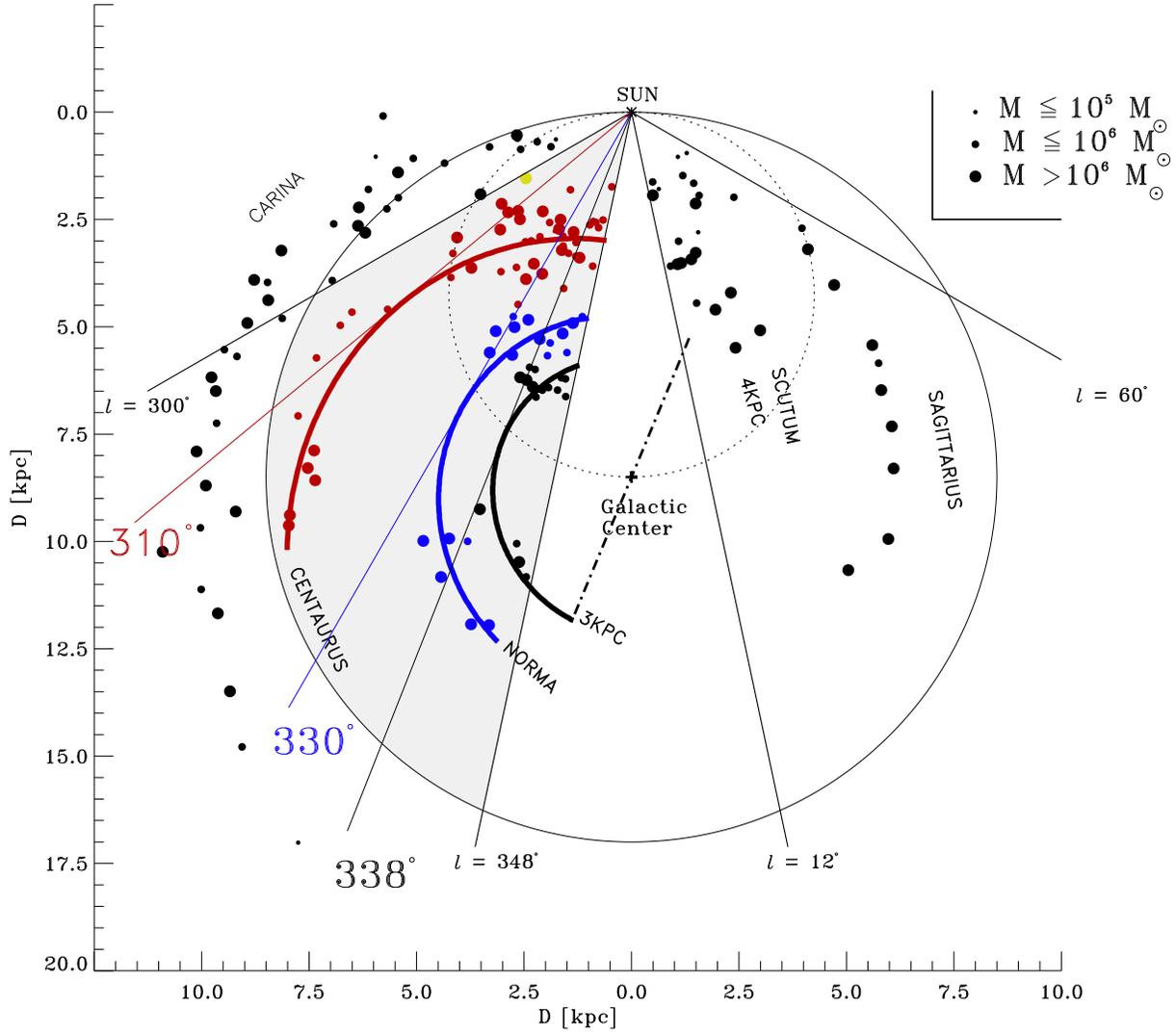}
\caption{\footnotesize{Spatial distribution (face-on view) of giant molecular clouds in the first and fourth Galactic quadrants. In the fourth Galactic quadrant, the 87 molecular complexes from Table \ref{tbl-2} are drawn within the covered area in this work (gray filled area between $l =$ 300\deg\textrm{ }and $l =$ 348\deg), and are associated by colors to their corresponding spiral arm, as explained in Figure \ref{fig9}. The size of a circle is related to the molecular mass of the cloud. Toward lower Galactic longitudes, the molecular complexes tracing the \emph{Carina} arm plotted as black filled circles are from \citet{grabelsky87}. In the first Galactic quadrant, between $l =$ 12\deg\textrm{ }and $l =$ 60\deg, the molecular complexes plotted as filled black circles are from \citet{dame86}. The dotted large circle represents the tangent region within the solar circle, and the dashed-dotted straight line represents the position of the Galactic bar taken from \citet{englmaier99}. The parameters for the three fitted spiral arms (seen as thick color lines) in our catalog are summarized in Table \ref{tbl-4}. The fit was done weighting each point by its Galactocentric radius error.}\label{fig11}}
\end{center}
\end{figure}

There is a clear correlation between the molecular gas and the \IRAS sources, in particular, toward the position of spiral arms traced by GMCs. The \IRAS sources utilized in the present work are plotted over the CO emission in the longitude-velocity diagram from the Columbia Survey in Figure \ref{fig1}. The reddish color scale represents the FIR fluxes of the UC HII regions. \IRAS sources with positive velocities were excluded from the present work since they are located outside the solar circle. From Figure \ref{fig1} the correlation between the gas and the position of the \IRAS sources results evident and, in particular, their concentration to the position of the spiral features, such as the prominent \emph{Centaurus} arm (near side from 305\deg\textrm{ }to 348\deg, and from $-$70 \kms\textrm{ }to $-$30 \kms, and far side from 35\deg\textrm{ }to 321\deg, and from $-$20 \kms\textrm{ }to 0 \kms). The 284 \IRAS sources are plotted in different velocity ranges, corresponding to spiral arms in the Southern Galaxy, in Figure \ref{fig10}. We see that most of the sites of massive star formation are well concentrated within $\pm$ 1\deg\textrm{ }of the Galactic plane and are well correlated with giant molecular clouds tracing the spiral arms. The \emph{3-kpc expanding} arm contains around 4\% of the UC HII regions in the present work. Concerning the \emph{Norma} arm, two prominent regions of massive star formation appear at 331\deg.5\textrm{ }and between 336\deg\textrm{ }and 338\deg. This spiral arm contains around 14\% of the \IRAS sources. Since the near side of the \emph{Centaurus} arm is the largest spiral segment in the longitude-velocity diagram, it contains most of the UC HII regions in our sample (57\% of the \IRAS sources). Important regions of massive star formation in this spiral arm appear between 310\deg\textrm{ }and 313\deg, at 331\deg, and between 332\deg\textrm{ }and 334\deg. The \emph{far} side of the \emph{Centaurus} arm contains around 4\% of the \IRAS sources. Although these sources have small FIR fluxes, they are very luminous since they are located at the far distance of the spiral arm.\\ 
      
The first step in estimating the massive star formation rate (MSFR) of the clouds in our catalog is related to the association of each \IRAS source to its parent GMC. The association between UC HII regions and GMCs is established using the following criteria: (a) the Galactic coordinates $(b,l)$ of the \IRAS source fall inside the spatial range defined for each GMC, and (b) the CS radial velocity of the \IRAS source falls within the 3$\times\sigma_{v}$ velocity range, where $\sigma_{v}$ is the velocity dispersion of the parent GMC. We associated 214 sources ($\sim$ 75\% of the total 284 \IRAS sources within the covered area in this work) with their parent GMCs. The remainder 49 \IRAS sources are either associated with less bright GMCs not considered in our catalog, or have peculiar velocities for their parental GMCs. Since these sources represent only $\sim$ 20\% of the total number of \IRAS sources, we consider the remaining 75\% as a fair representation of the massive star formation within giant molecular clouds in our catalog. An example of associated \IRAS sources to their parent GMC is found in Figure \ref{fig4_3}.\\

The FIR luminosity ($L_{FIR}$) of embedded stars in giant molecular clouds is summarized in Table \ref{tbl-5}. The first column represents the GMCs identification ($clouds$). The second column shows the number of UC HII associated to the cloud. The third column contains the total Far-Infrared flux $F_{IRAS}$ of the massive star forming regions associated to each giant molecular cloud. The FIR flux of each UC HII region was derived directly from the fluxes reported in the four bands of the IRAS point-like catalog (version 1) as follows:\\
  
\begin{equation}\label{eq16}
F_{IRAS} = 4\pi \sum_{\nu} \nu F_{\nu}.
\end{equation} 
The fourth column contains the Far-Infrared luminosity $L_{FIR}$ derived for each GMC as $L_{FIR} = F_{IRAS} D^{2}$ where $F_{IRAS}$ is the total Far-Infrared flux for each cloud and $D$ is its distance. The fifth column contains the massive star formation efficiency $\epsilon$ of GMCs described in section  \ref{effic}. In our sample, the total molecular mass of the GMCs which harbor at least one massive star formation region is $\sim$ 9.7 $\times$ 10$^{7}$ M$\solar$, which accounts for 85\% of the total molecular mass in form of GMC and from the 60 GMCs in Table \ref{tbl-5} with at least one UC HII region associated, 36 of them have molecular masses larger than 10$^{6}$ M$\solar$. In general, the most massive clouds exhibit most of the massive star formation in the Galactic disk. The total FIR luminosity in Table \ref{tbl-5} emitted by the UC HII regions associated with giant molecular clouds is 3.66 $\times$ 10$^{7}$ L$\solar$. From Table \ref{tbl-5}, the \emph{Norma} spiral arm concentrates most of the massive star formation in the Southern Galaxy. The eleven giant molecular clouds in this arm account for a total FIR luminosity of 1.85 $\times$ 10$^{7}$ L$\solar$, equivalent to the 50\% of the total FIR luminosity contained in GMCs. The GMC G337.750$+$0.000 (number 54 in Table \ref{tbl-2}) belonging to the far side of this arm is the most prominent massive star forming region in the inner fourth Galactic quadrant, with a FIR luminosity of 6.45 $\times$ 10$^{6}$ L$\solar$, around 35\% of the total FIR luminosity in the \emph{Norma} spiral arm.\\ 

%%%%%%%%%%%%%%%%%%%%%%%%%%%%%%%%%%%%%%%%%%%%%%%%%%%%%%%%%%%%%%%%%%%
\subsection{Massive Star Formation Rate Per Unit \HH Mass}

\begin{figure}
\begin{center}
\includegraphics[angle=90,height=0.6\textheight]{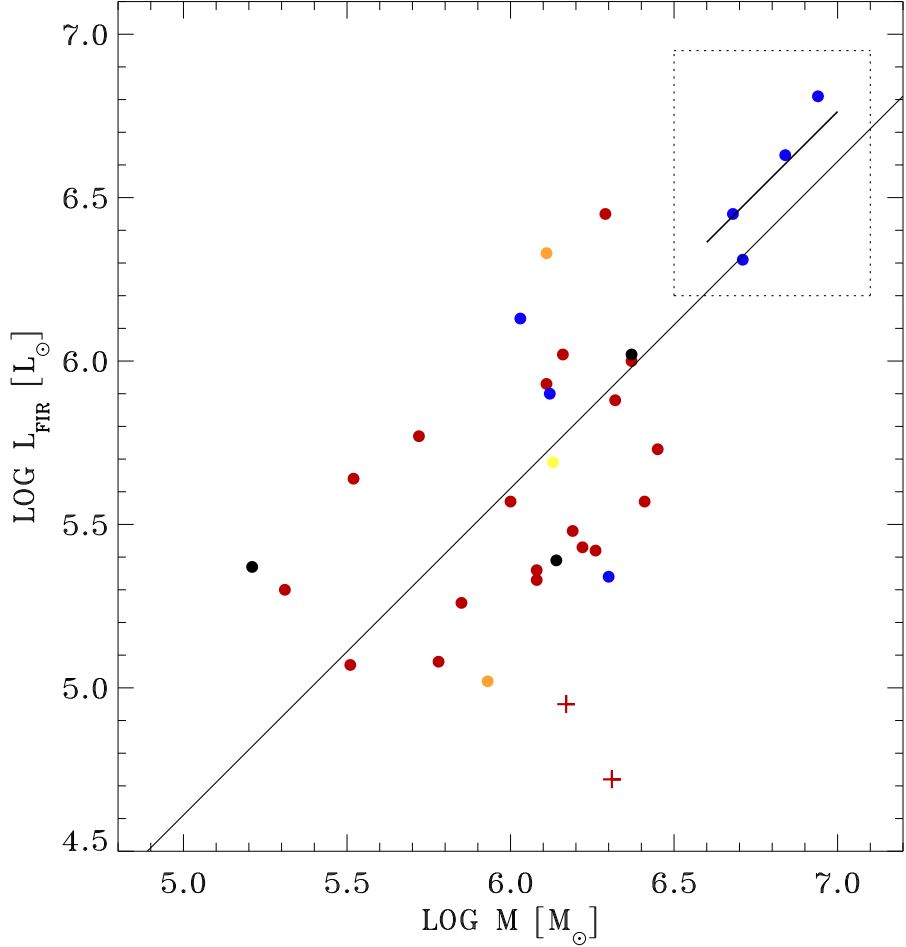}
\caption{Logarithm of the Far-Infrared luminosity in Table \ref{tbl-5} vs. molecular mass of GMCs. Only clouds with more than one \IRAS source associated are included in the figure. The straight line is a least-squares fit given by $\log L_{FIR} = -$0.39 + $\log$ $M(H_{2})$. Colors are associated to spiral arms features as explained in Figure \ref{fig9}. Red crosses are GMCs 9 and 14 in Table \ref{tbl-2} belonging to the \emph{Centaurus} spiral arm, and were not considered in the fit. The straight line in the upper right square is a least-squares fit given by $\log L_{FIR} = -$0.24 + $\log$ \MHH to the four most massive GMCs in the figure. We notice that the most intense massive star formation activity takes place in the \emph{Norma} spiral arm (blue filled circles).\label{fig12}} 
\end{center}
\end{figure}

We calculate the massive star formation rate per unit \HH mass defined as MSFR $= L_{FIR}/M$ for GMCs in our catalog, following \citet{bron00}. In Figure \ref{fig12}, the observational relationship between the FIR luminosity in Table \ref{tbl-5} and molecular masses for GMCs in Table \ref{tbl-2} is shown. For the plot, we used only FIR luminosities of GMCs from Table \ref{tbl-5} that have more than one UC HII region associated to reduce the statistical error of the relationship. The colors of the filled circles are related to the spiral arms in Figure \ref{fig9}.\\

The two red crosses in the plot are the GMC G309.125$-$0.375 and GMC G313.875$-$0.125 belonging to the \emph{Centaurus} spiral arm were left out of the fit in Figure \ref{fig12} due to their deviation from the general trend. For GMC G309.125$-$0.375 there are neither other \IRAS sources in (l,b,v) space close to the clouds nor non-detections in the CS data, so the low FIR luminosity seems to be real and not an artifact of the criteria used in associating UC HII regions to GMCs. For GMC G313.875$-$0.125, there are two \IRAS sources close to the cloud in (l,b,v) space, but do not fulfill the association criteria. The total IRAS FIR flux of the sources is 3293 L$\solar$ kpc$^{-2}$, which would increase $L_{FIR}$ by a factor 1.8. Nonetheless, this increase would not conciliate the position of the cloud within the general trend in Figure \ref{fig9}. Besides, no non-detections in the CS data are present. A possible reason for the deviation of the clouds from the trend in Figure \ref{fig9} could be related to some \textrm{``missing''} UC HII regions during the process of selecting them by FIR colors in the IRAS catalog. \\ 

The massive star formation rate is obtained from a least-squares fit to the proportionality relationship between the FIR luminosity and molecular mass of GMCs which yields $L_{FIR} =$ 0.41 $\pm$ 0.07 $M(H_{2})$. Additionally, a least-squares fit to the four most massive (\MHH $>$ 3 $\times$ 10$^{6}$ M$\solar$) and most active ($L_{FIR}$ $>$ 2 $\times$ 10$^{6}$ L$\solar$) star forming GMCs of the sample, GMC G328.250$+$0.375, GMC G331.500$-$0.125, GMC G336.875$+$0.125, and GMC G337.750$+$0.000 belonging to the \emph{Norma} spiral arm (blue filled circles in the region enclosed by the dotted lines) was performed yielding $L_{FIR} =$ 0.58 $\pm$ 0.09 $M(H_{2})$.\\

The clouds belonging to the far side of the \emph{Norma} arm are, within our catalog, the most massive and with the highest massive star formation rate in the Southern Galaxy. \citet{bron00} showed that the mean massive star formation rate per unit \HH mass is higher in the Southern Galaxy than in the Northern Galaxy, being around $\sim$ 0.21 L$\solar$/M$\solar$ for 0.2 $\leq$ R/R$\solar \leq $ 1; around $\sim$ 0.28 L$\solar$/M$\solar$ for 0.4 $\leq$ R/R$\solar \leq $ 0.8; and with a peak around $\sim$ 0.41 L$\solar$/M$\solar$ for 0.5 $\leq$ R/R$\solar \leq $ 0.6 (without correction for helium abundance). They suggested that such an enhancement of massive star formation appears to be dominated by the presence of the \emph{Norma} spiral arm, where a large fraction of the most FIR luminous sources are found.\\

In order to compare our results with those in \citet{bron00}, we first divide their estimates by $m(H_{2})^{\ast}/m(H_{2})=$ 2.72$m(H)/$2$m(H) =$ 1.36, where $m(H_{2})^{\ast}$ is the \HH molecular mass corrected for helium abundance utilized in this work. The clouds in Table \ref{tbl-5} are located between 0.37 $\leq$ R/R$\solar \leq $ 0.94. Their MSFR is about a factor two higher than one derived by \citet{bron00} for almost the same region of the Galactic disk ($\sim$ 0.21 L$\solar$/M$\solar$ for 0.4 $\leq$ R/R$\solar \leq $ 0.8). The four most massive GMCs in the sample are located between 0.55 $\leq$ R/R$\solar \leq $ 0.60 in Galactocentric radius and coincide with the position of the MSFR peak found by \citet{bron00}. Their MSFR is also about a factor of two higher than the one found by  \citet{bron00} at the peak of the distribution of embedded stars $\sim$ 0.30 L$\solar$/M$\solar$ between Galactocentric radii 0.5 $\leq$ R/R$\solar \leq $ 0.6. This shows that most massive star formation activity in the southern Galaxy not only occurs in the \emph{Norma} spiral arm, but in its most massive clouds. It is interesting to notice that, the factor two difference between the MSFRs in GMCs found here and by  \citet{bron00} for the average and peak values, is the same difference we have between the molecular mass in form of GMCs here and the axisymmetric model molecular mass estimated by \citet{bron88}. According to this evidence, it appears reasonable to use the massive star formation rate per unit \HH mass of the most massive clouds in the \emph{Norma} spiral arm as a standard scale to compare the massive star formation activity in the rest of the Galaxy.\\

The massive star formation rates estimated here and in \citet{bron00} are a strict lower limit because only UC HII regions are contributing to it. Since the diffuse FIR emission from dust contained in the clouds is not considered in the IRAS point-like source catalog, and some embedded OB star are probably illuminating dust regions with large angular sizes, escaping from the identification as ultra-compact sources, the total FIR flux derived for a particular cloud is a lower limit to the real value, giving a lower limit for the MSFR.\\

%%%%%%%%%%%%%%%%%%%%%%%%%%%%%%%%%%%%%%%%%%%%%%%%%%%%%%%%%%%%%%%%%%%
\subsection{Massive Star Formation Efficiency $\epsilon$ \label{effic}}

The values for the massive star formation efficiency of the GMCs in our catalog are given in the last column of Table \ref{tbl-5}. To derive the massive star formation efficiency $\epsilon$ we first transform the FIR luminosity of the clouds into the corresponding massive star formation rate $\dot{M}$ in units of M$\solar$yr$^{-1}$. Following \citet{luna06} we convert the FIR-Infrared luminosity of the clouds as follows:\\
  
\begin{equation}\label{eq17}
  \dot{M} = 6.5 \times 10^{-10} \left( \frac{L_{FIR}}{L_{\odot}} \right) \textrm{ } M_{\odot} yr^{-1} .
\end{equation}
Using the $\dot{M}$ computed in equation \ref{eq17}, the massive star formation efficiency for GMCs in our catalog can be determined as:\\

\begin{equation}\label{eq18}
\epsilon =  \tau_{OB} \frac{ \dot{M}}{M_{GMC}},
\end{equation}
where $\tau_{OB} = $ 10$^{8}$ yr, is assumed to be a typical lifetime for an OB star region \citep[and references therein]{luna06}. The estimated massive star formation efficiencies shown in Table \ref{tbl-5} are derived for the molecular mass of clouds in our catalog. All efficiencies derived here are of a few percents of the available molecular mass of the clouds, and the average value is 3\%. This is consistent with the values found in the literature \citep{maclow03,luna06,zinnecker07,mckee07}. Our estimates reinforce the common idea that GMCs  are inefficient in forming massive stars, implying that processes other than simple gravitational collapse must control the MSFR in the clouds.\\

%%%%%%%%%%%%%%%%%%%%%%%%%%%%%%%%%%%%%%%%%%%%%%%%%%%%%%%%%%%%%%%%%%%%%
%%%%%%%%%%%%%%%%%%%%%%%%%%%%%%%%%%%%%%%%%%%%%%%%%%%%%%%%%%%%%%%%%%%%%
%%%%%%%%%%%%%%%%%%%%%%%%%%%%%%%%%%%%%%%%%%%%%%%%%%%%%%%%%%%%%%%%%%%%%
%%%%%%%%%%%%%%%%%%%%%%%%%%%%%%%%%%%%%%%%%%%%%%%%%%%%%%%%%%%%%%%%%%%%%

\section{SUMMARY\label{concl}}

Using the Columbia - U. de Chile CO Survey of the Southern Milky Way we present the first catalog of GMCs in the fourth Galactic quadrant, within the solar circle. Their average size, linewidth, and mass are estimated and their statistical properties are determined.\\  

\textit{GMCs Identification and Physical Properties}\\
We identify 92 molecular complexes in the inner Southern Galaxy and remove the two-fold distance ambiguity for 87 of them. The total molecular mass of GMCs is \MHH $=$ 1.14 $\pm$ 0.05 $\times$ 10$^{8}$ M$\solar$. The evidence in our catalog from the observational size-to-linewidth (\dv $ \propto R^{\alpha}$) and size-to-\HH volume density (\nh $ \propto R^{-\beta}$)  relationships suggests that GMCs are close to virial equilibrium ($\alpha + \beta/2 =$ 0.95). The mass spectrum of the clouds (slope $\gamma =$ 1.50 $\pm$ 0.40) shows that most of the molecular mass in concentrated to the high mass end of the distribution. From the total molecular mass in our sample, 60\% is located at the near distance while the remaining 40\% is at the far distance. The completeness limit of the molecular mass spectrum is at 75\% of the total mass.\\ 

\textit{Total GMCs Mass and Axisymmetric Model}\\
The total molecular mass of GMCs in our catalog accounts for 40\% of the molecular mass derived from the axisymmetric analysis of the \HH volume density in the fourth Galactic quadrant, while in previous studies it was only 20\%. From the virial analysis, we estimate an average \COH conversion factor GMCs $\chi_{GMCs} = 1.71 \pm 0.09 \times 10^{20}$ [K \kms]$^{-1}$ cm$^{-2}$ which is consistent with the Galactic average value $\chi = 1.56 \pm 0.05 \times 10^{20}$ [K \kms]$^{-1}$ cm$^{-2}$ \citep{hunter97}. The average Galactic value of the conversion factor in \citet{hunter97} is estimated for both, cloud and inter-cloud medium simultaneously. A higher conversion factor for GMCs and a lower conversion factor for the inter-cloud medium could also reproduce the estimated average Galactic value and increase the molecular mass of GMCs reducing the difference even further.\\ 

\textit{Large Scale Spiral Structure}\\
We trace here three large scale spiral arms within the solar circle: the \emph{Centaurus}, \emph{Norma}, and \emph{3-kpc expanding} arms. The  \emph{Carina} arm is not traced because of its wide latitude. After fitting a logarithmic spiral arm model for each of the arms, tangent directions are found to be at roughly 310\deg, 330\deg, and 338\deg\textrm{ }respectively. These values are consistent with previous determinations of the tangent directions done by \citet{alvarez90} and \citet{bron92}. The \emph{Centaurus} arm is the most prominent feature in the southern Galaxy. Its pitch angle $p =$ 13\deg.4\textrm{ }is large compare to the pitch angles of the \emph{Norma} ($p =$ 6\deg.6) and \emph{3-kpc} ($p =$ 5\deg.6) arms. The spatial maps of the arms show that the molecular gas contained in GMCs and UC HII regions (\IRAS sources) are tightly correlated, being the \emph{Norma} spiral arm the feature with the most massive clouds in our sample. The positions of the far clouds in the \emph{3-kpc expanding} arm converge to the location of the far side of the arm traced in \citet{dame2008} down to $l = $ 348\deg.\\
 
\textit{Massive Star Formation Rate for GMCs}\\
Massive stars form in GMCs. Making use of the \CS survey \citep{bron96} toward IRAS point-like sources with FIR colors characteristic of UC HII regions \citep{wood89b} and a new unpublished \CS survey, we analyzed massive star formation GMCs. The total molecular mass of GMCs with massive star formation activity accounts for 85\% of the total molecular mass in our sample, and 60\% of them have masses larger than 6 $\times$ 10$^{6}$ M$\solar$. From the spiral features identified in our catalog, the \emph{Norma} spiral arm is the most active in forming massive stars. The GMCs in this arm have a total FIR luminosity of 1.85 $\times$ 10$^{7}$ L$\solar$, equivalent to 50\% of the total FIR luminosity in our sample. The GMC G337.750$+$0.000 is the most prominent massive star forming region in the inner southern Galaxy, with a FIR luminosity of 6.45 $\times$ 10$^{6}$ L$\solar$, equivalent to 35\% of the total FIR luminosity in the arm. The observational evidence is consistent with the idea that, in general, the most massive clouds exhibit most of the massive star formation in the Galactic disk.\\

The massive star formation rate per unit \HH mass (MSFR) and the massive star formation efficiency $\epsilon$ are estimated for GMCs in our catalog. From the molecular mass of GMCs in our sample, the massive star formation rate is MSFR $=$ 0.41 $\pm$ 0.06 L$\solar$/M$\solar$. A separate estimation for the most massive and most active massive star forming GMCs in our sample, GMC G328.250$+$0.375, GMC G331.500$-$0.125, GMC G336.875$+$0.125, and GMC G337.750$+$0.000 belonging to the \emph{Norma} spiral arm, yields MSFR $=$ 0.58 $\pm$ 0.09 L$\solar$/M$\solar$.\\

The MSFR in GMCs is about a factor two higher than the values derived for the Galactic plane $\sim$ 0.21 L$\solar$/M$\solar$ between 0.4 $\leq$ R/R$\solar \leq $ 0.8 and peak value $\sim$ 0.31 L$\solar$/M$\solar$ for 0.5 $\leq$ R/R$\solar \leq $ 0.6 (where part of the \emph{Norma} arm is located) derived by \citet{bron00} after a correction for Helium abundance is applied to their results. The discrepancy could be explained in terms of the molecular mass difference between GMCs in the present catalog and the molecular mass derived from the axisymmetric model analysis done by  \citet{bron88}. We derive an average massive star formation efficiency $\epsilon$ for GMCs of $\sim$ 3\% percent of their available molecular mass.\\

%%%%%%%%%%%%%%%%%%%%%%%%%%%%%%%%%%%%%%%%%%%%%%%%%%%%%%%%%%%%%%%%%%%%%
%%%%%%%%%%%%%%%%%%%%%%%%%%%%%%%%%%%%%%%%%%%%%%%%%%%%%%%%%%%%%%%%%%%%%
%%%%%%%%%%%%%%%%%%%%%%%%%%%%%%%%%%%%%%%%%%%%%%%%%%%%%%%%%%%%%%%%%%%%%
%%%%%%%%%%%%%%%%%%%%%%%%%%%%%%%%%%%%%%%%%%%%%%%%%%%%%%%%%%%%%%%%%%%%%

\acknowledgments
We gratefully acknowledge Prof. Patrick Thaddeus for his seminal contributions to this work. We thank Hector Alvarez and Abraham Luna for their early help with the analysis. We remember Prof. Jorge May (R.I.P.) for his unconditional dedication to this project. P.G. and L.B acknowledge support from Basal Center For Astrophysics And Associated Technologies PFB-06.

\appendix

\section{AXISYMMETRIC MODEL SUBTRACTION}\label{app1}

\subsection{The Axisymmetric Model}

In order to isolate the largest molecular clouds in the fourth Galactic quadrant from the CO \back emission in which they are immersed, similar to the work of \citet{dame86}, an axisymmetric model of CO \back emission was generated and subtracted to the full 3-dimensional $(v,b,l)$ observed CO dataset. We assumed that the \back emission follows the same radial distribution of the molecular mass in the fourth Galactic quadrant determined by \citet{bron88}. Following \citet{dame86} the \back level was adjusted so that 65\% of the non local ($|v| < $ 20 \kms) emission was removed from the survey.\\  

\begin{figure}
\begin{center}
\includegraphics[angle=180,width=0.58\textwidth]{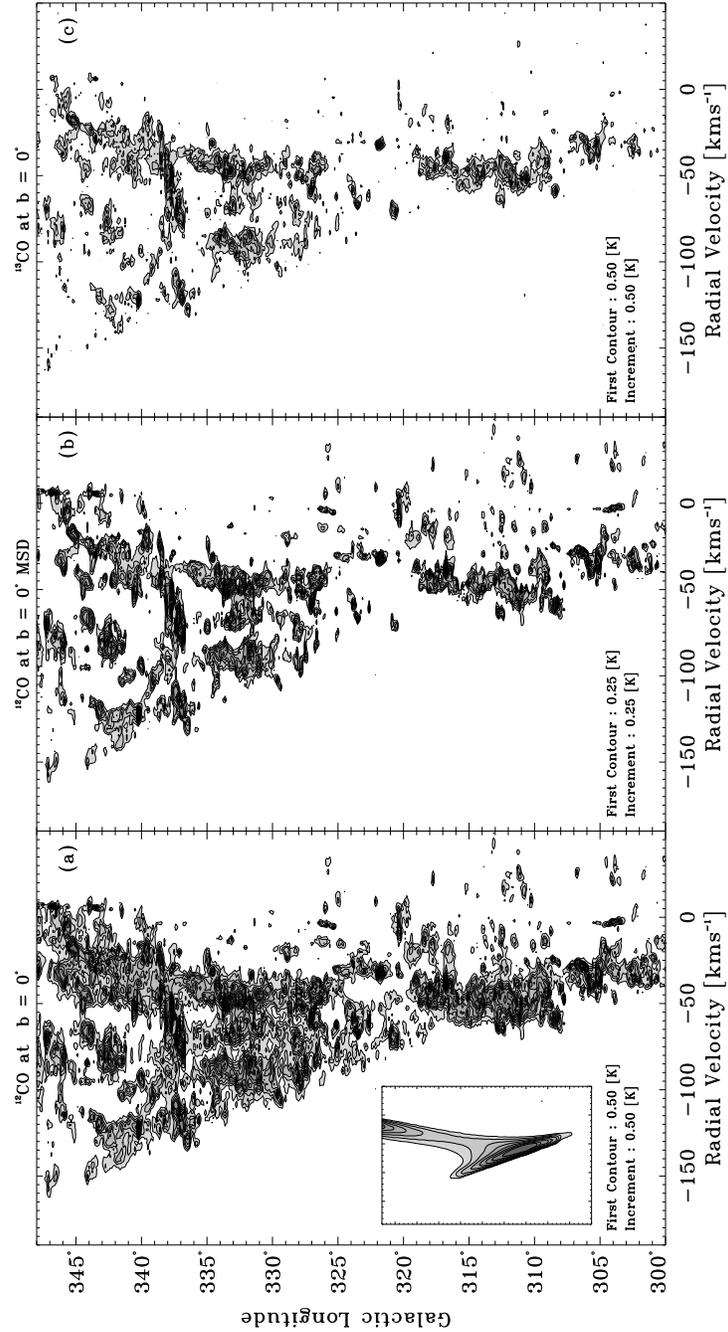}
\caption{\footnotesize{Longitude-velocity diagrams of CO and $^{13}$CO emission at $b =$ 0\deg\textrm{ } in the fourth Galactic quadrant. The panels represent: (a) CO observations the Columbia CO Survey. The insert on the lower left corner shows the axisymmetric model subtracted to the CO data. The first contour in the model insert is at $0.06$ K degree, and contour intervals are at $0.01$ K degree; (b) model subtracted CO emission; and (c) $^{13}$CO Observations of the Galactic plane \citep{bron13co88}. The subtraction of the axisymmetric model from the CO data dramatically improves the similitude between the $^{13}$CO (tracer of higher molecular densities than CO) and the CO data.}\label{fig13}} 
\end{center}
\end{figure}

On large scales, the effect of subtracting the axisymmetric model from the CO dataset can be observed in Figure \ref{fig13}. The figure shows standard longitude-velocity diagrams at Galactic latitude $b = 0^{\circ}$ of CO and $^{13}$CO emission. The different panels represent: (a) CO(1-0) emission from the Columbia CO Survey, (b) CO(1-0) emission of the Columbia CO Survey with the axisymmetric model subtracted, and (c) \CCO emission from \citet{bron13co88}. The insert on the left lower corner in panel (a) represents the axisymmetric model subtracted from the CO data to produce the longitude-velocity diagram in panel (b). The emission of the rare $^{13}$CO isotope is less saturated than the CO line, being a better molecular mass tracer at high densities ($\sim 10^{3}$ cm$^{-3}$), and its narrow line facilitates separation of individual clouds along the velocity axis \citep{dickman78,frerking82,liszt84,liszt95}. From Figure \ref{fig13} results evident that the subtraction of the \back model from the CO dataset dramatically improves the similarity between the $^{13}$CO and $^{12}$CO longitude-velocity diagrams, suggesting that such a subtraction effectively distinguish the clouds from the CO \back in which they are immersed.\\   

\begin{figure}
\begin{center}
\includegraphics[angle=90]{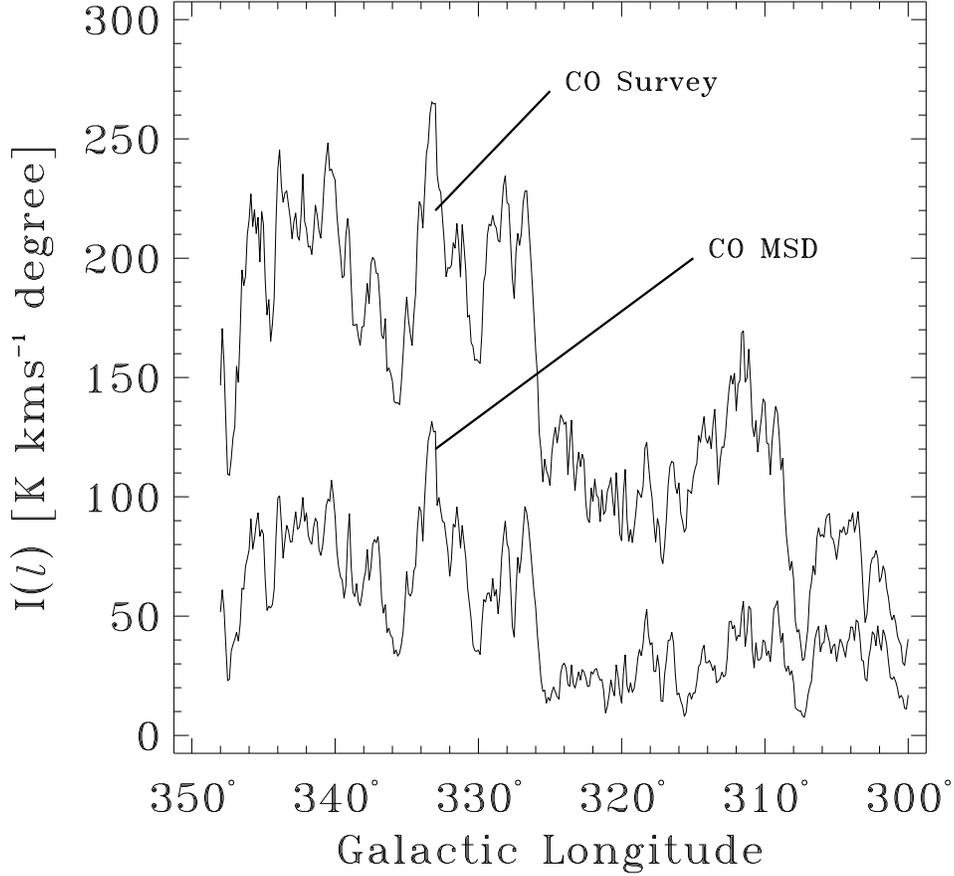}
\caption{Intensity distribution along Galactic longitude ($I(l) = \int T_{A} dvdb$) of the Columbia Survey (CO raw data), and the model subtracted dataset (MSD). Although a large fraction of the CO emission was removed from the observed dataset by the model subtraction (63\% of the non local, $|v| < $ 20 \kms, emission), the intensity structure of the emission is preserved, and features like tangent directions toward spiral arms around 309\deg\textrm{ }(\emph{Centaurus} arm), 328\deg\textrm{ }(\emph{Norma} arm), and 337\deg\textrm{ }(\emph{3-kpc expanding} arm) \citep{bron92} appear in both datasets. \label{fig14}} 
\end{center}
\end{figure}

The large scale features are preserved after the model subtraction. The total CO intensity with no correction for the main beam efficiency ($I = \sum T_{A}(v,l,b)  \Delta v  \Delta b  \Delta  l$, with $ \Delta v = $1.3 \kms, and $ \Delta b  = \Delta  l = $ 0\deg.125) is 6.9 $\times$ 10$^{3}$ K \kms\textrm{ }degree$^{2}$ and 2.4 $\times$ 10$^{3}$ K \kms\textrm{ }degree$^{2}$ for the observed and subtracted model datasets, respectively, indicating that 65\% of the total emission was removed. The subtraction of the model from the observed CO  dataset does not modify the overall shape of the intensity distribution, preserving the large scale features of the emission. In Figure \ref{fig14} the CO intensity distribution as a function of Galactic longitude ($I(l) =  \sum T_{A}(v,l,b)  \Delta v  \Delta b $) is presented for the model subtracted (CO MSD) and the observed (CO Survey) datasets. The tangent directions to spiral arms at 309\deg, 328\deg, and 337\deg\textrm{ }\citep{bron92} are present in both datasets.\\ 

\subsection{Impact of the Model Subtraction on Individual Clouds}
\begin{figure}
\begin{center}
\includegraphics[angle=180]{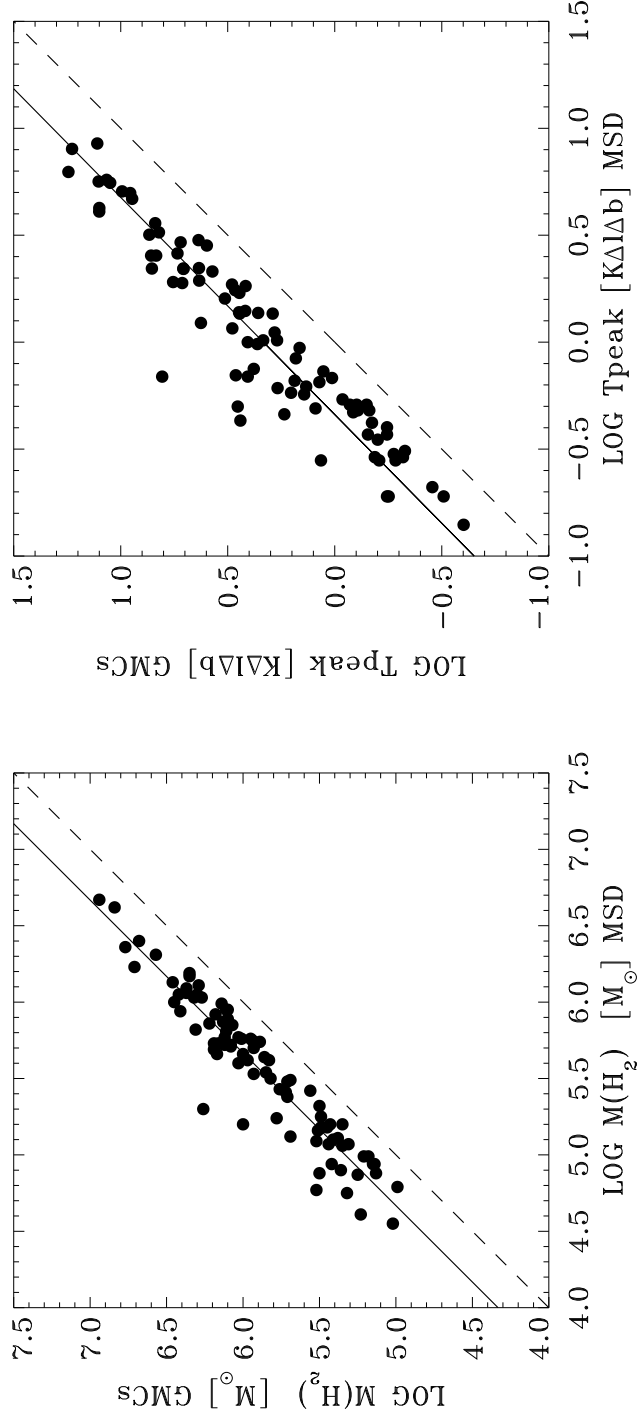}
\caption{\scriptsize{Left panel: molecular mass of GMCs versus  the molecular mass obtained from Gaussian fits in the subtracted model dataset (MSD). The solid straight line represents a proportionality fit of the form $\log$\MHH$_{GMCs}$ = 0.33 +  $\log$\MHH$_{GMCs}$. Right panel: peak antenna temperature of GMCs versus the  peak antenna temperature obtained from Gaussian fits in the MSD. The solid straight line represents a power law fit of the form $\log Tpeak(GMCs)$ = 0.33 +  0.98$\log Tpeak(MSD)$. All values are summarized in Table \ref{tbl-app}. \label{fig15_tpeak_mass}}} 
\end{center}
\end{figure}

We investigate variations in the molecular mass $M(H_{2})$ due to changes in the peak antenna temperature $Tpeak$ from the recovery of the CO flux lost during the subtraction of the \back axisymmetric model from the observed dataset. In the following, we refer to the values calculated in the model subtracted dataset as  ``MSD"  and the values determined for the clouds in our catalog as  ``GMCs." The values of $Tpeak$ and \MHH for both datasets are presented in Table \ref{tbl-app}. The relationship for each quantity between both datasets is shown in Figure \ref{fig15_tpeak_mass}. A least-squares fit between $Tpeak(MSD)$ versus $Tpeak(GMCs)$, and $M(H_{2})$(MSD) versus $M(H_{2})$(GMCs)  yields:

		\begin{equation}\label{eq24}
		\Delta v(FWHM)_{GMCs} = \Delta v(FWHM)_{MSD}\textrm{ }\textrm{(by construction)},
		\end{equation}

		\begin{equation}\label{eq25}
 		Tpeak_{GMCs} = ( 2. 14 \pm 0.10 ) \textrm{ }Tpeak_{MSD} ^{0.98 \pm 0.04}, 
		\end{equation}

		\begin{equation}\label{eq26}
 		M(H_{2})_{GMCs} = ( 2. 14 \pm 0.10 ) \textrm{ }M(H_{2})_{MSD}.
		\end{equation}

From Figure \ref{fig15_tpeak_mass} we see that $Tpeak$ changes linearly from the model subtracted dataset to the GMCs catalog. The differences between the subtracted model dataset and the values of the GMCs in our catalog are only due to the CO flux recovery. After we have accounted for the loss of CO flux (see section \ref{param}), the (average) $Tpeak$ value increase by a factor 2.14 as shown in equation \ref{eq25}, and this is directly mirrored in the increase of molecular mass shown in equation \ref{eq26}.\\ 

\section{NOTES ON INDIVIDUAL GMCs.}\label{app2}

In this appendix we review the resolution for the two-fold distance ambiguity for giant molecular clouds in our sample. After making an exhaustive search in the literature we use nine criteria to distinguish between near and far distances. In most cases, more than one criterion is used to remove the distance ambiguity and when all methods converge to the same solution, we assign the near or far distance to the cloud. The criteria are the following:\\

\begin{enumerate}[(a)]

\item Spatial Association with optical objects of the RCW catalog \citep{153} and visual counterparts for HII regions \citep{caswell87}: usually and because of the high extinction in optical wavelengths, optical emission is detectable only for objects closer than 6 kpc within the Galactic plane locating any cloud, with an optical counterpart, almost certainly at the near distance.\\

\item \IRAS sources with already removed distance ambiguity associated to a GMC: in this case, the distance ambiguity resolution of the \IRAS source is assigned to the cloud, but the kinematic distance derived from the rotation curve is used in estimating all the physical quantities of the parent GMC.\\

\item Presence (or absence) of absorption features of species such as formaldehyde (H$_{2}$CO) or the hydroxyl radical (OH) against the H$\alpha$ continuum emission of HII regions \citep{caswell87}, or HI absorption features in molecular clouds of the cold ($10$ - $30$ K) atomic hydrogen against the warm ($100$ - $10^{4}$ K) 21 cm continuum emission across the Galactic plane  \citep{roman2009} . In the first method, depending on the radial velocity of the recombination and absorption features, and on the CO radial velocity $V_{lsr}$ of the cloud, it is possible to distinguish between far o near distance to the cloud in several cases. For instance, if the absorption feature occurs at a radial velocity close to $V_{lsr}$ (within 2$\times\sigma_{v}$) and it is smaller (less negative) than the radial velocity of the recombination line, the absorption can be interpreted as the cloud being in front of the HII region absorbing the continuum emission most probably at the near distance. On the other hand, if the LSR velocity of the absorption feature does not coincide with the LSR velocity of the cloud, this suggests the far distance as the most probable distance to the cloud. Because of the large linewidths of the HI recombination line (\dv $\sim$ 20 - 30 \kms), the absorption features fall sometimes inside the recombination linewidth, making it hard to distinguish whether the emission originates in a different location in the Galactic disk than that of the recombination line. In those cases, the distance resolution solved by this technique should be taken with caution. The second method consist of comparing the HI self-absorption of the cold gas against the background of HI warm continuum emission in the Galactic plane with the \CCO emission at the same LSR velocity \citep{busfield2006,roman2009}. If HI self-absorption is identified to be at the same radial velocity of the \CCO velocity for a particular source, the source lies almost certainly at the near distance, since being at the far distance, the absorption feature would be ``filled up'' with the background of HI continuum emission along the line of sight.\\

\item Observational size-to-linewidth relationship (\emph{first Larson's Law}): the existence of an observational size-to-linewidth relationship for GMCs \citep{solomon87,scoville87,dame86} can be used in some cases to distinguish between the near and far distances. Since the physical radius of a cloud is proportional to the distance, locating a cloud at the wrong position might yield an unrealistically large (more than 200 pc) or small (less than 10 pc) physical radius for the GMC. Concerning the velocity structure of the clouds,  very massive GMCs at the far distance present an elongated structure along the radial velocity axis and a very reduced angular extension. An example of this can be found in the latitude-velocity maps in \citet{bron88b} at 337\deg.250, between $v  = -90$ \kms\textrm{ }and $v  = -30$ \kms. If a cloud is located at a given distance (near or far), the physical radius of the cloud should be consistence with the overall trend in the observational size-to-linewidth relationship.\\    

\item Distance off the Galactic plane following the molecular gas distribution of \citet{bron88}: the position of a certain cloud relative to the Galactic plane in azimuth direction can be used in some cases to distinguish between near and far distances. We examined the perpendicular projection of the distance to the clouds respect to the plane, and compared it with the distribution of the molecular gas in \citet{bron88}. The near distance is assigned to clouds which, being located at the far distance, strongly deviate ($\sim$ 3$\sigma_{z}$) from the molecular gas distribution of the Galactic plane in the azimuth direction.\\ 

\item Continuity of the near or far side of spiral arms: in a few cases, well defined segments of spiral arms might also indicate the location of a particular cloud in the Galactic plane. Since spiral arms are seen as lanes of emission in the longitude-velocity diagram, and the near and far sides of the arms are usually separated by a few tens \kms, clouds located at the far and near distances of the same arm are expected to be separated in LSR velocity well enough to be distinguish from each other.\\

\item CO radial velocity of a cloud lies close ($|v| < 10$ \kms) to the tangential LSR velocity: since the difference between near and far distances decreases toward the tangent point to the line of sight, clouds that are close enough to the tangent LSR velocity at a given Galactic longitude $l$ \citep{alvarez90} are located at the tangent distance, i.e., at the distance where the galactocentric radius is minimum ($R_{gal} = R\solar\sin l$).\\

\end{enumerate}

We explore the consistency of these criteria by comparing our distance determination results with those obtained for 6.7 GHz methanol (CH$_{3}$OH) masers in the work of \citet{green_mcclure2011} using the HI self-absorption method. We associated the methanol masers to their parent GMCs following the criteria explained for the association of the \IRAS sources to GMCs. We only considered methanol masers with the most solid distance ambiguity resolution (labeled with ``a'' in column 5 of Table 2 in their work) that fall within the solar circle, and within the range covered by the present work: $l = $ 300\deg\textrm{ }- 348\deg\textrm{ }and $b = \pm$ 2\deg.\\

We found 20 GMCs with at least one methanol maser associated in our catalog. These GMCs in Table \ref{tbl-2} are: clouds 18, 28, 45, 49, 51, 63, 72, and 78, for which the maser information is fully consistent with the distance assigned to the corresponding GMC; clouds 9, 34, 35, 38, and 76, for which the methanol maser information is contradictory with our distance estimate but that can be explained in terms of the sensitivity of our work to clouds less massive at the far distance; clouds 57, 64, and 73, associated with the near side of the \emph{3-kpc expanding} arm for which the masers information puts them at the far distance; and clouds 14, 21, 33, and 54, for which there is an unexplained discrepancy between the methanol maser distance determination and our distance estimates. We include a detail discussion of the methanol maser information for each GMC in the following notes on individual clouds.\\

\begin{itemize}

%NCAR-1:
\item 1: The \IRAS sources G300.969$+$1.152 and G301.116$+$0.966 associated to this cloud have as visual counterparts the objects RCW65 (G301.0$+$1.2) and RCW66 (G301.1$+$0.9) \citep{80,14,caswell87} locating the cloud at the near distance. The radial velocity of this cloud ($V_{lsr} = -$39.1 \kms) was corrected by $+$12.2 \kms\textrm{ }in order to take into account the anomalous terminal CO velocities between $l =$ 300\deg\textrm{ }- 312\deg\textrm{ }in the fourth Galactic quadrant as described in \citet{alvarez90}.\\  

%NCAR-2:
\item 2: The \IRAS sources G305.085$+$0.062, and G305.194$+$0.036 are associated with the optical object RWC74 (G305.1$+$0.15) \citep{14,caswell87,46,5,25} locating the cloud at the near distance. We did not correct the radial velocity of the cloud ($V_{lsr} = -$36.4 \kms) by the anomalous $+$12.2 \kms\textrm{ }radial velocity since it lies far ($>$ 15 \kms) from the terminal velocity $v_{term} = -$55 \kms\textrm{ }\citep{alvarez90} at this Galactic longitude ($l =$ 305\deg.250).\\

%FCEN-6: 
\item 3: At the position of the cloud, no observations of absorption or recombination line emission were found in the literature. Locating the cloud at the near distance (1.9 kpc) yields a very small physical radius ($\sim$ 10 pc). In this case, the absence of an optical counterpart at such close distance is unlikely. Placing the cloud at the far distance yields a more consistent physical radius ($\sim$ 42 pc) with the size-to-linewidth relationship. We adopt the far distance to the cloud.\\

%NCAR-3:
\item 4: The HII region G305.678$+$1.607 has a visual optical counterpart \citep{caswell87}. The high latitude of the cloud ($b = +$1\deg.250) also suggests the near distance. The CO radial velocity ($V_{lsr} = -$51.9 \kms) was corrected by $+$12.2 \kms\textrm{ }in order to take into account the anomalous terminal CO velocities between $l =$  300\deg\textrm{ }- 312\deg\textrm{ }as described in \citet{alvarez90}. After this correction, we noticed that its galactocentric radius ($R_{gal} =$  6.88 kpc) is slightly smaller than the galactocentric radius at the tangent point of its line of sight ($l =$ 305\deg.750) given by $R\solar\sin (l) =$ 6.90 kpc. Since the last is not allowed under the assumption of pure circular motion, the tangent galactocentric radius was chosen and the corresponding heliocentric distance assigned to the cloud.\\    

%FCEN-0:
\item 5: Based on \CCO and HI absorption observations, \citet{busfield2006} assigned the far distance to the \IRAS source G306.313$-$0.347 associated with this cloud. Since the HI data present no absorption at the \CCO radial velocity of this source, the location of the cloud at the far distance is almost certain.\\

%FCEN-7:
\item 6: The HII region G308.092$-$0.432 presents deep \HHCO absorption at $-$57 \kms\textrm{ }and at $-$13 \kms, and recombination line emission at $-$17 \kms. The absorption features and the absence of an optical counterpart places the source almost certainly at the far distance \citep{caswell87} and, by spatial and velocity coincidence, its parent GMC.\\

%NCEN-1:
\item 7: The radial velocity of this cloud ($V_{lsr} = -$59.5 \kms) was corrected by $+$12.2 \kms\textrm{ }in order to take into account the anomalous terminal CO velocities between $l =$  300\deg\textrm{ }- 312\deg\textrm{ }as described in \citet{alvarez90}. After this correction, we noticed that its galactocentric radius ($R_{gal} =$  6.59 kpc) is slightly smaller than the galactocentric radius at the tangent point of its line of sight ($l =$ 308\deg.375) given by $R\solar\sin (l) =$ 6.66 kpc. Since the last is not allowed under the assumption of pure circular motion, the tangent galactocentric radius was chosen and the corresponding heliocentric distance assigned to the cloud.\\

%FCEN-1:
\item 8: Given the large velocity dispersion of the cloud (\dv = 12.8 \kms), putting the cloud at the far distance yields a physical radius ($\sim$ 65 pc) more consistent with the observational size-to-linewidth relationship (\emph{first Larson's Law}) than the physical radius estimated from the near distance ($\sim$ 30 pc). The continuity of the far side of the \emph{Centaurus} spiral arm also suggests the far distance. Therefore, the far distance is assigned to the cloud.\\

%NCEN-2:
\item 9: The HII region G309.057$+$0.186 has a visual counterpart and exhibits recombination line emission at $-$47 \kms\textrm{ }and no absorption, putting the cloud almost certainly at the near distance \citep{caswell87}. The 6.7 GHz methanol maser 308.754$+$0.549 is apparently associated to this cloud. The removal of its distance ambiguity done by \citet{green_mcclure2011} seems to contradict our distance determination for the GMC. Nonetheless, there is an explanation for the discrepancy. The maser is very close to a local maximum, far from the bulk of the integrated CO emission in the GMC's spatial map. The local maximum can be attributed to a less massive cloud at the far distance (if the distance determination of the associated maser is right) whose CO emission overlaps in phase space with the CO emission of the GMC at the near distance. Since we are sensitive to only the most massive GMCs at the far distance, this is to be expected in some cases. In this sense, there is no contradiction with our determination of the GMC distance, because the maser is associated to a different and less massive cloud at the far distance that we can not properly isolate with the ASM subtraction technique applied in this work.\\

%NCEN-3:
\item 10: The HII region G310.994$+$0.389 is associated with the catalog object RCW82 (G311.00$+$0.40) \citep{caswell87}, locating the cloud at the near distance. The \IRAS source G311.625$+$0.291 (radial velocity from the CS(2-1) line is $-$56.2 \kms) associated with this cloud presents \HHCO absorption at $-$54 \kms\textrm{ }and $-$49 \kms, and recombination line emission at $-$61 \kms\textrm{ }also placing the cloud almost certainly at the near distance. The radial velocity of this cloud ($V_{lsr} = -$53.2 \kms) was corrected by $+$12.2 \kms\textrm{ }in order to take into account the anomalous terminal CO velocities between $l =$ 300\deg\textrm{ }- 312\deg\textrm{ }as described in \citet{alvarez90}.\\  

%NCEN-3.1:
\item 11: The \IRAS source G311.899$+$0.084 associated to the cloud presents a HI self-absorption feature at the same radial velocity of the recombination line emission locating the cloud almost certainly at the near distance \citep{93,46,1}. The \IRAS source G312.109$+$0.309 (radial velocity from the CS(2-1) line is $-$48.4 \kms) associated to this cloud presents \HHCO absorption and recombination line emission at $-$49 \kms\textrm{ }\citep{caswell87,93,46,1}. Since no other absorption feature is present along the line of sight, this suggests the near distance. Given the velocity dispersion of the cloud (\dv = 9.4 \kms), putting the cloud at the near distance yields a physical radius (59 pc) more consistent with the observational size-to-linewidth relationship (\emph{first Larson's Law}) than the physical radius estimated from the far distance (103 pc). Therefore, the near distance is assigned to the cloud.\\

%FCEN-2:
\item 12: Given the velocity dispersion of the cloud (\dv = 10.9 \kms), putting the cloud at the far distance yields a physical radius ($\sim$ 62 pc) much more consistent with the observational size-to-linewidth relationship (\emph{first Larson's Law}) than the physical radius estimated from the near distance ($\sim$ 6 pc). Locating the cloud at the near distance yields an heliocentric distance of 0.9 kpc to the cloud. In this case, the absence of an optical counterpart at such close distance is unlikely, suggesting also the far distance. The continuity of the far side of the \emph{Centaurus} spiral arm also suggests the far distance. Therefore, the far distance is assigned to the cloud.\\

%NCEN-3.2:
\item 13: To assign a heliocentric distance to the cloud, we notice that its CO radial velocity ($V_{lsr} = -$61.6 \kms) is very close to the CO terminal velocity ($V_{lsr} = -$65 \kms) at the line of sight $l =$ 312\deg.500 \citep{alvarez90}. Its galactocentric radius ($R_{gal} =$ 6.07 kpc) is slightly smaller than the galactocentric radius at the tangent point given by $R\solar\sin (l) =$ 6.27 kpc. Since the last is not allowed under the assumption of pure circular motion, we extend the Galactic longitude range suggested by \citet{alvarez90} and apply the $+$12.2 \kms\textrm{ }correction to the radial velocity of the cloud. The new galactocentric radius $R_{gal} =$ 6.43 kpc is slightly larger than the one at the tangent point of the line of sight. Since the calculated galactocentric radii are very close before and after the correction (less than 5\% difference), the galactocentric radius tangent to the line of sight was chosen, and the corresponding heliocentric distance assigned to the cloud.\\

%NCEN-4:
\item 14: Given the velocity dispersion of the cloud (\dv $=$ 9.3 \kms), putting this cloud at the far distance implies a large physical radius (148 pc), deviating the cloud from the observational size-to-linewidth relationship (\emph{first Larson's Law}). The continuity of the near side of the \emph{Centaurus} arm also suggests the near distance to this cloud. We also notice that clouds at distances further than 4 kpc usually subtend angular scales smaller than 1\deg\textrm{ }\citep{green2011} while the angular size of the cloud in Galactic longitude is larger than 2\deg\textrm{ }and in Galactic latitude it is larger than 1\deg. The 6.7 GHz methanol maser 314.320$+$0.112 is associated with this cloud. The removal of its distance ambiguity done by \citet{green_mcclure2011} contradicts our distance determination for its parent GMC. The inspection of the spatial distribution of the CO emission as well as the longitude-velocity diagram does not show any particular strucutre at the positon of the maser that could be associated with a less massive cloud at the far distance. In this case, there is no obvious explanation to account for the contradictory evidence. We stick to our result and adopt the near distance to the GMC.\\

%NCEN-4.1:
\item 15: The HII regions G314.183$+$0.314 and G314.228$+$0.473 present \HHCO absorptions features both at $-$61 \kms\textrm{ }and recombination line emission at $-$62 \kms\textrm{ }and  $-$63 \kms. We notice that the terminal velocity at $l =$ 314\deg.250\textrm{ }is around $-$67 \kms\textrm{ }\citep{alvarez90}, implying that the HII regions are either at the near or tangent distance. Since the absorption features at $-$61 \kms\textrm{ }are close to the CO velocity of the cloud ($V_{lsr} = -$57.6 \kms) and fall within its extension in radial velocity (\dv $\sim$ 8 \kms), we interpret this as the cloud being in front of the HII regions producing the absorption features, locating the cloud at the near distance.\\
  
%FCEN-3: 
\item 16: At the position of the cloud, no observations of absorption or recombination line emission were found in the literature. We assign the far distance to the cloud based on observational size-to-linewidth relationship (\emph{first Larson's Law}). Putting the cloud at the near distance yields a very small physical radius ($\sim$ 15 pc) deviating the cloud far away from the general trend (\dv $=$ 11.0 \kms). We also notice that, in general, clouds at the far distance tend to have a small solid angle and small Galactic latitude (because of the distance projection) but still, if the cloud is massive, a large velocity dispersion (\dv $>$ 10 \kms). This is in fact the case for this cloud.\\

%FCEN-4:
\item 17: Based on \CCO and HI absorption observations \citet{busfield2006} assigned the far distance to the MYSO G318.2650$-$00.1269 which coincides in phase space with this cloud. Since the HI data present no absorption at the \CCO radial velocity of the MYSO, the location of this source and its parent cloud at the far distance is almost certain. Based also on HI absorption data,  \citet{ur2011} place the HII region G318.7251$-$0.2241 at the far distance. The observational size-to-linewidth relationship (\emph{first Larson's Law}) also suggests the far distance. Putting the cloud at the near distance yields a very small physical radius ($\sim$ 13 pc) deviating the cloud far away from the general trend (\dv $=$ 12.8 \kms).\\

%NCEN-5:
\item 18: The HII region G320.153$+$0.780 is associated with the optical object RCW87 (G320.20$+$0.80) \citep{caswell87,25}. The HII regions G319.874$+$0.770 and G316.808$-$0.037 present visual optical counterparts \citep{caswell87,68}. Base on \CCO and HI absorption observations, \citet{busfield2006} solved the distance ambiguity for the \IRAS source G316.810$-$0.070 associated to this cloud, putting it at the near distance. Based on HI absorption data, \citet{ur2011} place the HII regions G317.4122$+$0.1050 and G318.7748$-$0.1513 at the near distance. The 6.7 GHz methanol maser 318.050$+$0.087 is associated to this cloud. The removal of its distance ambiguity done by \citet{green_mcclure2011} confirms our distance determination for its parent GMC.\\

%FCEN-5.1:
\item 19: The HII region G319.157$-$0.423 and G319.380$-$0.025 listed by \citet{caswell87} present \HHCO absorption features at $-$20 \kms, and recombination line emission at $-$22 \kms\textrm{ }and $-$14 \kms, respectively. No optical counterparts were found. The authors put these sources at the far distance. Since the \HHCO absorption radial velocity coincides with the CO radial velocity of the cloud ($V_{lsr} = -$20.3 \kms), we locate the cloud at the far distance. Based on HI absorption data,  \citet{ur2011} place the HII regions G318.9148$-$0.1647, G319.1632$-$0.4208, and G319.3622$+$0.0126  at the far distance.\\

%FCEN-5.2:
\item 20: We assign the far distance to the cloud based on the following: given the velocity dispersion of the cloud (\dv $=$ 14.9 \kms), putting this cloud at the near distance implies an unrealistic small physical radius (5 pc), deviating the cloud from the observational size-to-linewidth relationship (\emph{first Larson's Law}).\\

%FCEN-5.4:
\item 21: The HII region G320.317$-$0.208 (possible the same object as the \IRAS G320.316$-$0.180) presents \HHCO absorption at $-$11.4 \kms\textrm{ }and recombination line emission at $-$11 \kms\textrm{ }\citep{caswell87}. The same occurs with the HII region G320.379$+$0.139 presenting \HHCO absorption and recombination line emission at the same velocity $ -$3 \kms\textrm{ }\citep{caswell87}. These observations and the lack of an optical counterpart  places the cloud almost certainly at the far distance. Putting the cloud at the near distance leads to an unrealistic physical radius of $\sim$ 4 pc. The 6.7 GHz methanol maser 320.123$-$0.504 is associated with this cloud. The removal of its distance ambiguity done by \citet{green_mcclure2011} contradicts our distance determination for its parent GMC. We notice that the maser coincide spatially and spectrally with the \IRAS G320.124$-$0.501 associated to this cloud. In the longitude-velocity diagram, there is no other CO emission feature around this GMC to which the maser could be associated, so the evidence suggests that the maser is indeed at the far distance. Given the strong evidence we have to put the cloud at the far distance, we believe that the distance determination of the maser may be wrong, so we stick to our distance estimate with the GMC at the far distance.\\

%NCEN-7.2:   
\item 22: The \IRAS source G320.252$-$0.296 associated with this cloud presents \HHCO absorption at $-$13 \kms\textrm{ }and recombination line emission at $-$68 \kms\textrm{ }\citep{caswell87}. The absence of absorption features between the \IRAS radial velocity and the terminal velocity at its line of sight ($v_{term} \sim -$76 \kms) suggests the near distance for the source. For the \IRAS source G320.332$-$0.307, \citet{gyg76} places it at the near distance applying the criterion of the continuity of the arms. We adopt the near distance to the cloud.\\

%NCEN-7.1:
\item 23: The HII regions G321.038$-$0.519 and G321.105$-$0.549 have visual optical counterparts \citep{caswell87,25}. The \IRAS source G321.057$-$0.520 is associated with the optical object RCW91 (G321.20$-$0.50) \citep{80,caswell87} locating the cloud at the near distance.\\  

%NCEN-6:
\item 24: The HII region G324.192$+$0.109 present several \HHCO absorptions behind $-$92 \kms\textrm{ }\citep{caswell87} with no features between the recombination line velocity and the terminal velocity locating this region at the near distance. One of the \HHCO absorption features of this region occurs at $-$29 \kms, very close to the CO radial velocity of the cloud ($V_{lsr} = -$31.6 \kms) locating the cloud in front of the HII region at the near distance. We also notice that, given the velocity dispersion of the cloud (\dv $=$ 4.7 \kms), putting this cloud at the far distance implies an unrealistic large physical radius (280 pc), deviating the cloud from the general trend observational size-to-linewidth relationship (\emph{first Larson's Law}). The large angular extension of the cloud also supports the near distance.\\

%NCEN-8:
\item 25: The HII region G322.153$+$0.613 is associated with the optical object RCW92 (G322.20$+$0.60) locating the cloud at the near distance \citep{caswell87,25}. Based on HI absorption data,  \citet{ur2011} place the HII region G322.1729$+$0.6442 at the near distance.\\   

%NCEN-9:
\item 26: The \citet{129} paper is devoted to the \IRAS source G323.459$-$0.079, associated with this cloud. They place the source at the near distance, so we adopt the near distance to its parent cloud.\\

%NCEN-10.2:
\item 27: Given the velocity dispersion of the cloud (\dv = 8.0 \kms), putting the cloud at the near distance yields a physical radius ($\sim$ 44 pc) more consistent with the observational size-to-linewidth relationship (\emph{first Larson's Law}) than the physical radius estimated from the far distance ($\sim$ 121 pc). The continuity of the near side of the \emph{Centaurus} spiral arm also suggests the near distance to this cloud.\\

%NCEN-10.1:
\item 28: \citet{137} is devoted to the \IRAS source G323.740$-$0.254 associated with this cloud. From observations of the Bracket-$\gamma$ recombination line of atomic hydrogen in the near-infrared the authors place the source at the near distance. \citet{139} reports CH$_{3}$OH emission of two close sources: G323.740$-$0.263 ($-$51.2 \kms) and G323.741$-$0.263 ($-$57.1 \kms). Because of the high luminosity resulting at the far distance, the authors place the source ``almost certainly'' at the near distance. The HII region G324.192$+$0.109 present several \HHCO absorptions behind $-$92 \kms\textrm{ }\citep{caswell87} with no features between the recombination line velocity and the terminal velocity locating this region at the near distance. One of the \HHCO absorption features occurs at $-$50 \kms, very close to the CO radial velocity of the cloud ($V_{lsr} = -$45.3 \kms, \dv $=$ 10.7 \kms) locating the cloud in front of the HII region at the near distance. The 6.7 GHz methanol maser 323.740$-$0.263 is associated to this cloud. The removal of its distance ambiguity done by \citet{green_mcclure2011} confirms our distance determination for its parent GMC.\\

%NCEN-11.1:
\item 29: The HII regions G326.230$+$0.976 and G326.315$+$0.689 and the \IRAS source G326.447$+$0.906 are associated with the optical object RCW94 (G326.20$+$0.90) \citep{gyg76,80,30,caswell87,83}. The \IRAS sources G326.655$+$0.592, G326.726$+$0.613, and the HII region G326.645$+$0.589 are associated with the optical object RCW95 (G326.70$+$0.80) \citep{gyg76,caswell87,98,106,124,25}. Concerning these sources, \citet{106} gives the HI spectrum at position (326\deg.6,$+$0\deg.6) showing absorption at $-$47 \kms. The authors place the source at the near distance. \citet{98} reports OH absorption at $-$45 \kms, at the transitions 1665 MHz, 1667 MHz, and 1720 MHz, and emission at the same velocity at 1612 MHz. There is also absorption at $-$21 \kms\textrm{ }at 1665 MHz and 1667 MHz. This reference places the source at the near distance. Based on HI absorption data,  \citet{ur2011} place the HII region G326.7249$+$0.6159 at the near distance. \\ 

%NCEN-11.2:
\item 30: The HII region G327.313$-$0.536 is associated with the optical object RCW97 (G327.30$-$0.55) putting the cloud at the near distance \citep{caswell87}. Based on HI absorption data, \citet{ur2011} place the HII region G327.3017$-$0.5382 at the near distance.\\

%NCEN-12:
\item 31: The HII region G326.441$-$0.396 presents \HHCO absorption at $-$44 \kms\textrm{ }and recombination line emission at $-$61 \kms\textrm{ }\citep{caswell87} locating the source at the near distance (at the $l =$ 326\deg.500 line of sight, the terminal velocity is $\sim$ $-$100 \kms\textrm{ }as shown by \citet{alvarez90}). The HII region G326.660$-$0.471 presents no \HHCO absorption but recombination line emission at $-$57 \kms\textrm{ }\citep{caswell87} locating the source almost certainly at the near distance. The HII region G326.959$-$0.031 presents \HHCO absorption at $-$57.3 \kms\textrm{ }and $-$45.5 \kms, and recombination line emission at $-$64 \kms\textrm{ }\citep{caswell87} locating the source at the near distance (at the $l =$ 327\deg.000 line of sight, the terminal velocity is $\sim$ $-$107 \kms\textrm{ }as shown by \citet{alvarez90}). Since the CO radial velocity of the cloud is $V_{lsr} = -$62.3 \kms\textrm{ }(\dv $=$ 10.8 \kms), we associate the previous HII regions to the cloud and put it at the near distance. Based on HI absorption data,  \citet{ur2011} place the HII region G326.4719$-$0.3777 at the near distance.\\

%NCEN-13:
\item 32: The HII region G328.593$-$0.518 associated to the optical object RCW99 (G328.57$-$0.53) which locates this source at the near distance \citep{caswell87}. The source presents \HHCO absorptions features at $-$71.5 \kms\textrm{ }and $-$46.1 \kms, and recombination line emission at $-$51 \kms. We notice that the absorption feature coincides with the CO radial velocity of the cloud $V_{lsr} = -$70.7 \kms. Although the absorption feature at $-$70 \kms\textrm{ }occurs at larger velocities (more negative) than the one of the recombination line emission, the full width at half maximum of the recombination line (\dv $=$ 27 \kms) shows that the absorption feature falls within 2$\times\sigma_{v}$ of its velocity profile. Therefore, such feature can not be assigned directly to the cloud in front of the HII region. Based on HI absorption data, \citet{ur2011} place the HII region G327.7579$-$0.3515 at the near distance. Given the velocity dispersion of the cloud (\dv $=$ 11.3 \kms), putting this cloud at the far distance implies a large physical radius (137 pc), deviating the cloud from the general trend of the observational size-to-linewidth relationship (\emph{first Larson's Law}). The continuity of the near side of the \emph{Centaurus} spiral arm also suggests the near distance.\\  

%NCEN-14:
\item 33: The HII region G328.593$-$0.518 is associated to the optical object RCW99 (G328.57$-$0.53) which locate this source at the near distance \citep{caswell87}. The source presents \HHCO absorptions features at $-$71.5 \kms\textrm{ }and $-$46.1 \kms, the latter coinciding with the CO velocity of the cloud ($V_{lsr} = -$45.0 \kms), putting the cloud in front of the HII region at the near distance. For the \IRAS source G328.191$-$0.571 associated to this cloud, \citet{gyg76} reported the distance to the exiting stars (2.6 kpc) at position (328\deg.1,$-$0\deg.5) locating this cloud at the near distance. Using recombination line velocities ($v_{rrl}$) from continuum sources and HI absorptions along the same line of sight,  \citet{jones2012} locates the continuum source G328.567$-$0.533 ($v_{rrl} = -51$ \kms) associated to the cloud at the near distance. The 6.7 GHz methanol masers 329.029$-$0.205, 329.031$-$0.198, and 329.066$-$0.308 are associated with this cloud. The removal of its distance ambiguity done by \citet{green_mcclure2011} contradicts our distance determination for the parent GMC. We notice that the maser 329.066$-$0.308 coincides spatially and spectrally with the \IRAS G329.066$-$0.307 associated to the GMC. The inspection of the spatial distribution of the CO emission as well as the longitude-velocity diagram shows a strucutre at the positon of the masers that could represent a different cloud at the far distance overlapping with this GMC in phase space. The fact that the masers are all around the same CO emission contour, far from the peak of the CO emission in the CO spatial map of the cloud, suggests that they could be indeed tracing another cloud. Since our evidence to put the GMC at the near distance is strong, but the sources found in the literature to assign the near distance go only up to $l = $ 328\deg.576, we can not discard that the masers located at larger longitudes are indeed tracing a different GMC.\\ 

%NORM-1:
\item 34: The HII region G328.310$+$0.448 presents \HHCO absorption at $-$92 \kms\textrm{ }and $-$52 \kms, with recombination line emission at $-$97 \kms\textrm{ }\citep{caswell87}. Because of its proximity to the terminal velocity ($\sim$ $-$111 \kms\textrm{ }as shown by \citet{alvarez90}) and the lack of absorption features toward higher (more negative) velocities, the source is located either at the tangent or at the near distance. Since the CO radial velocity of the cloud ($V_{lsr} =$ $-$92.0 \kms) coincides with the absorption feature, we assign the near distance to the cloud. Using recombination line velocities ($v_{rrl}$) from continuum sources and HI absorptions along the same line of sight,  \citet{jones2012} locates the continuum source G328.311$+$0.433 ($v_{rrl} = -97$ \kms) associated to the cloud at the near distance. The 6.7 GHz methanol masers 327.590$-$0.094 and 327.618$-$0.111 are apparently associated to this cloud. The removal of their distance ambiguity done by \citet{green_mcclure2011} seems to contradict our distance determination for the GMC. Nonetheless, there is an explanation for the discrepancy. The masers are very close to a local maximum, far from the bulk of the integrated CO emission in the GMC's spatial map. The local maximum can be attributed to a less massive cloud at the far distance (if the distance determination of the associated masers are right) whose CO emission overlaps in phase space with the CO emission of the GMC at the near distance. Since we are sensitive to only the most massive GMCs at the far distance, this is to be expected in some cases. In this sense, there is no contradiction with our determination of the GMC distance, because the masers are associated to a different and less massive cloud at the far distance that we can not properly isolate with the ASM subtraction technique applied in this work.\\

%NCEN-16.2:
\item 35: In the case of this cloud, none of the previous methods could distinguish between near and far distance, and therefore, no distance was assigned to the cloud. The 6.7 GHz methanol maser 329.469$+$0.503 is apparently associated to this cloud. The removal of its distance ambiguity done by \citet{green_mcclure2011} seems to contradict our distance determination for the GMC. Nonetheless, there is an explanation for the discrepancy. The maser is very close to a local maximum, far from the bulk of the integrated CO emission in the GMC's spatial map. The local maximum can be attributed to a less massive cloud at the far distance (if the distance determination of the associated maser is right) whose CO emission overlaps in phase space with the CO emission of the GMC at the near distance. Since we are sensitive to only the most massive GMCs at the far distance, this is to be expected in some cases. In this sense, there is no contradiction with our determination of the GMC distance, because the maser is associated to a different and less massive cloud at the far distance that we can not properly isolate with the ASM subtraction technique applied in this work.\\

%NCEN-16.3:
\item 36:  In the case of this cloud, none of the previous methods could distinguish between near and far distance, and therefore, no distance was assigned to the cloud.\\

%NORM-2.1:
\item 37: The HII region G329.353$+$0.144 presents \HHCO absorption at $-$106 \kms and  $-$85 \kms, recombination line emission at $-$107 \kms, and HI absorption at $-$109 \kms\textrm{ }\citep{caswell87,30} locating this source either at the near or tangent distance. The same occurs for the HII region G329.489$+$0.207 which presents \HHCO absorption at $-$100 \kms, recombination line emission at $-$102 \kms, and HI absorption at $-$109 \kms\textrm{ }\citep{caswell87,30}. For both sources, \citet{caswell87} assigned them the tangent distance. Since the CO radial velocity of the cloud $V_{lsr} = -$99.0 \kms\textrm{ }coincides with the absorption feature at $-$100 \kms, we locate the cloud in front of the HII region, at the near distance. Using recombination line velocities ($v_{rrl}$) from continuum sources and HI absorptions along the same line of sight,  \citet{jones2012} locates the continuum sources G329.478$+$0.211 ($v_{rrl} = -102$ \kms) and G329.333$+$0.144 ($v_{rrl} = -107$ \kms) at the tangent point, behind the cloud.\\

%NCEN16.4:
\item 38: The HII region G329.353$+$0.144 presents \HHCO absorption at $-$106 \kms\textrm{ }and $-$85 \kms\textrm{ }and recombination line emission at $-$107 \kms. Since the terminal velocity at $l =$ 329\deg.5 is around $-$110 \kms\textrm{ }\citep{alvarez90}, \citet{caswell87} locate the source at the tangent distance. The absorption at $-$85 \kms\textrm{ }falls within the CO velocity distribution of the cloud ($V_{lsr} = -$80.9 \kms, \dv $=$ 9.4 \kms), putting the cloud in front of the source at the near distance. Using recombination line velocities ($v_{rrl}$) from continuum sources and HI absorptions along the same line of sight,  \citet{jones2012} locates the continuum sources G329.478$+$0.211 ($v_{rrl} = -102$ \kms) and G329.333$+$0.144 ($v_{rrl} = -107$ \kms) at the tangent point, behind the cloud. The 6.7 GHz methanol maser 329.622$+$0.138 is apparently associated to this cloud. The removal of its distance ambiguity done by \citet{green_mcclure2011} seems to contradict our distance determination for the GMC. The maser mid-velocity $v_{mid}  = -$84.7 \kms\textrm{ }is almost the same as the velocity of the peak emission in the spectrum $v_{peak} = -$85.0 \kms\textrm{ }which, in turn, is identical to the folmaldeyde absorption at $-$85.0 \kms. \citet{green_mcclure2011} put the maser at the far distance due to the lack of HI self-absorption at the mid-velocity. Nonetheless, the HI spectrum of this maser presents a sharp decrease in brigthness temperature around $\pm$5 \kms\textrm{ }of $v_{mid}$\textrm{ }that the authors do not attribute to be an absorption feature. We believe that our distance determination for the GMC is solid and that the methanol maser may have been incorrectly assigned to the far distance due to the complex HI spectrum it shows.\\

%NCEN16.1:
\item 39:  In the case of this cloud, none of the previous methods could distinguish between near and far distance, and therefore, no distance was assigned to the cloud.\\

%NORM-2.2:
\item 40: The location of the cloud in Galactic latitude ($b =$ 1\deg.000) favors the near distance. Putting the cloud at the far distance yields a distance off the plane of 146 pc toward positive latitudes. This is in contradiction with the molecular gas distribution at the galactocentric radius of the cloud $R_{gal} =$ 4.66 kpc (for R$\solar$ = 8.5 kpc). The centroid of the molecular gas distribution lies at z $\sim$ 45 pc, with a half width half maximum extension of  $\sim$ 81 pc \citep{bron88} at this galactocentric radius. We adopt the near distance to the cloud.\\   

%NCEN-17:
\item 41: The HII region G332.148$-$0.446 has an optical counterpart suggesting the near distance \citep{caswell87}. The \IRAS source G330.883$-$0.369 is discussed in \citet{134} as having an anomalous velocity. It is placed at 5.0 kpc, corresponding to the near distance. The \IRAS source G331.169$-$0.455 \citet{92} shows observations of 4830 MHz \HHCO absorption pointing to 331\deg.1$-$0\deg.4. They report also 1667 MHz OH emission at $-$41.8 \kms. This reference places the source at the near distance. The \IRAS source G331.333$-$0.339 \citet{1} presents a HI spectrum which shows deep absorption features at about $-$69 \kms\textrm{ }and $-$47 \kms. In an individual note the authors discuss ``a single weak possible absorption feature at $-$94 \kms'', however the authors favor the near kinematic distance. Based on HI absorption data, \citet{ur2011} place the HII regions G330.8708$-$0.3715, G331.1194$-$0.4955, and G331.4181$-$0.3546 at the near distance. Using recombination line velocities ($v_{rrl}$) from continuum sources and HI absorptions along the same line of sight,  \citet{jones2012} locates the continuum source G330.867$-$0.367 ($v_{rrl} = -56$ \kms), G330.678$-$0.389 ($v_{rrl} = -61$ \kms), and G331.122$-$0.533 ($v_{rrl} = -68$ \kms) associated to the cloud at the near distance.\\

%NCEN-18:
\item 42: The HII region G331.026$-$0.152 presents \HHCO absorption features at $-$92.2 \kms\textrm{ }and $-$46.7 \kms\textrm{ }and recombination line emission at $-$89 \kms\textrm{ }\citep{caswell87,25}, putting the source at the far distance. According to the pure circular motion assumption, since the recombination line emission occurs at $-$89 \kms, the absorption feature at $-$46.7 \kms\textrm{ }must be located at the near distance. We identify this absorption feature as being produced by the cloud ($V_{lsr} = -$45.3 \kms, and \dv $=$ 8.6 \kms) along the line of sight of the HII region, locating the cloud at the near distance.\\   

%NORM-4:
\item 43: \citet{1} presents a HI spectrum for the \IRAS source G331.552$-$0.115 showing absorption at $-$90 \kms\textrm{ }and H109$\alpha$ emission at $-$89 \kms. The authors discuss extensively this source and place it at near distance. The HII region  G331.517$-$0.069 has absorption features at  $-$99.8 \kms\textrm{ }and $-$89.3 \kms\textrm{ }and recombination line emission at $-$89 \kms\textrm{ }\citep{caswell87}. Based on HI absorption data, they put the region at the near distance. \citet{93} presents OH spectra for the same source showing emission between $-$87 \kms\textrm{ }and $-$95 \kms\textrm{ }specially at 1665 MHz. Using recombination line velocities ($v_{rrl}$) from continuum sources and HI absorptions along the same line of sight,  \citet{jones2012} locates the continuum sources  G331.522$-$0.078 ($v_{rrl} = -89$ \kms),  G331.055$-$0.144 ($v_{rrl} = -89$ \kms), G331.278$-$0.189 ($v_{rrl} = -85$ \kms), and G331.055$-$0.222 ($v_{rrl} = -89$ \kms), associated to the cloud at the tangent point. This is due to the their adopted tolerance for the difference between the terminal and LSR velocity of the source (25 \kms) to assign it to the tangent point, which is larger than the one we use in this work (10 \kms) . We adopt the near distance to the cloud.\\ 

%NCEN-20.1:
\item 44: The HII region G332.978$+$0.792, associated to this cloud presents \HHCO absorption at $-$48.4 \kms\textrm{ }and recombination line emission at $-$52 \kms\textrm{ }\citep{caswell87}. The absence of absorption features between the emission and terminal velocities places the source almost certainly at the near distance. Based on HI absorption data, \citet{ur2011} place the HII regions G333.0058$+$0.7707 and G333.0162$+$0.7615 at the near distance. Using recombination line velocities ($v_{rrl}$) from continuum sources and HI absorptions along the same line of sight,  \citet{jones2012} locates the continuum source G332.978$+$0.767 ($v_{rrl} = -52$ \kms) associated to the cloud at the near distance.\\   

%NCEN-20:
\item 45: The HII region G332.662$-$0.607 is associated to the optical RCW106 (G332.90$-$0.60) locating the cloud at the near distance \citep{83,14}.  Based on HI absorption data, \citet{ur2011} place the HII regions G332.1544$-$0.4487 to G332.8256$-$0.5498, G333.0145$-$0.4438, and from G333.1306$-$0.4275 to G333.6032$-$0.2184 at the near distance. The observational size-to-linewidth relationship (\emph{first Larson's Law}) also suggests the near distance. Putting the cloud at the far distance yields an unrealistic large physical radius ($\sim$ 218 pc) deviating the cloud far away from the general trend (\dv $=$ 8.9 \kms). Using recombination line velocities ($v_{rrl}$) from continuum sources and HI absorptions along the same line of sight,  \citet{jones2012} locates the continuum sources G333.133$-$0.433 ($v_{rrl} = -52$ \kms), G332.833$-$0.555 ($v_{rrl} = -57$ \kms), and G332.656$-$0.611 ($v_{rrl} = -48$ \kms), associated in phase space to the cloud at the near distance, and the G333.689$-$0.444 ($v_{rrl} = -50$ \kms), also associated in phase space to the cloud at the far distance. Most of the evidence favors the near distance to the cloud. The 6.7 GHz methanol masers 332.352$-$0.117 and 332.960$+$0.135 are associated to this cloud. The removal of their distance ambiguity done by \citet{green_mcclure2011} confirms our distance determination for their parent GMC.\\

%NORM-3:
\item 46: The region HII G333.168$-$0.081 presents \HHCO absorption features at $-$88.6 \kms\textrm{ }and $-$46.6 \kms, and recombination line emission at $-$91 \kms\textrm{ }\citep{caswell87}. The lack of absorption features at higher (more negative) velocities, put this source at the near distance, with this cloud in front of it absorbing at $-$88.6 \kms, the same CO radial velocity of the cloud ($V_{lsr} = -$88.1 \kms). For the \IRAS source G333.168$-$0.104 \citet{133} reports 1612 MHz OH emission at $-$87 \kms\textrm{ }and absorption at $-$41 \kms\textrm{ }at position (333\deg.1,$-$0\deg.1) suggesting the near distance for the source. The authors note that this source is associated with an extensive cloud, which is also detected at 100 adjacent positions. Using recombination line velocities ($v_{rrl}$) from continuum sources and HI absorptions along the same line of sight,  \citet{jones2012} locates the continuum source G333.167$-$0.100 ($v_{rrl} = -91$ \kms) at the tangent point, just behind the cloud.\\   

%NORM-5:
\item 47: Because of the proximity of its CO radial velocity ($V_{lsr} = -$105.0 \kms) to the terminal velocity at $l =$ 334\deg\textrm{ }($v_{term} = -$116 \kms\textrm{ }as shown by \citet{alvarez90}), the difference between the far and near distances to this cloud (2.7 kpc) is not as large as in other parts of the longitude velocity diagram. The observational size-to-linewidth relationship and the continuity of the near side of the \emph{Norma} spiral arm place the cloud at the near distance. Using recombination line velocities ($v_{rrl}$) from continuum sources and HI absorptions along the same line of sight,  \citet{jones2012} locates the continuum source G333.167$-$0.100 ($v_{rrl} = -91$ \kms) at the tangent point, that would lie in front of the cloud. We adopt the near distance to the cloud.\\   

%NORM-F21:
\item 48: The HII region G334.173$+$0.068 presents \HHCO absorption at $-$86 \kms\textrm{ }and $-$40 \kms\textrm{ }and recombination line emission at $-$70 \kms\textrm{ }\citep{caswell87}. The presence of an absorption feature toward the terminal velocity locates the source at the far distance. Since the cloud is located along the line of sight of this HII region at $V_{lsr} =  -$62.2 \kms, the absence of absorption features within the velocity dispersion of the cloud locates it behind the HII region at the far distance. We also notice that, in general, clouds at the far distance tend to have a small solid angle and small Galactic latitude (because of the distance projection) but still, if the cloud is massive, a large velocity dispersion (\dv $>$ 10 \kms). This is in fact the case for this cloud.\\    

%NCEN-19:
\item 49:  The HII region G335.751$+$0.110 and G335.978$+$0.185 present \HHCO absorption features at $-$38 \kms\textrm{ }and $-$37 \kms, and recombination line emission $-$51 \kms\textrm{ }and  $-$79 \kms\textrm{ }respectively \citep{caswell87}. The absence of absorption features between the recombination line velocity and terminal velocity at the line of sight ($\sim$ $-$131 \kms\textrm{ }at $l =$ 336\deg\textrm{ }as shown by \citet{alvarez90}) strongly supports the near distance for these sources and the position of the cloud in front of them absorbing against their continuum emission ($V_{lsr} = -$40.4 \kms, and \dv $=$ 8.2 \kms), locating the cloud at the near distance. Putting the cloud at the far distance yields an unrealistic large physical radius ($\sim$ 258 pc) deviating the cloud far away from the general trend of the size-to-linewidth relationship (\dv $=$ 8.2 \kms). The 6.7 GHz methanol maser 335.060$-$0.427 is associated to this cloud. The removal of its distance ambiguity done by \citet{green_mcclure2011} confirms our distance determination for its parent GMC.\\

%NCEN-22.1:
\item 50: The cloud is located far off the Galactic plane, at $b = -$0\deg.875. Putting this cloud at the far distance implies a distance off the plane of 182 pc toward negative latitudes. \citet{bron88} estimated the distribution of the molecular layer for the southern Galaxy. At the galactocentric radius of this cloud ($R_{gal} =$ 5.4 kpc), the molecular ring centroid lies at z = $-$5 pc, with a half width half maximum extension of $\sim$ 68 pc. Following the molecular gas distribution, a far distance to this cloud results unlikely. Therefore, we adopt the near distance in this case.\\ 

%NORM-F6.1:
\item 51: The \IRAS source G336.872$-$0.014 presents OH absorption at $-$120 \kms, HI absorption at $-$128 \kms, and recombination line emission at $-$78.9 \kms\textrm{ }locating the source at the far distance \citep{94,105,25}. The \IRAS source G337.121$-$0.174 presents \HHCO absorption at $-$119.6 \kms\textrm{ }and at $-$73.7 \kms, OH absorption at $-$118 \kms\textrm{ }(and around this velocity), HI absorption at $-$116 \kms\textrm{ }and recombination line emission at $-$73 \kms\textrm{ }suggesting the far distance \citep{caswell87,44,1,25}. The presence of absorption features in several lines between the recombination line velocity and the tangent velocity at $l =$ 339\deg\textrm{ }($-$128 \kms\textrm{ }as shown by \citet{alvarez90}) in these two sources locates this cloud almost certainly at the far distance. Based on HI absorption data, \citet{ur2011} place the HII regions G336.8324$+$0.0301, G337.0047$+$0.3226, G337.7091$+$0.0932, and G337.1218$-$0.1748 at the far distance. Using recombination line velocities ($v_{rrl}$) from continuum sources and HI absorptions along the same line of sight,  \citet{jones2012} locates the continuum sources G337.711$+$0.089 ($v_{rrl} = -76.7$ \kms), G337.122$-$0.178 ($v_{rrl} = -73$ \kms), G336.911$-$0.155 ($v_{rrl} = -73$ \kms), and G336.844$-$0.000 ($v_{rrl} = -79$ \kms) associated to the cloud at the far distance. The 6.7 GHz methanol masers 337.202$-$0.094 and 337.258$-$0.101 are associated to this cloud. The removal of their distance ambiguity done by \citet{green_mcclure2011} confirms our distance determination for their parent GMC.\\

%NCEN-22.2:
\item 52: The cloud is located far off the Galactic plane, at $b = -$1\deg.125. Putting this cloud at the far distance implies a distance off the plane of 242 pc toward negative latitudes. \citet{bron88} estimated the distribution of the molecular layer for the southern Galaxy. At the galactocentric radius of this cloud ($R_{gal} =$ 5.6 kpc), the molecular ring centroid lies at z = $-$18 pc, with a half width half maximum extension of  $\sim$ 76 pc. Following the molecular gas distribution, a far distance to this cloud results unlikely. Therefore, we adopt the near distance in this case.\\ 

%3KPC-1:
\item 53: For the \IRAS source G336.888$+$0.049 associated to this cloud, \citet{94} reports 1665 MHz OH absorption at $-$120 \kms\textrm{ }and $-$53.2 \kms\textrm{ }for position (336\deg.8,$+$0\deg.05). Considering the CS radial velocity of this source at $-$121.2 \kms\textrm{ }we locate it at the near distance. \citet{gyg76} discusses the three close sources G336.8$+$0.0, G336.9$-$0.1, and G337.1$-$0.2 (``group 4'') and places them at the near distance. For the \IRAS source G336.993$-$0.023 associated to this cloud, \citet{20} reports CH$_{3}$OH emission at $-$126 \kms\textrm{ }and adopts a distance of 9.3 kpc (close to the tangent point, for R$\solar$ = 10 kpc and $\Theta\solar$ = 250 \kms), probably because of the resulting spectral type of the exciting star. Using recombination line velocities ($v_{rrl}$) from continuum sources and HI absorptions along the same line of sight,  \citet{jones2012} locates the continuum source G337.533$-$0.311 ($v_{rrl} = -101$ \kms) associated to the cloud at the tangent point. The source lies just above the lowest intensity contour, so the spatial association with the cloud is dubious. We adopt the near distance to the cloud.\\   

%NORM-F6.2:
\item 54: The HII region G338.014$-$0.121 presents \HHCO absorption at $-$88 \kms\textrm{ }and at $-$49 \kms\textrm{ }and recombination line emission at $-$54 \kms\textrm{ }\citep{caswell87}. The HII region G338.131$-$0.173 presents \HHCO absorption at $-$89.1 \kms\textrm{ }and $-$42.7 \kms\textrm{ }and recombination line emission at $-$53 \kms. \citet{caswell87} uses the \HHCO data from \citet{80} and \citet{81} to study the posicion of the HII region G337.665$-$0.048 which presents \HHCO  absorption at $-$94.4 \kms\textrm{ }and at $-$47.5 \kms\textrm{ }and recombination line emission at $-$55 \kms. In these three cases, the presence of \HHCO absorption features at velocities higher (more negative) than the recombination line velocity put these sources at the far distance, and because of the spatial and velocity coincidence of the sources with the cloud, we assigned the far distance to the cloud. \citet{fish03} presents a HI spectrum at position (337\deg.7,$-$0\deg.05) (coinciding with the \IRAS G337.703$-$0.052 associated to this cloud), showing absorption at $-$50 \kms\textrm{ }and $-$106 \kms. The authors assigned the far distance to the source. Based on HI absorption data, \citet{ur2011} place the HII regions G337.6651$-$0.1750, and G337.7051$-$0.0575 at the far distance. Using recombination line velocities ($v_{rrl}$) from continuum sources and HI absorptions along the same line of sight,  \citet{jones2012} locates the continuum sources G338.122$-$0.189 ($v_{rrl} = -53$ \kms), G338.078$+$0.011 ($v_{rrl} = -47$ \kms), G337.711$-$0.056 ($v_{rrl} = -50$ \kms), G337.622$-$0.067 ($v_{rrl} = -55$ \kms), G337.667$-$0.167 ($v_{rrl} = -53$ \kms), and G337.289$-$0.122 ($v_{rrl} = -54$ \kms), associated to the cloud at the far distance. The 6.7 GHz methanol maser 337.632$-$0.079 is associated with this cloud. The removal of its distance ambiguity done by \citet{green_mcclure2011} contradicts our distance determination for its parent GMC. The maser is spatially close to the associated \IRAS G337.616$-$0.061, but $\sim$ 10 \kms\textrm{ }appart in LSR velocity. The inspection of the spatial distribution of the CO emission as well as the longitude-velocity diagram does not show any particular structure at the positon of the maser that could be associated with a less massive cloud at the near distance. In this case, there is no obvious explanation to account for the discrepancy. Since our evidence to adopt the far distance to the GMC is strong, we stick to our result in this case.\\

%NORM-7.3:
\item 55: The HII region G337.548$-$0.304 presents \HHCO absorption at $-$62 \kms\textrm{ }and $-$35 \kms\textrm{ }and recombination line emission at $-$101 \kms\textrm{ }\citep{caswell87}. The absence of absorption features at larger velocities suggests the near distance for the source.  Because of the spatial and velocity coincidence of the HII region with the cloud, we assign the near distance to the GMC. Using recombination line velocities ($v_{rrl}$) from continuum sources and HI absorptions along the same line of sight,  \citet{jones2012} locates the continuum source G337.533$-$0.311 ($v_{rrl} = -101$ \kms) at the tangent point, just behind of the cloud.\\

%3KPC-2a:
\item 56: The cloud is located far off Galactic the plane, at $b = -$1\deg.000. Putting this cloud at the far distance (9.4 kpc) implies a distance off the plane of 164 pc toward negative latitudes (and 206 pc at the far distance of 11.8 kpc suggested by \citet{dame2008} for the far side of the arm). \citet{bron88} estimated the distribution of the molecular layer for the southern Galaxy. At the galactocentric radius of this cloud ($R_{gal} =$ 3.5 kpc), the molecular ring centroid lies at z = $-$12 pc, with a half width half maximum extension of  $\sim$ 70 pc. Following the molecular gas distribution, a far distance to this cloud results unlikely. Therefore, we adopt the near distance in this case.\\ 

%3KPC-2b:
\item 57: The HII region G339.089$-$0.216 associated to the cloud presents no absorption features but recombination line emission at $-$120 \kms\textrm{ }and is placed at the tangent distance by \citet{caswell87}. For the \IRAS source G338.569$-$0.145, \citet{133} presents 1612 MHz OH spectrum at position (338\deg.5,$-$0\deg.2), showing emission at $-$128 \kms\textrm{ }and no absorption features. We also notice that the CO tangent velocity at the line of sight $l =$ 338\deg.500 is $-$126 \kms\textrm{ }\citep{alvarez90}, very close to the CO radial velocity of the cloud ($V_{lsr} = -$119.1 \kms). We adopt the near distance to the cloud. The 6.7 GHz methanol maser 339.053$-$0.315 is associated to this cloud. Using the HI self-absorption method, \citet{green_mcclure2011} put the source at the far distance. This contradicts our determination of the distance to this GMC. The cloud is clearly associated to the near side of the \emph{3-kpc expanding} arm in the longitude-velocity diagram. Since it is known that the arm is expanding, the assumption that presence/absence of a HI self-absorption feature at the same LSR velocity would removed the distance ambiguity (near or far) of the masers at this Galacocentric radius is not clear. Given the LSR expantion velocity of both arm sides (near and far), by defnition, there is no material at the same LSR velocity to compare with. Hence, there will be always an absence of HI self-absorption that will put the masers artificially at the far distance. Also, the invariant manifolds models of \citet{romero2011b} reproduce very well the near side of the \emph{3-kpc expanding} arm in the longitude-velocity diagram, region that coincides with the position of this GMC.\\

%NCEN-24:
\item 58: We assign the near distance to the cloud based on observational size-to-linewidth relationship (\emph{first Larson's Law}). Putting the cloud at the far distance yields a large physical radius ($\sim$ 103 pc) deviating the cloud from the general trend of the relationship (\dv $=$ 6.8 \kms).\\

%NCEN-23:
\item 59: The HII region G338.742$+$0.641 has a visual optical counterpart \citep{caswell87}. HI absorption data of the source HII region G338.943$+$0.604 place this cloud at the near distance \citep{caswell87}. For the \IRAS source G338.851$+$0.406, \citet{44} presents the OH spectrum showing emission at $-$62 \kms, $-$60 \kms, and $-$40 \kms\textrm{ }at 1665 MHz, 1612 MHz, and 1720 MHz, respectively, and absorption at $-$2 \kms\textrm{ }at 1667 MHz. The absence of absorption toward the terminal velocity ($-$128 \kms\textrm{ }as shown by \citep{alvarez90}) at the line of sight $l = $ 339\deg.000 locates this source at the near distance. Based on HI absorption data, \citet{ur2011} place the HII region G338.9217$+$0.6233 at the near distance. Using recombination line velocities ($v_{rrl}$) from continuum sources and HI absorptions along the same line of sight,  \citet{jones2012} locates the continuum source G338.922$+$0.622 ($v_{rrl} = -63$ \kms) and G338.711$-$0.644 ($v_{rrl} = -62$ \kms) associated to the cloud at the near distance.\\

%NORM-F7.1:PERTECENE AL FAR SIDE THE 3 KPC
\item 60: We based our determination of the near or far distance to this cloud by inspection of the CO latitude-velocity maps in \citet{bron88b}. We notice that, in general, clouds at the far distance tend to have a small solid angle and small Galactic latitude (because of the distance projection) but still, if the cloud is massive, a large velocity dispersion (\dv $\sim$ 10 \kms). The elongated structure in radial velocity of this cloud is very similar to those ``far'' clouds with better determination of their distance ambiguity. It is also interesting to notice that placing the cloud at the far distance, there is a much better agreement between the molecular and viral masses of the cloud, within a factor of two, while at the near distance, the difference between the masses is increased to a factor of 4. The location of the cloud in the longitude-velocity diagram does not allow to distinguish between far and near distances mainly because of the position of the ``far distance'' clouds of the \emph{3-kpc expanding} arm. Although the continuity of the near side of the \emph{Norma} spiral arm suggests the near side to the cloud, it cannot be ruled out that the cloud belongs to the far side of the  \emph{3-kpc expanding} arm. Since the near and far sides of the \emph{3-kpc expanding} arm are separated by more than 100 \kms\textrm{ }\citep{dame2008}, some clouds at the far distance are expected to cross the \emph{Norma} and \emph{Centaurus} spiral arms along the radial velocity axis above Galactic longitude $l = $ 337\deg\textrm{ }(tangent point). We choose the far distance to the cloud, but the near distance can not be ruled out completely.\\

%NORM-F17:
\item 61: Using recombination line velocities ($v_{rrl}$) from continuum sources and HI absorptions along the same line of sight,  \citet{jones2012} locates the continuum source G339.289$+$0.233 ($v_{rrl} = -71$ \kms) associated to the cloud at the far distance. Given the velocity dispersion of the cloud (\dv = 9.3 \kms), putting the cloud at the far distance yields a physical radius ($\sim$ 61 pc) more consistent with the observational size-to-linewidth relationship (\emph{first Larson's Law}) than the physical radius estimated from the near distance ($\sim$ 29 pc). Therefore, the far distance is assigned to the cloud. We also notice that, in general, clouds at the far distance tend to have a small solid angle and small Galactic latitude (because of the distance projection) but still, if the cloud is massive, a large velocity dispersion (\dv $\sim$ 10 \kms). The elongated structure in radial velocity of this cloud is very similar to those ``far'' clouds with better determination of their distance ambiguity.\\

%NCEN-26.4:
\item 63: The HII region G339.128$-$0.408 associated to the cloud has a visual counterpart and presents recombination line emission at $-$37  \kms\textrm{ }but no absorption features \citep{caswell87}, putting the cloud at the near distance. Given the velocity dispersion of the cloud (\dv = 7.2 \kms), putting the cloud at the near distance yields a physical radius ($\sim$ 31 pc) much more consistent with the observational size-to-linewidth relationship (\emph{first Larson's Law}) than the physical radius estimated from the far distance ($\sim$ 156 pc). The cloud is located far off the Galactic plane at $b = -$1\deg.250. Putting this cloud at the far distance implies a distance off the plane of 288 pc toward negative latitudes. \citet{bron88} estimated the distribution of the molecular layer for the southern Galaxy. At the galactocentric radius of this cloud ($R_{gal} =$ 6.0 kpc), the molecular ring centroid lies at z = $-$13 pc, with a half width half maximum extension of  $\sim$ 53 pc. Following the molecular gas distribution, a far distance to this cloud results unlikely. Therefore, we adopt the near distance in this case. The 6.7 GHz methanol maser 340.034$-$1.110 is associated to this cloud. The removal of its distance ambiguity done by \citet{green_mcclure2011} confirms our distance determination for its parent GMC.\\      

%NCEN-25
\item 65: The HII region G341.050$-$0.100 has a visual counterpart and presents \HHCO absorption at $-$42 \kms\textrm{ }\citep{caswell87}, very close to the CO radial velocity of the cloud ($V_{lsr} = -$42.9 \kms) locating it at the near distance. The \IRAS source G340.053$-$0.237 associated to this cloud presents absorption at $-$53.2 \kms\textrm{ }and $-$35.3 \kms\textrm{ }and recombination line emission at $-$52 \kms\textrm{ }\citep{80,caswell87}. Considering that the CS line velocity for this source is at $-$53.1 \kms\textrm{ }and the absence of absorption features between the emission and terminal velocities, the source is located at the near distance. Based on HI absorption data, \citet{ur2011} place the HII region G340.2480$-$0.3725 at the near distance. Using recombination line velocities ($v_{rrl}$) from continuum sources and HI absorptions along the same line of sight,  \citet{jones2012} locates the continuum source G341.211$-$0.233 ($v_{rrl} = -39$ \kms) and G340.278$-$0.200 ($v_{rrl} = -43$ \kms) associated to the cloud at the near distance. Putting this cloud at the far distance implies an unrealistic physical radius of 221 pc, displacing the cloud far away of the observational size-to-linewidth relationship (\dv = 12.3 \kms).\\

%NORM-7.2:
\item 66: Based on \CCO and H I absorption data, \citet{busfield2006} locates three MYSOs G339.6221$-$00.1209 ($-$93.5 \kms), G339.8712$-$00.6701 ($-$91.89 \kms), and G340.5454$-$00.3754 ($-$90.3 \kms) at the far distance. Given the velocity dispersion of the cloud (\dv = 5.8 \kms), putting the cloud at the near distance yields a physical radius ($\sim$ 69 pc) by far more consistent with the observational size-to-linewidth relationship (\emph{first Larson's Law}) than the physical radius estimated from the far distance ($\sim$ 127 pc). The location of the cloud in the longitude-velocity diagram does not allow to distinguish between far and near distances mainly because of the position of the ``far distance'' clouds of the \emph{3-kpc expanding} arm. Although the continuity of the near side of the \emph{Norma} spiral arm suggests the near distance to the cloud, it cannot be ruled out that the cloud actually belongs to the far side of the \emph{3-kpc expanding} arm. Since the near and far sides of the \emph{3-kpc expanding} arm are separated by more than 100 \kms\textrm{ }\citep{dame2008}, some clouds at the far distance are expected to cross the \emph{Norma} and \emph{Centaurus} spiral arms along the radial velocity axis above Galactic longitude $l =$ 337\deg\textrm{ }(tangent point). We adopt the near distance to his cloud, but given the contradictory evidence, the far distance can not be ruled out.\\

%NCEN-26.3:
\item 67: The HII region G340.777$-$1.008 is associated to the optical object RCW110 (G340.79$-$1.01) \citep{caswell87,14}. Using recombination line velocities ($v_{rrl}$) from continuum sources and HI absorptions along the same line of sight,  \citet{jones2012} locates the continuum source G340.789$-$1.022 ($v_{rrl} = -25$ \kms) associated to the cloud at the near distance. The cloud is located far off the Galactic plane at $b = -$1\deg.000. Putting this cloud at the far distance implies a distance off the plane of 232 pc toward negative latitudes. \citet{bron88} estimated the distribution of the molecular layer for the southern Galaxy. At the galactocentric radius of this cloud ($R_{gal} =$ 6.0 kpc), the molecular ring centroid lies at z = $-$13 pc, with a half width half maximum extension of $\sim$ 53 pc. Following the molecular gas distribution, a far distance to this cloud results unlikely. Therefore, we adopt the near distance in this case.\\

%NORM-7.4:
\item 68: The observational size-to-linewidth relationship suggests the near distance because it improves the place of the cloud in the general trend of the relationship. Also the continuity on the near side of the \emph{Norma} spiral arm also suggests the near distance. We locate locate this cloud at the near distance.\\  

%NCEN26.5
\item 69:  In the case of this cloud, none of the previous methods could distinguish between near and far distance, and therefore, no distance was assigned to the cloud.\\

%NCEN-26.2:
\item 72: In \citet{busfield2006}, the authors identified three MYSOs G342.2263$-$00.3801 ($-$25.9 \kms), G342.3877$+$00.0742 ($-$25.9 \kms), and G342.3891$-$00.0723 ($-$21.5 \kms) to be at the near distance. These three sources coincide in space and radial velocity ($V_{lsr} = -$26.6 \kms) with this cloud. Given the velocity dispersion of the cloud (\dv = 5.4 \kms), putting it at the near distance yields a physical radius ($\sim$ 30 pc) by far more consistent with the observational size-to-linewidth relationship (\emph{first Larson's Law}) than the physical radius estimated from the far distance ($\sim$ 152 pc). The 6.7 GHz methanol maser 342.446$-$0.072 is associated to this cloud. The removal of its distance ambiguity done by \citet{green_mcclure2011} confirms our distance determination for its parent GMC.\\

%3KPC4.2:
\item 73: Using recombination line velocities ($v_{rrl}$) from continuum sources and HI absorptions along the same line of sight, \citet{jones2012} locates the continuum source G342.255$+$0.300 ($v_{rrl} = -122$ \kms) at the tangent point. This would put the cloud at the tangent point. However, their adopted velocity range (25 \kms) toward the terminal velocity  to assign a source to the tangent point is larger than the one we use in this work (10 \kms) . We stick to our limit and, following the discussions for clouds 62, 64, 70, 71, 77, 83, 84, 89, and 91 at the end of the appendix, adopt the near distance to the cloud. The 6.7 GHz methanol maser 342.251$+$0.308 is associated to this cloud. Using the HI self-absorption method, \citet{green_mcclure2011} put the source at the far distance. This contradicts our determination of the distance to this GMC. The cloud is clearly associated to the near side of the \emph{3-kpc expanding} arm in the longitude-velocity diagram. Since it is known that the arm is expanding, the assumption that presence/absence of a HI self-absorption feature at the same LSR velocity would removed the distance ambiguity (near or far) of the masers at this Galacocentric radius is not clear. Given the LSR expantion velocity of both arm sides (near and far), by defnition, there is no material at the same LSR velocity to compare with. Hence, there will be always an absence of HI self-absorption that will put the masers artificially at the far distance. Also, the invariant manifolds models of \citet{romero2011b} reproduce very well the near side of the \emph{3-kpc expanding} arm in the longitude-velocity diagram, region that coincides with the position of this GMC.\\

%NORM-F16:
\item 74: Based on \CCO and HI absorption data, \citet{busfield2006} assigned the far distance to the \IRAS source G342.954$-$0.311 associated with this cloud. Since the HI data present no absorption at the \CCO radial velocity of source, its location at the far distance is almost certain.\\

%NCEN-26.6:
\item 75: In the case of this cloud, none of the previous methods could distinguish between near and far distance, and therefore, no distance was assigned to the cloud.\\

%NORM-8:
\item 76: Based on \CCO and HI data, \citet{busfield2006} locate the MYSO G342.3877$+$00.0742 ($-$77.5 \kms) at the near distance, but the HI absorption feature used to discriminate between near or far distance is less prominent than the 20\% of the surrounding continuum emission, making the determination of the distance dubious according to the author. Locating this cloud at the far distance yields a large physical radius of $\sim$ 170 pc. We also notice that clouds at distances further than 4 kpc usually subtend angular scales smaller than 1\deg\textrm{ }\citep{green2011} while the angular size of the cloud in Galactic longitude is larger than 2\deg\textrm{ }and in Galactic latitude it is larger than 1\deg. The observational size-to-linewidth relationship suggests the near distance because it improves the place of the cloud in the general trend of the relationship as well as the continuity of the near side of the \emph{Norma} spiral arm. On the other hand, the same authors place three MYSOs G342.3877$+$00.0742 ($-$84.8 \kms), G342.3891$-$00.0723 ($-$85.5 \kms), and G342.8089$-$00.3825 ($-$80.6 \kms) at the far distance. The MYSOs coincide in position and radial velocity with the GMC, suggesting the far distance to the cloud. If we adopted the far distance to this cloud, it would become the most massive GMC in our catalog, with a molecular mass of \MHH $=$ 8.7 $\times$ 10$^{6}$ M$\solar$, the same as GMC G337.750$+$0.000 at the far side of the \emph{Norma} spiral arm. Nonetheless, its FIR Luminosity would be very low around $L_{FIR} =$ 2.7 $\times$ 10$^{5}$ L$\solar$, which is only 4\% of the FIR luminosity of the Norma GMC. This would yield and extremely massive GMC with almost no massive star formation activity while clouds of the same order of molecular mass in our catalog are very active in forming massive stars. This difference is counterintuitive with what we know from GMCs. We believe, it is unlikely that such a massive GMC could have almost no UC H II regions tracing its massive star formation activity, as it is the case for the Norma clouds, and therefore we favor the near distance assignement to this cloud. The 6.7 GHz methanol maser 341.367$+$0.336 is apparently associated to this cloud. The removal of its distance ambiguity done by \citet{green_mcclure2011} seems to contradict our distance determination for the GMC. Nonetheless, there is an explanation for this. The maser is very close to a local maximum, far from the bulk of the integrated CO emission in the GMC's spatial map. The local maximum can be attributed to a less massive cloud at the far distance (if the distance determination of the associated maser is right) whose CO emission overlaps in phase space with the CO emission of the GMC at the near distance. Since we are sensitive to only the most massive GMCs at the far distance, this is to be expected in some cases. In this sense, there is no contradiction with our determination of the GMC distance, because the maser is associated to a different and less massive cloud at the far distance that we can not properly isolate with the ASM subtraction technique applied in this work.\\

%NCEN-26.1:
\item 78: The HII region G343.150$-$0.402 presents \HHCO absorption at $-$23 \kms\textrm{ }and recombination line emission at $-$33 \kms\textrm{ }\citep{caswell87}. The lack of absorption features between the recombination and terminal velocities locates the source almost certainly at the near distance. The absorption feature at $-$23 \kms\textrm{ }coincides with the CO velocity dispersion of the cloud ($V_{lsr} = -$26.0 \kms, \dv $=$ 10.0 \kms), locating the cloud in front of the HII region at the near distance. The 6.7 GHz methanol maser 343.756$-$0.163 is associated to this cloud. The removal of its distance ambiguity done by \citet{green_mcclure2011} confirms our distance determination for its parent GMC.\\

%NORM-9: 
\item 79: The HII region G344.439$+$0.048 associated to this cloud presents \HHCO absorption at $-$65 \kms\textrm{ }and $-$62 \kms\textrm{ }and recombination line emission at $-$67 \kms. The absence of absorption features toward the terminal velocity locates the cloud at the near distance \citep{caswell87}. Based on HI absorption data, \citet{ur2011} place the HII regions G344.3976$+$0.0533 and G344.4257$+$0.0451 at the near distance.\\\\ 

%NORM-F29:
\item 80: The HII region G346.206$-$0.071 located at the near distance lies in front of this cloud and presents no absorption features within the velocity range of the cloud implying that the cloud has to be located behind the HII region, at the far distance (see discussion for cloud 73). Nonetheless, putting this cloud at the near distance is more in agreement with the general trend of the observational size-to-linewidth relationship. We adopt the far distance to this cloud, but keeping in mind that the near distance can not be completely discarded.\\       

%NORM-15:
\item 81: Given the velocity dispersion of the cloud (\dv = 4.3 \kms), putting the cloud at the near distance yields a physical radius ($\sim$ 32 pc) much more consistent with the observational size-to-linewidth relationship (\emph{first Larson's Law}) than the physical radius estimated from the far distance ($\sim$ 68 pc). Therefore, the near distance is assigned to the cloud.\\ 

%NORM-12 3KPC-FAR)
\item 82: We based our determination of far distance to this cloud by inspection of the \CO latitude-velocity maps in \citet{bron88b}. Although the extension in velocity of the cloud is small, it still presents an elongated structure similar to those ``far'' clouds with better determination of their distance ambiguity. The ratio between virial and molecular mass at the near ($\sim$ 2.3) and the far ($\sim$ 1.3) distance give a very weak indication that the cloud could be placed at the far distance. It is interesting to note that, assuming that the distances for clouds 82, 88, and 92 are correct, there seems to be a lane of clouds between $l =$ 345\deg\textrm{ }- 348\deg, and $v = -$95 \kms\textrm{ }- $-$60 \kms\textrm{ }located at the far distance, possible tracing the far side of the \emph{3-kpc expanding} arm.\\

%NCEN-27:
\item 85: The \IRAS source G345.400$-$0.941 is associated to the optical object RCW117 (G345.50$-$1.00) \citep{caswell87,44}. \citet{1} and \citet{82} present the HI spectrum showing absorption extending to $-$34 \kms. There is a peak of emission at $-$100 \kms\textrm{ }in the spectrum of both references which confirms the near distance. \citet{124} and \citet{134} discuss this source in the context of explaining anomalous motions of HI clouds. \citet{134} places the source at 4.7 kpc which corresponds to the near distance.\\

%NCAR-28:
\item 86: The HII regions G345.031$+$1.540, G345.231$+$1.035, G345.308$+$1.471, and G345.404$+$1.406 associated to the cloud posses visual optical counterparts \citep{25,caswell87} locating them at the near distance. The \IRAS source G345.208$+$1.028 and G345.393$+$1.399 are associated with the supernova remnant CTB 35 (part of RCW116) and RCW116 (G345.00$+$1.70), respectively \citep{1,gyg76,80,121}. According to the evidence, the near distance to the cloud is almost certain.\\ 

%NCEN-30:
\item 87: The HII region G346.206$-$0.071 presents no absorption features but recombination line emission at $-$108 \kms\textrm{ }\citep{caswell87}. Since the CO terminal velocity is $-$171 \kms\textrm{ }at the line of sight $l  =$ 346\deg\textrm{ }\citep{alvarez90}, it is unlikely that, being this cloud at the far distance ($\sim$ 11.2 kpc), presents no absorption features across the Galactic plane, favoring the near distance for this source. Putting this cloud at the far distance yields a physical radius $\sim$ 96 pc and displaces the position of the cloud from the general trend in the observational \emph{first Larson's law} (\dv $=$ 7.7 \kms). We adopt the near distance to this cloud.\\
%PARECE LEJANA

%NORM-10 (3KPC-FAR)
\item 88: The HII region G346.206$-$0.071 presents no \HHCO absorption features but recombination line emission at $-$108 \kms\textrm{ }favoring the far distance. Also, given the large velocity dispersion of the cloud (\dv = 13.9 \kms), putting the cloud at the far distance yields a physical radius ($\sim$ 90 pc) much more consistent with the observational size-to-linewidth relationship (\emph{first Larson's Law}) than the physical radius estimated from the near distance ($\sim$ 47 pc). We also notice that, in general, clouds at the far distance tend to have a small solid angle and small Galactic latitude (because of the distance projection) but still, if the cloud is massive, a large velocity dispersion (\dv $\sim$ 10 \kms). The elongated structure in radial velocity of this cloud is very similar to those ``far'' clouds with better determination of their distance ambiguity. We adopt the far distance to the cloud.\\

%NORM-14: 
\item 90: On the one hand, given the velocity dispersion of the cloud (\dv = 7.4 \kms), putting the cloud at the near distance yields a physical radius ($\sim$ 50 pc) much more consistent with the observational size-to-linewidth relationship (\emph{first Larson's Law}) than the physical radius estimated from the far distance ($\sim$ 118 pc). On the other hand, putting the cloud at the far distance implies a distance off the plane of 202 pc toward positive latitudes. \citet{bron88} estimated the distribution of the molecular layer for the southern Galaxy. At the galactocentric radius of this cloud ($R_{gal} =$ 3.9 kpc), the molecular ring centroid lies at z = 2 pc, with a half width half maximum extension of  $\sim$ 83 pc. Following the molecular gas distribution, a far distance to this cloud results unlikely. Therefore, we adopt the near distance in this case.\\   

%NORM-11 (3KPC-FAR
\item 92: The HII region G347.386$+$0.266 presents no \HHCO absorption features but recombination line emission at $-$97 \kms. Given the terminal velocity at the line of sight of this source ($\sim$ $-$200 \kms, estimated by simple inspection of the longitude-velocity diagram of the CO emission) and the absence of absorption features, we locate the HII region at the near distance. Since the HII region coincides with position of the cloud on the sky, the absence of absorption features suggests the far distance to the cloud. We notice that, in general, clouds at the far distance tend to have a small solid angle and small Galactic latitude (because of the distance projection) but still, if the cloud is massive, a large velocity dispersion (\dv $\sim$ 10 \kms). The elongated structure in radial velocity of this cloud is very similar to those ``far'' clouds with better determination of their distance ambiguity. We adopt the far distance to the cloud.\\

% 62 (3KPC-2c), 64 (3KPC-3), 70 (3KPC-4.4), 71 (3KPC-4.3), 77 (3KPC-4.1), 83 (3KPC-6), 84 (3KPC-5), 89 (3KPC-7), and 91 (3KPC-8)     
\item GMCs belonging to the \emph{3-kpc expanding arm}: 62, 64, 70, 71, 77, 83, 84, 89, and 91: In order to distinguish between far and near distances for these clouds, we take advantage of the well known $\sim$ $-$53 \kms\textrm{ }expansion velocity of the \emph{3-kpc expanding} spiral arm. The position of these clouds in the longitude-velocity diagram clearly corresponds to the location of the arm in the fourth Galactic quadrant as it has been traced by the 21 cm line observations in previous surveys \citep{rougoor60}. \citet{dame2008} traced, for the first time, the far side of the spiral arm between Galactic longitudes $l = $ 13\deg\textrm{ }and $l = -$12\deg. The far side of the arm presents also an expanding velocity of $+$56 \kms\textrm{ }being remarkably similar to its near counterpart, and indicating a more than 100 \kms\textrm{ }separation between the near and far sides of the arm, at least, in the range of Galactic longitudes covered in their work. The emission from both sides of the arm behaves as parallel lanes of emission across Galactic longitude. \citet{dame2008} fit this emission radial velocity to be a function of Galactic longitude yielding $v = -$53.1 $+$ 4.16$l$ for the near side and, $v =$ 56 $+$ 4.08$l$ for the far side. Concerning the near side fit, we note that it agrees extremely well with the position of these clouds in our catalog tracing this arm up to $l =$ 342\deg.375. At smaller longitudes, the position of the clouds in phase space deviates from this trend. Due to the good spatial and velocity correspondence between the near side of the \emph{3-kpc expanding} arm and the clouds 62, 64, 70, 77, 83, 84, 89, and 91 in our catalog, we are confident in our location of these clouds at the near distance. In the case of cloud 64, the 6.7 GHz methanol maser 340.118$-$0.021, 340.182$-$0.047, and 340.249$-$0.04 are associated to it. Using the HI self-absorption method, \citet{green_mcclure2011} put the masers at the far distance. This contradicts our determination of the distance to this GMC. The cloud is clearly associated to the near side of the \emph{3-kpc expanding} arm in the longitude-velocity diagram. Since it is known that the arm is expanding, the assumption that presence/absence of a HI self-absorption feature at the same LSR velocity would removed the distance ambiguity (near or far) of the masers at this Galacocentric radius is not clear. Given the LSR expantion velocity of both arm sides (near and far), by defnition, there is no material at the same LSR velocity to compare with. Hence, there will be always an absence of HI self-absorption that will put the masers artificially at the far distance. Also, the invariant manifolds models of \citet{romero2011b} reproduce very well the near side of the \emph{3-kpc expanding} arm in the longitude-velocity diagram, region that coincides with the position of this GMC.\\

\end{itemize}

\clearpage

%%%%%%%%%%%%%%%%%%%%%%FIGURAS CATALOGO GMCs%%%%%%%%%%%%%%%%%%%%%%%%
%DIAGRAMA LV DATOS CO
\clearpage

%DIAGRAMA LV NUBES
\clearpage

%COMPOSITE SPECTRUM: nube norm3-d
%FIGURA 3
\clearpage

%MAPA LB NUBE 43, NORM4
\clearpage

%DELTA V vs R
\clearpage

%DENSIDAD vs R GMCs
\clearpage

%MVIR vs LCO
\clearpage

%MASS SPECTRUM SM
\clearpage

%%%%%%%%%%%%%%%%%%%%%%FIGURAS BRAZOS ESPIRALES %%%%%%%%%%%%%%%%%%%%%%%%

%DIAGRAMA LV CAJA COLORES
\clearpage

%MAPA LB BRAZOS
\clearpage

%FACE ON MODELO BRAZOS
\clearpage

%%%%%%%%%%%%%%%%%%%%%%%%%MASSIVE STAR FORMATION%%%%%%%%%%%%%%%%%%%%%%%

%MSFR LFIR vs MASAS
\clearpage

%%%%%%%%%%%%%%%%%%%%%%%%%FIRUGA APPENDIX A%%%%%%%%%%%%%%%%%%%%%%%

%DIAGRAMA LV 13CO Y 12CO
\clearpage

%INTESIDAD MODELO Y RAW DATA
\clearpage

%TPEAK and MOLECULAR MASS  GMCs AND SM
\clearpage

%%%%%%%%%%%%%%%%%%%%%%%%%%%%%%%%%%%%%%%%%%%%%%%%%%%%%%%%%%%%

\clearpage
\begin{deluxetable}{ccc}
%\tabletypesize{\small}
%\tablewidth{0.5\textwidth}
\tablecolumns{3}
\tablecaption{The Columbia - U. de Chile Deep CO Survey of the Southern Milky Way.\label{tbl-1}}

\startdata
\hline \hline \\
\multicolumn{1}{l}{Galactic longitude} & 
\multicolumn{1}{l}{300\deg { }to 348\deg}  &  
\multicolumn{1}{l}{ } \\ 

\multicolumn{1}{l}{Galactic latitude} & 
\multicolumn{1}{l}{ $-$2\deg { }to +2\deg}  &  
\multicolumn{1}{l}{ } \\
 
\multicolumn{1}{l}{Velocity coverage}   & 
\multicolumn{1}{l}{$-$166 \kms\textrm{ }to +166 \kms} &  
\multicolumn{1}{l}{(300\deg $\leq l \leq$ 335\deg)}  \\

\multicolumn{1}{l}{ }                    & 
\multicolumn{1}{l}{ $-$180 \kms\textrm{ }to +153 \kms} &  
\multicolumn{1}{l}{(335\deg $\leq l \leq$ 345\deg)}  \\

\multicolumn{1}{l}{ }                   & 
\multicolumn{1}{l}{$-$218 \kms\textrm{ }to +144 \kms} &  
\multicolumn{1}{l}{(345\deg $\leq l \leq$ 348\deg)}  \\

\multicolumn{1}{l}{Sampling interval} & 
\multicolumn{1}{l}{0$^\circ$.125} &  
\multicolumn{1}{l}{(0$^\circ$.00 $ \leq |b| \leq$ 0$^\circ$.75)} \\

\multicolumn{1}{l}{ } & 
\multicolumn{1}{l}{0$^\circ$.250} &  
\multicolumn{1}{l}{(0$^\circ$.75 $ \leq |b| \leq$ 2$^\circ$.00)} \\

\multicolumn{1}{l}{Telescope HPBW } & 
\multicolumn{1}{l}{ 0$^\circ$.147} &  
\multicolumn{1}{l}{ } \\

\multicolumn{1}{l}{Velocity resolution} & 
\multicolumn{1}{l}{ 1.3 \kms} &  
\multicolumn{1}{l}{ } \\

\multicolumn{1}{l}{Sensitivity\tablenotemark{a}} & 
\multicolumn{1}{l}{ $\Delta$ T$_{rms} \leq$ 0.13 K} &  
\multicolumn{1}{l}{ } \\

\multicolumn{1}{l}{Main Beam efficiency} & 
\multicolumn{1}{l}{$\eta = 0.82$ }& 
\multicolumn{1}{l}{ } \\

\enddata

\tablecomments{A detailed description of the Columbia Southern Deep CO Survey of the Milky Way can be found in \citet{bron88,bron88b}.}

\tablenotetext{a}{ At velocity resolution of 1.3 \kms}

\end{deluxetable}

\clearpage

\begin{deluxetable}{lcccccccrc}
\tabletypesize{\footnotesize}
\tablewidth{1.\textwidth}

\tablecolumns{10}
\tablecaption{Giant Molecular Clouds in the Fourth Galactic Quadrant, Within the Solar Circle.\label{tbl-2}}

\tablehead{

\multicolumn{1}{l}{\textbf{cloud}}        	& 
\multicolumn{1}{c}{\textbf{$l$}}                & 
\multicolumn{1}{c}{\textbf{$b$}}          	&  
\multicolumn{1}{c}{\textbf{$V_{lsr}$}}     	& 
\multicolumn{1}{c}{\textbf{$\Delta v$}}   	&
\multicolumn{1}{c}{\textbf{$D$}}          	&  
\multicolumn{1}{c}{\textbf{$D.R.$}}       	&  
\multicolumn{1}{r}{\textbf{$R$}}          	&  
\multicolumn{1}{c}{\textbf{$M_{virial}$}}   	& 
\multicolumn{1}{c}{\textbf{$M(H_{2})$}}   	\\

\multicolumn{1}{l}{ }                  	& 
\multicolumn{1}{c}{ }                  	& 
\multicolumn{1}{c}{ }                  	& 
\multicolumn{1}{c}{ }                  	& 
\multicolumn{1}{c}{\textbf{$(FWHM)$}}  	&
\multicolumn{1}{c}{ }                  	&  
\multicolumn{1}{c}{ }                  	&  
\multicolumn{1}{r}{ }                  	& 
\multicolumn{1}{c}{ }                  	& 
\multicolumn{1}{c}{\textbf{GMCs} }    	\\

\multicolumn{10}{c}{}\\ 

\multicolumn{1}{l}{}                                	& 
\multicolumn{1}{c}{\textbf{(\deg)}}         		& 
\multicolumn{1}{c}{\textbf{(\deg)}}          		& 
\multicolumn{1}{c}{\textbf{(\kms)}}           		& 
\multicolumn{1}{c}{\textbf{(\kms)}}            		&
\multicolumn{1}{c}{\textbf{(kpc)}}              	&  
\multicolumn{1}{c}{\textbf{} }                     	&  
\multicolumn{1}{r}{\textbf{(pc)}}                   	& 
\multicolumn{1}{c}{\textbf{$\log (M/M_{\odot})$}}    	& 
\multicolumn{1}{c}{\textbf{$\log (M/M_{\odot})$}}   	\\}

\startdata
 1  & 302.125   & $+$0.750  &   $-$26.9\tablenotemark{*}	&  12.8  	&  2.9  	& N\tablenotemark{a,b}  		&  74  	&    6.41  &    6.13 	 	\\
 2  & 305.250   & $+$0.375  &   $-$35.1 				&  11.4  	&  3.7 	& N\tablenotemark{a,b}  		&  54    	&    6.17  &    6.16  		\\
 3  & 305.625   & $-$0.625   &   $-$20.7  				&   4.7   	&  8.0   	& F\tablenotemark{d}  		&  41  	&    5.29  &    5.45		\\
 4  & 305.750   & $+$1.250  &   $-$39.7\tablenotemark{*}  &   9.6   	&  5.0  	& T\tablenotemark{a,e,g}  		&  75 	&    6.16  &    6.10 		\\
 5  & 306.250   & $-$0.375   &   $-$18.7     				&   6.2  	&  8.4  	& F\tablenotemark{b} 		&  45  	&    5.56  &    5.49 		\\
 6  & 308.000   & $-$0.375   &   $-$13.2  				&   4.7  	&  9.3  	& F\tablenotemark{c}  		&  32  	&    5.18  &    5.13 		\\
 7  & 308.375   & $+$0.000  &   $-$47.3\tablenotemark{*}  &  10.8  	&  5.3  	& T\tablenotemark{g}  		&  69  	&    6.23  &    5.89 		\\
 8  & 309.000   & $-$0.250   &   $-$10.8  				&  12.8  	&  7.3  	& F\tablenotemark{d,f}  		&  65  	&    6.35  &    5.93 		\\
 9  & 309.125   & $-$0.375   &   $-$40.2  				&  11.5  	&  3.7  	& N\tablenotemark{a}  		&  55  	&    6.19  &    6.17		\\
10  & 311.125  & $+$0.125  &   $-$41.0\tablenotemark{*}	&  12.6  	&  3.5  	& N\tablenotemark{a,c}  		&  66  	&    6.34  &    6.41 		\\
11  & 311.875  & $+$0.125  &   $-$47.1  				&   9.4  	&  4.1  	& N\tablenotemark{b,c,d}  	&  58  	&    6.04  &    6.19		\\
12  & 312.375  & $+$0.125  &   $-$11.5  				&  10.9  	& 10.5  	& F\tablenotemark{d}  		&  62  	&    6.19  &    6.00 		\\
13  & 312.500  & $+$0.125  &   $-$49.4\tablenotemark{*}	&   6.5  	&  5.7  	& T\tablenotemark{g}  		&  35  	&    5.50  &    5.69 		\\
14  & 313.875  & $-$0.125   &   $-$44.9  				&   9.3  	&  3.6  	& N\tablenotemark{d,f}  		&  65  	&    6.07  &    6.31 		\\
15  & 314.250  & $+$0.250  &   $-$57.6  				&   8.0  	&  5.2  	& N\tablenotemark{c}  		&  48  	&    5.81  &    6.26 		\\
16  & 316.875  & $+$0.250  &   $-$21.3  				&  11.0  	& 10.8  	& F\tablenotemark{d}  		& 100  	&    6.41  &    6.42 		\\
17  & 317.750  & $+$0.000  &   $-$17.6  				&  12.8  	& 11.2  	& F\tablenotemark{c,d}  		& 106  	&    6.56  &    6.46 		\\
18  & 318.250  & $-$0.375   &   $-$41.8  				&  12.5  	&  3.1  	& N\tablenotemark{a,b,c}  	&  73  	&    6.38  &    6.45 		\\
19  & 319.375  & $-$0.125   &   $-$20.3  				&   7.5  	& 11.3  	& F\tablenotemark{c}  		& 100  	&    6.07  &    6.18 		\\
20  & 319.750  & $-$0.375   &    $-$8.4  				&  14.9  	& 12.3  	& F\tablenotemark{d}  		&  82  	&    6.58  &    6.27 		\\
21  & 320.375  & $+$0.125  &    $-$7.8  				&   9.0  	& 12.5  	& F\tablenotemark{c,d}  		&  82  	&    6.14  &    6.11		\\
22  & 320.750  & $-$0.375   &   $-$66.3  				&  11.7  	&  4.8  	& N\tablenotemark{b}  		&  32  	&    5.97  &    6.00 		\\
23  & 320.750  & $-$0.250   &   $-$55.3  				&   5.6  	&  3.9  	& N\tablenotemark{a,b}  		&  34  	&    5.36  &    5.52 		\\
24  & 321.875  & $+$0.000  &   $-$31.6  				&   4.7  	&  2.3  	& N\tablenotemark{c,d}  		&  58  	&    5.43  &    5.76 		\\
25  & 322.125  & $+$0.625  &   $-$54.8  				&   4.1  	&  3.8  	& N\tablenotemark{a,c}  		&  23  	&    4.92  &    5.32 		\\
26  & 323.500  & $+$0.000  &   $-$65.5  				&   4.0  	&  4.5  	& N\tablenotemark{b}  		&  33  	&    5.04  &    5.50 		\\
27  & 323.500  & $+$0.625  &   $-$45.3  				&  10.7  	&  3.2  	& N\tablenotemark{d,f}  		&  22  	&    5.72  &    5.23		\\
28  & 323.750  & $-$0.250   &   $-$52.3  				&   8.0  	&  3.6  	& N\tablenotemark{b,c}  		&  43  	&    5.77  &    5.78 		\\
29  & 326.625  & $+$0.625  &   $-$42.1  				&   9.6  	&  3.0  	& N\tablenotemark{a,b,c}  	&  56  	&    6.04  &    6.11 		\\
30  & 327.250  & $-$0.500   &   $-$46.7  				&   6.0  	&  3.3  	& N\tablenotemark{a,c}  		&  16  	&    5.08  &    4.99 		\\
31  & 327.250  & $-$0.250   &   $-$62.3  				&  10.2  	&  4.2  	& N\tablenotemark{c}  		&  52  	&    6.06  &    6.08 		\\
32  & 327.750  & $-$0.375   &   $-$70.7  				&  11.3  	&  4.6  	& N\tablenotemark{c,d,f}  		&  65  	&    6.24  &    6.22 		\\
33  & 328.250  & $-$0.500   &   $-$45.0  				&  11.9  	&  3.2  	& N\tablenotemark{a,b,c}  	&  54  	&    6.21  &    6.08 		\\
34  & 328.250  & $+$0.375  &   $-$92.0  				&  18.4  	&  6.0  	& N\tablenotemark{c}  		&  86  	&    6.78  &    6.71 		\\
35  & 329.250  & $+$0.750  &   $-$66.2  				&   5.5  	&\dots  	&\dots 					&\dots       & \dots     & \dots     		\\
36  & 329.375  & $-$0.250   &   $-$74.3  				&   7.7  	&\dots  	&\dots 					&\dots       & \dots     & \dots   		\\
37  & 329.500  & $+$0.125  &   $-$99.0  				&  12.1  	&  6.5  	& N\tablenotemark{c}  		&  49 	&    6.18  &    6.03 		\\
38  & 329.500  & $+$0.500  &   $-$80.9  				&   9.4  	&  5.2  	& N\tablenotemark{c}  		&  52  	&    5.99  &    5.93		\\
39  & 329.625  & $+$0.125  &   $-$65.3  				&   6.5  	&\dots  	&\dots 					&\dots 	& \dots     & \dots   		\\
40  & 330.000  & $+$1.000  &   $-$86.2  				&   4.7  	&  5.5  	& N\tablenotemark{e}  		&  21  	&    4.98  &    5.02 		\\
41  & 331.125  & $-$0.500   &   $-$65.0  				&  10.6  	&  4.3  	& N\tablenotemark{a,b,c}  	&  78  	&    6.26  &    6.37		\\
42  & 331.125  & $+$0.000  &   $-$45.3  				&   8.6  	&  3.3  	& N\tablenotemark{c}  		&  43  	&    5.83  &    5.82 		\\
43  & 331.500  & $-$0.125   &   $-$92.0  				&  15.9  	&  5.7  	& N\tablenotemark{b,c}  		&  92  	&    6.69  &    6.68		\\
44  & 333.000  & $+$0.750  &   $-$48.0  				&  5.0  	&  3.5  	& N\tablenotemark{c}  		&  34  	&    5.25  &    5.41		\\
45  & 333.250  & $-$0.375   &   $-$50.1  				&   8.9  	&  3.6  	& N\tablenotemark{a,c,d}  	&  67  	&    6.05  &    6.29		\\
46  & 333.625  & $-$0.125   &   $-$88.1  				&   9.6  	&  5.4  	& N\tablenotemark{b,c}  		&  75  	&    6.17  &    6.30 		\\
47  & 333.875  & $-$0.375   &  $-$105.0  				&  12.2  	&  6.3  	& N\tablenotemark{d,f}  		&  70  	&    6.34  &    6.19 		\\
48  & 334.125  & $+$0.500  &   $-$62.2  				&  12.2  	& 11.1  	& F\tablenotemark{c,d}  		&  95  	&    6.47  &    6.57		\\
49  & 334.250  & $-$0.125   &   $-$40.4  				&   8.2  	&  3.1  	& N\tablenotemark{c,d}  		&  64  	&    5.96  &    6.12 		\\
50  & 336.000  & $-$0.875   &   $-$48.3  				&   7.4  	&  3.6  	& N\tablenotemark{e}  		&  23  	&    5.44  &    5.25 		\\
51  & 336.875  & $+$0.125  &   $-$74.2  				&  16.0  	& 10.8  	& F\tablenotemark{b,c}  		& 121  	&    6.81  &    6.84 		\\
52  & 337.000  & $-$1.125   &   $-$42.2  				&   7.4  	&  3.3  	& N\tablenotemark{e}  		&  29  	&    5.53  &    5.49 		\\
53  & 337.250  & $+$0.000  &  $-$116.1  				&  14.2  	&  6.7  	& N\tablenotemark{c}  		&  75  	&    6.50  &    6.37 		\\
54  & 337.750  & $+$0.000  &   $-$55.1  				&  18.2  	& 11.7  	& F\tablenotemark{b,c}  		& 104  	&    6.86  &    6.94 		\\
55  & 338.000  & $-$0.125   &   $-$94.1  				&  16.2  	&  5.7  	& N\tablenotemark{c}  		&  57  	&    6.50  &    6.03		\\
56  & 338.250  & $-$1.000   &  $-$111.5  				&   7.4  	&  6.4  	& N\tablenotemark{e}  		&  37  	&    5.63  &    5.36  		\\
57  & 338.625  & $-$0.125   &  $-$119.1  				&   8.4  	&  6.7  	& N\tablenotemark{b,c}  		&  69  	&    6.01  &    6.07 		\\
58  & 338.875  & $-$0.625   &   $-$45.7  				&   6.8  	&  3.6  	& N\tablenotemark{d}  		&  30  	&    5.47  &    5.44		\\
59  & 339.000  & $+$0.625  &   $-$60.5  				&  10.6  	&  4.4  	& N\tablenotemark{a,c}  		&  35  	&    5.93  &    5.71 		\\
60  & 339.125  & $+$0.000  &  $-$100.3  				&   9.3  	&  9.9  	& F\tablenotemark{d}  		&  76  	&    6.14  &    6.01		\\
61  & 339.125  & $+$0.250  &   $-$78.9  				&   9.3  	& 10.7  	& F\tablenotemark{c,d}  		&  60  	&    6.04  &    5.97		\\
62  & 339.500  & $+$0.125  &  $-$111.6  				&   6.1  	&  6.4  	& N\tablenotemark{f}  		&  31  	&    5.38  &    5.21		\\
63  & 339.750  & $-$1.250   &   $-$30.7  				&   7.2  	&  2.8  	& N\tablenotemark{a,d,e}  	&  33  	&    5.56  &    5.51 		\\
64  & 340.250  & $+$0.000  &  $-$122.1  				&   8.6  	&  6.8  	& N\tablenotemark{f}  		&  77  	&    6.07  &    6.14		\\
65  & 340.375  & $-$0.375   &   $-$42.9  				&  12.3  	&  3.6  	& N\tablenotemark{a,c,d}  	&  63  	&    6.30  &    6.32 		\\
66  & 340.625  & $-$0.625   &   $-$90.8  				&   5.8  	&  5.7  	& N\tablenotemark{d}  		&  68  	&    5.68  &    5.86 		\\
67  & 340.750  & $-$1.000   &   $-$28.9  				&   5.5  	&  2.7  	& N\tablenotemark{a,e}  		&  25  	&    5.20  &    5.31 		\\
68  & 341.000  & $+$0.000  &  $-$101.1  				&   6.5  	&  6.0  	& N\tablenotemark{d,f}  		&  49  	&    5.64  &    5.43 		\\
69  & 341.375  & $+$0.250  &   $-$24.3  				&   8.6  	&\dots  	&\dots 					&\dots 	& \dots     & \dots  		\\
70  & 341.500  & $-$0.125   &  $-$129.3  				&  11.2  	&  7.0  	& N\tablenotemark{f}  		&  45  	&    6.08  &    5.69 		\\
71  & 341.500  & $+$0.000  &  $-$121.9  				&   6.5  	&  6.8  	& N\tablenotemark{f}  		&  37  	&    5.51  &    5.35 		\\
72  & 342.125  & $+$0.500  &   $-$26.6  				&   5.4  	&  2.7  	& N\tablenotemark{c,d}  		&  30  	&    5.27  &    5.35 		\\
73  & 342.250  & $+$0.250  &  $-$122.8  				&   8.3  	&  6.8  	& N\tablenotemark{f}  		&  47  	&    5.83  &    5.56 		\\
74  & 342.625  & $+$0.125  &   $-$41.5  				&   5.1  	& 12.5  	& F\tablenotemark{b}  		&  87  	&    5.67  &    6.12 		\\
75  & 342.750  & $-$0.500   &   $-$27.0  				&   7.7  	&\dots  	&\dots 					&\dots 	& \dots     & \dots  		\\
76  & 342.750  & $+$0.000  &   $-$79.8  				&  13.2  	& 5.4  	& N\tablenotemark{d,f}  		& 75  	&    6.44  &    6.35 		\\
77  & 343.250  & $+$0.125  &  $-$120.1 			 	&  10.3  	&  6.7  	& N\tablenotemark{f}  		&  66  	&    6.17  &    5.95		\\
78  & 344.125  & $-$0.625   &   $-$26.0  				&  10.0  	&  2.8  	& N\tablenotemark{c}  		&  38  	&    5.90  &    5.85 		\\
79  & 344.500  & $+$0.125  &   $-$68.2  				&  12.0  	&  5.1  	& N\tablenotemark{c}  		&  67  	&    6.31  &    6.10 		\\
80  & 344.500  & $+$0.125  &   $-$44.1  				&   9.6  	& 12.4  	& F\tablenotemark{c}  		& 131  	&    6.41  &    6.77 		\\
81  & 345.000  & $-$0.250   &   $-$85.9  				&   4.3  	&  5.8  	& N\tablenotemark{d}  		&  36  	&    5.16  &    5.18 		\\
82  & 345.125  & $-$0.250   &   $-$93.8  				&   4.7  	& 10.4  	& F\tablenotemark{d}  		&  52  	&    5.38  &    5.50 		\\
83  & 345.125  & $+$0.125  &  $-$119.6  				&  14.2  	&  6.7  	& N\tablenotemark{f}  		&  59  	&    6.40  &    5.83 		\\
84  & 345.250  & $-$1.000   &  $-$108.0  				&   7.4  	&  6.4  	& N\tablenotemark{f}  		&  37  	&    5.62  &    5.15 		\\
85  & 345.250  & $-$0.750   &   $-$22.1  				&   7.0  	&  2.6  	& N\tablenotemark{a,b}  		&  32  	&    5.52  &    5.52 		\\
86  & 345.250  & $+$1.000  &   $-$13.7  				&  10.1  	&  1.8  	& N\tablenotemark{a,b}  		&  34  	&    5.87  &    5.72		\\
87  & 345.875  & $+$0.000  &   $-$36.0  				&   7.7  	&  3.7  	& N\tablenotemark{c,d}  		&  27  	&    5.53  &    5.42 		\\
88  & 346.000  & $+$0.000  &   $-$79.8  				&  13.9  	& 10.8	& F\tablenotemark{c,d}  		&  90  	&    6.56  &    6.35 		\\
89  & 346.125  & $-$0.125   &  $-$104.9  				&   8.3  	&  6.4  	& N\tablenotemark{f}  		&  26  	&    5.57  &    5.14 		\\
90  & 346.500  & $+$1.000  &   $-$57.6  				&   7.4  	&  4.9  	& N\tablenotemark{d,e}  		&  49  	&    5.75  &    5.71		\\
91  & 347.000  & $+$0.250  &  $-$118.4  				&   7.4  	&  6.8  	& N\tablenotemark{f}  		&  49  	&    5.76  &    5.38 		\\
92  & 347.250  & $+$0.000  &   $-$68.8  				&   9.3  	& 11.1  	& F\tablenotemark{c,d}  		&  61  	&    6.04  &    5.94 		\\
\enddata

\tablenotetext{(*)}{.- The CO radial velocity of the cloud was corrected by $+$12.2 \kms\textrm{ }in order to take into account the unusual velocity excess toward terminal velocities up to galactocentric longitude 312\deg\textrm{ }reported by \citet{alvarez90}.}

%\tablenotetext{\textrm{ }}{.- Distances in parenthesis were estimated by correcting the CO radial %velocity of clouds belonging to the near side of the 3-kpc expanding arm by 53.1 \kms, according %to the linear fit done by \citet{dame2008}, and extended in the present work until galactic %longitude 342\deg. The CO radial velocities of clouds in the near side of the arm but at lower %galactic longitudes were not corrected.}

\tablenotetext{(\dots)}{.- The two-fold distance ambiguity could not be removed for these clouds.}

\tablecomments{The following letters represent the method in which the two-fold distance ambiguity was removed:}

\tablenotetext{a}{.- Spatial association with optical objects from the RWC catalog \citep{153} 
or visual optical counterparts \citep{caswell87}.}

\tablenotetext{b}{.- \IRAS source associated to the cloud with distance ambiguity already removed.}

\tablenotetext{c}{.- Presence or absence of absorption features from species like \HHCO or OH against the H$\alpha$ continuum emission from HII regions, or cold ($10$ - $30$ K) HI absorption against the warm ($100$ - $10^{4}$ K) HI continuum background.}

\tablenotetext{d}{.- Observational size-to-linewidth relationship (\emph{Larson's  Law}).}

\tablenotetext{e}{.- Latitude criterion.}

\tablenotetext{f}{.- Continuity of spiral arm.}

\tablenotetext{g}{.- CO radial velocity of the cloud close ($|v| < 10$ \kms) to the tangential velocity.}

\end{deluxetable}

\clearpage
\begin{table}
\tabletypesize{\huge}
\begin{center}
\caption{Fitted Parameters for the Logarithmic Spiral Arms Model in the Fourth Galactic Quadrant.\label{tbl-4}}

\begin{tabular}{|lccc|}

\tableline\tableline

\multicolumn{1}{|l}{\textbf{Spiral Arm}}  & 
\multicolumn{1}{c}{\textbf{$r_{\circ}$}}  & 
\multicolumn{1}{c}{\textbf{$p$}}          & 
\multicolumn{1}{c|}{\textbf{Tangent}}    \\
    
\multicolumn{1}{|l}{\textbf{ }}        & 
\multicolumn{1}{c}{\textbf{(kpc)}}    & 
\multicolumn{1}{c}{\textbf{(\deg)}}   & 
\multicolumn{1}{c|}{\textbf{(\deg)}}   \\

\tableline

Centaurus & 5.40 $\pm$ 0.14 & 13.4 $\pm$ 2.0 & 310 \\

Norma     & 3.72 $\pm$ 0.16 & 6.6 $\pm$ 2.3 & 330 \\

3-kpc     & 2.75 $\pm$  0.16& 5.6 $\pm$ 3.0 & 338 \\
\tableline 

\end{tabular}

\end{center}
\end{table}

\clearpage
%\twocolumn
%\setlength{\columnseprule}{1pt}
%\setlength{\columnsep}{5cm}
\linespread{1.2}

\begin{deluxetable}{ccrcc|ccrcc}
\tabletypesize{\scriptsize}
\tablewidth{0.94\textwidth}

\tablecolumns{10}
\tablecaption{FIR Luminosity and Massive Star Formation Efficiency for GMCs.\label{tbl-5}}

\tablehead{
\multicolumn{1}{l}{\textbf{cloud}}               &
\multicolumn{1}{c}{\textbf{\# sources}}     	& 
\multicolumn{1}{c}{\textbf{$F_{IRAS}$}}   	& 
\multicolumn{1}{c}{\textbf{$L_{FIR}$}}     	& 
\multicolumn{1}{c}{\textbf{$\epsilon$}}    	&
\multicolumn{1}{|l}{\textbf{cloud}}        	&
\multicolumn{1}{c}{\textbf{\# sources}}   	& 
\multicolumn{1}{c}{\textbf{$F_{IRAS}$}}   	& 
\multicolumn{1}{c}{\textbf{$L_{FIR}$}}    	& 
\multicolumn{1}{c}{\textbf{$\epsilon$}}  	\\
    
\multicolumn{5}{l}{ } &
\multicolumn{5}{|l}{ } \\

\multicolumn{1}{c}{ }                                   &                                
\multicolumn{1}{r}{ }                                   &  
\multicolumn{1}{c}{\textbf{($L_{\odot} kpc^{-2}$)}}       & 
\multicolumn{1}{c}{\textbf{$\log (L/L_{\odot})$}}        & 
\multicolumn{1}{c}{\textbf{($\%$)}}                    &
\multicolumn{1}{|c}{ }                                 &                                
\multicolumn{1}{r}{ }                                  &  
\multicolumn{1}{c}{\textbf{($L_{\odot} kpc^{-2}$)}}   	& 
\multicolumn{1}{c}{\textbf{$\log (L/L_{\odot})$}}     	& 
\multicolumn{1}{c}{\textbf{($\%$)}}                   \\}

\startdata

1  & 10	& 57344	& 5.69	& 2.4           & 37	& 4  & 32191  &	6.13 & 8.1 \\
2  & 11	& 76319	& 6.02	& 4.8           & 38	& 4  & 13121  &	\dots&\dots\\
3  & 1	& 306	& 4.30	& 0.5           & 41	& 7  & 54382  &	6.00 & 2.8 \\
4  & 1	& 1763	& 4.64	& 0.2           & 42	& 1  & 1462   &	4.19 & 0.2 \\
5  & 1	& 1693  & 5.08	& 2.5           & 43	& 10 & 85494  &	6.45 & 3.8 \\
8  & 3	& 1957	& 5.02	& 0.8           & 44	& 1  & 8433   &	5.00 & 2.5 \\
9  & 3	& 6501	& 4.95	& 0.4           & 45	& 14 & 218249 &	6.45 & 9.2 \\
10 & 8	& 30355	& 5.57  & 0.9           & 46	& 2  & 7361   &	5.34 & 0.7 \\
11 & 3	& 17899	& 5.48	& 1.3           & 49	& 1  & 3017   &	4.46 & 0.1 \\
13 & 1	& 1343	& 4.65	& 0.6           & 50	& 1  & 4492   &	4.76 & 2.1 \\
14 & 3	& 4059	& 4.72	& 0.2           & 51	& 2  & 36413  &	6.63 & 4.0 \\
15 & 3	& 9828	& 5.42	& 0.9           & 53	& 5  & 23309  &	6.02 & 2.9 \\
16 & 1	& 1825	& 5.33	& 0.5           & 54	& 8  & 46858  &	6.81 & 4.8 \\
17 & 1	& 797	& 5.00	& 0.2           & 57	& 1  & 4807   &	5.34 & 1.2 \\
18 & 11	& 56481	& 5.73	& 1.3           & 59	& 1  & 3119   &	4.77 & 0.7 \\
19 & 1	& 5874	& 5.88	& 3.2           & 61	& 1  & 2210   &	5.41 & 1.8 \\
20 & 1	& 748	& 5.05	& 0.4           & 62	& 2  & 5697   &	5.37 & 9.4 \\
21 & 2	& 13753	& 6.33	& 10.7          & 63	& 2  & 15141  &	5.07 & 2.4 \\
22 & 2	& 16087	& 5.57	& 2.4           & 64	& 3  & 5286   &	5.39 & 1.1 \\
23 & 1	& 6657	& 5.02	& 2.0           & 65	& 11 & 59526  &	5.88 & 2.4 \\
24 & 1	& 321	& 3.24	& 0.0           & 67	& 2  & 26713  &	5.30 & 6.4 \\
26 & 1	& 11122	& 5.35	& 4.7           & 69	& 1  & 1272   &	\dots&\dots\\
28 & 3	& 9132	& 5.08	& 1.3           & 74	& 2  & 5097   &	5.90 & 3.9 \\
29 & 7	& 93905	& 5.93	& 4.2           & 76	& 1  & 2304   &	4.83 & 0.2 \\
31 & 7	& 12227	& 5.33	& 1.1           & 78	& 5  & 22821  &	5.26 & 1.7 \\
32 & 3	& 12550	& 5.43	& 1.1           & 79	& 1  & 13478  &	5.55 & 1.8 \\
33 & 5	& 22562	& 5.36	& 1.3           & 80	& 1  & 254    &	4.59 & 0.0 \\
34 & 7	& 55326	& 6.31	& 2.6           & 85	& 2  & 63312  &	5.64 & 8.5 \\
35 & 1	& 1565	& \dots	& \dots         & 86	& 12 & 174892 &	5.77 & 7.2 \\
36 & 2	& 6612	& \dots	& \dots         & 88	& 1  &	2196  &	5.41 & 0.7 \\

\enddata
\end{deluxetable}
\newpage

\clearpage
\textheight 22cm
\textwidth 16.5cm

\linespread{1.1}

\begin{deluxetable}{lcccccc}
\tabletypesize{\footnotesize}
\tablewidth{0.78\textwidth}
\tablecolumns{7}
\tablecaption{Peak Antenna Temperature and Molecular Mass Derived from the Model Subtracted Dataset (MSD) and GMCs.\label{tbl-app}}

\tablehead{

\multicolumn{3}{l}{\textbf{ }}                         & 
\multicolumn{2}{c}{\textbf{MSD}}       	               & 
\multicolumn{2}{c}{\textbf{GMCs}}                      \\

\multicolumn{7}{c}{}\\ 

\multicolumn{1}{l}{\textbf{cloud}}                     	& 
\multicolumn{1}{c}{\textbf{$l$}}                        & 
\multicolumn{1}{c}{\textbf{$b$}}                        & 
\multicolumn{1}{c}{\textbf{$T_{peak}$}}          	&  
\multicolumn{1}{c}{\textbf{$M(H_{2})$}}          	& 
\multicolumn{1}{c}{\textbf{$T_{peak}$}}          	&  
\multicolumn{1}{c}{\textbf{$M(H_{2})$}}          	\\

\multicolumn{7}{c}{}\\ 

\multicolumn{1}{l}{}                                		& 
\multicolumn{1}{c}{\textbf{(\deg)}}                 		& 
\multicolumn{1}{c}{\textbf{(\deg)}}                 	   	& 
\multicolumn{1}{c}{\textbf{($K \Delta l\Delta b$)}} 		& 
\multicolumn{1}{c}{\textbf{$\log (M/M_{\odot})$}}           	& 
\multicolumn{1}{c}{\textbf{($K \Delta l\Delta b$)}}           	& 
\multicolumn{1}{c}{\textbf{$\log (M/M_{\odot})$}}               	\\}

\startdata
1  &    302.125  & $+$0.750  &       4.98  &       5.87  &       9.03  &       6.13     \\
 2  &    305.250  & $+$0.375  &       2.54  &       5.73  &       6.83  &       6.16     \\
 3  &    305.625  & $-$0.625  &       0.37  &       5.18  &       0.70  &       5.45     \\
 4  &    305.750  & $+$1.250  &       2.83  &       5.95  &       3.96  &       6.10     \\
 5  &    306.250  & $-$0.375  &       0.30  &       5.25  &       0.53  &       5.49     \\
 6  &    308.000  & $-$0.375  &       0.14  &       4.88  &       0.25  &       5.13     \\
 7  &    308.375  & $+$0.000  &       1.36  &       5.74  &       1.95  &       5.89     \\
 8  &    309.000  & $-$0.250  &       0.54  &       5.70  &       0.92  &       5.93     \\
 9  &    309.125  & $-$0.375  &       2.21  &       5.66  &       7.15  &       6.17     \\
10  &    311.125  & $+$0.125  &       4.24  &       5.94  &      12.59  &       6.41     \\
11  &    311.875  & $+$0.125  &       2.54  &       5.73  &       7.21  &       6.19     \\
12  &    312.375  & $+$0.125  &       0.28  &       5.66  &       0.62  &       6.00     \\
13  &    312.500  & $+$0.125  &       0.46  &       5.12  &       1.72  &       5.69     \\
14  &    313.875  & $-$0.125  &       4.08  &       5.82  &      12.60  &       6.31     \\
15  &    314.250  & $+$0.250  &       0.69  &       5.30  &       6.40  &       6.26     \\
16  &    316.875  & $+$0.250  &       0.66  &       6.05  &       1.54  &       6.42     \\
17  &    317.750  & $+$0.000  &       0.62  &       6.13  &       1.36  &       6.46     \\
18  &    318.250  & $-$0.375  &       6.25  &       6.00  &      17.55  &       6.45     \\
19  &    319.375  & $-$0.125  &       0.65  &       5.92  &       1.18  &       6.18     \\
20  &    319.750  & $-$0.375  &       0.35  &       6.03  &       0.63  &       6.27     \\
21  &    320.375  & $+$0.125  &       0.48  &       5.95  &       0.69  &       6.11     \\
22  &    320.750  & $-$0.375  &       0.43  &       5.20  &       2.76  &       6.00     \\
23  &    320.750  & $-$0.250  &       0.50  &       4.77  &       2.84  &       5.52     \\
24  &    321.875  & $+$0.000  &       8.01  &       5.43  &      16.88  &       5.76     \\
25  &    322.125  & $+$0.625  &       0.69  &       4.75  &       2.55  &       5.32     \\
26  &    323.500  & $+$0.000  &       0.70  &       4.88  &       2.90  &       5.50     \\
27  &    323.500  & $+$0.625  &       0.28  &       4.61  &       1.16  &       5.23     \\
28  &    323.750  & $-$0.250  &       1.23  &       5.24  &       4.22  &       5.78     \\
29  &    326.625  & $+$0.625  &       5.56  &       5.81  &      11.20  &       6.11     \\
30  &    327.250  & $-$0.500  &       0.73  &       4.79  &       1.13  &       4.99     \\
31  &    327.250  & $-$0.250  &       2.20  &       5.72  &       5.09  &       6.08     \\
32  &    327.750  & $-$0.375  &       2.21  &       5.86  &       5.10  &       6.22     \\
33  &    328.250  & $-$0.500  &       3.18  &       5.71  &       7.35  &       6.08     \\
34  &    328.250  & $+$0.375  &       1.91  &       6.23  &       5.69  &       6.71     \\
35  &    329.250  & $+$0.750  &       0.35  &       \dots  &       0.96  &       \dots     \\
36  &    329.375  & $-$0.250  &       0.40  &       \dots &       0.98  &       \dots     \\
37  &    329.500  & $+$0.125  &       0.58  &       5.60  &       1.60  &       6.03     \\
38  &    329.500  & $+$0.500  &       1.00  &       5.53  &       2.55  &       5.93     \\
39  &    329.625  & $+$0.125  &       0.31  &       \dots  &       1.01  &       \dots     \\
40  &    330.000  & $+$1.000  &       0.19  &       4.55  &       0.57  &       5.02     \\
41  &    331.125  & $-$0.500  &       4.68  &       6.09  &       8.84  &       6.37     \\
42  &    331.125  & $+$0.000  &       2.60  &       5.50  &       5.43  &       5.82     \\
43  &    331.500  & $-$0.125  &       3.60  &       6.40  &       6.90  &       6.68     \\
44  &    333.000  & $+$0.750  &       1.60  &       5.10  &       3.26  &       5.41     \\
45  &    333.250  & $-$0.375  &       8.49  &       6.11  &      12.87  &       6.29     \\
46  &    333.625  & $-$0.125  &       2.93  &       6.05  &       5.25  &       6.30     \\
47  &    333.875  & $-$0.375  &       0.75  &       5.69  &       2.39  &       6.19     \\
48  &    334.125  & $+$0.500  &       1.02  &       6.31  &       1.86  &       6.57     \\
49  &    334.250  & $-$0.125  &       5.64  &       5.77  &      12.68  &       6.12     \\
50  &    336.000  & $-$0.875  &       0.57  &       4.87  &       1.39  &       5.25     \\
51  &    336.875  & $+$0.125  &       1.70  &       6.62  &       2.79  &       6.84     \\
52  &    337.000  & $-$1.125  &       1.38  &       5.18  &       2.80  &       5.49     \\
53  &    337.250  & $+$0.000  &       1.36  &       6.06  &       2.78  &       6.37     \\
54  &    337.750  & $+$0.000  &       1.40  &       6.67  &       2.62  &       6.94     \\
55  &    338.000  & $-$0.125  &       0.84  &       5.77  &       1.52  &       6.03     \\
56  &    338.250  & $-$1.000  &       0.19  &       4.90  &       0.56  &       5.36     \\
57  &    338.625  & $-$0.125  &       1.37  &       5.85  &       2.28  &       6.07     \\
58  &    338.875  & $-$0.625  &       0.98  &       5.07  &       2.30  &       5.44     \\
59  &    339.000  & $+$0.625  &       1.11  &       5.48  &       1.91  &       5.71     \\
60  &    339.125  & $+$0.000  &       0.47  &       5.76  &       0.82  &       6.01     \\
61  &    339.125  & $+$0.250  &       0.29  &       5.62  &       0.65  &       5.97     \\
62  &    339.500  & $+$0.125  &       0.29  &       4.99  &       0.48  &       5.21     \\
63  &    339.750  & $-$1.250  &       1.94  &       5.16  &       4.30  &       5.51     \\
64  &    340.250  & $+$0.000  &       1.83  &       5.99  &       2.61  &       6.14     \\
65  &    340.375  & $-$0.375  &       5.07  &       6.03  &       9.85  &       6.32     \\
66  &    340.625  & $-$0.625  &       1.75  &       5.64  &       2.92  &       5.86     \\
67  &    340.750  & $-$1.000  &       2.14  &       5.07  &       3.73  &       5.31     \\
68  &    341.000  & $+$0.000  &       0.51  &       5.20  &       0.85  &       5.43     \\
69  &    341.375  & $+$0.250  &       0.75  &       \dots  &       1.73  &       \dots     \\
70  &    341.500  & $-$0.125  &       0.42  &       5.49  &       0.67  &       5.69     \\
71  &    341.500  & $+$0.000  &       0.40  &       5.20  &       0.57  &       5.35     \\
72  &    342.125  & $+$0.500  &       2.22  &       5.06  &       4.31  &       5.35     \\
73  &    342.250  & $+$0.250  &       0.51  &       5.42  &       0.71  &       5.56     \\
74  &    342.625  & $+$0.125  &       0.49  &       5.72  &       1.23  &       6.12     \\
75  &    342.750  & $-$0.500  &       1.92  &       \dots  &       3.61  &       \dots     \\
76  &    342.750  & $+$0.000  &       3.00  &       6.19  &       4.33  &       6.35     \\
77  &    343.250  & $+$0.125  &       0.94  &       5.76  &       1.46  &       5.95     \\
78  &    344.125  & $-$0.625  &       3.26  &       5.54  &       6.63  &       5.85     \\
79  &    344.500  & $+$0.125  &       1.86  &       5.89  &       3.02  &       6.10     \\
80  &    344.500  & $+$0.125  &       1.16  &       6.36  &       3.01  &       6.77     \\
81  &    345.000  & $-$0.250  &       0.51  &       4.99  &       0.79  &       5.18     \\
82  &    345.125  & $-$0.250  &       0.31  &       5.32  &       0.47  &       5.50     \\
83  &    345.125  & $+$0.125  &       0.48  &       5.62  &       0.78  &       5.83     \\
84  &    345.250  & $-$1.000  &       0.21  &       4.94  &       0.35  &       5.15     \\
85  &    345.250  & $-$0.750  &       1.89  &       5.09  &       5.16  &       5.52     \\
86  &    345.250  & $+$1.000  &       5.74  &       5.41  &      11.64  &       5.72     \\
87  &    345.875  & $+$0.000  &       0.61  &       4.94  &       1.85  &       5.42     \\
88  &    346.000  & $+$0.000  &       0.68  &       6.17  &       1.03  &       6.35     \\
89  &    346.125  & $-$0.125  &       0.19  &       4.94  &       0.31  &       5.14     \\
90  &    346.500  & $+$1.000  &       1.02  &       5.38  &       2.16  &       5.71     \\
91  &    347.000  & $+$0.250  &       0.28  &       5.11  &       0.52  &       5.38     \\
92  &    347.250  & $+$0.000  &       0.37  &       5.75  &       0.57  &       5.94     \\
\enddata

\end{deluxetable}

\newpage


\begin{thebibliography}{}
\bibitem[Allen(1973)]{allen73} Allen, C.~W.\ 1973, Astrophysical Quantities (3rd ed.; London:Athlone)
\bibitem[Alvarez et al.(1990)]{alvarez90} Alvarez, H., May, J., \& Bronfman, L.\ 1990, \apj, 348, 495 
\bibitem[Bash \& Leisawitz(1985)]{134} Bash, F.~N., \& Leisawitz, D.\ 1985, \aap, 145, 127 
\bibitem[Bertoldi \& McKee(1992)]{bertoldi1992} Bertoldi, F., \& McKee, C.~F.\ 1992, \apj, 395, 140 
\bibitem[Beuther et al.(2012)]{beuther2012} Beuther, H., Tackenberg, J., Linz, H., et al.\ 2012, \apj, 747, 43 
\bibitem[Blitz et al.(2007)]{blitz07} Blitz, L., Fukui, Y., Kawamura, A., et al.\ 2007, 2007prpl conf, 81 
\bibitem[Blitz \& Rosolowsky(2004)]{blitz04} Blitz, L., \& Rosolowsky, E.\ 2004, ArXiv Astrophysics e-prints, arXiv:astro-ph/0411520  
\bibitem[Brand \& Blitz(1993)]{brandandblitz93} Brand, J., \& Blitz, L.\ 1993, \aap, 275, 67 
\bibitem[Bronfman(1986)]{bron86} Bronfman, A.~L.\ 1986, Ph.D.~Thesis, Columbia University, New York.  
\bibitem[Bronfman(1992)]{bron92} Bronfman, L. 1992, in The center, bulge and disk of the Milky Way, ed. L. Blitz, p. 131
\bibitem[Bronfman et al.(1989)]{bron88b} Bronfman, L., Alvarez, H., Cohen, R.~S., \& Thaddeus, P.\ 1989, \apjs, 71, 481 
\bibitem[Bronfman et al.(1988a)]{bron13co88} Bronfman, L., Bitran, M., \& Thaddeus, P.\ 1988a, LNP, 315, 318   
\bibitem[Bronfman et al.(2000)]{bron00} Bronfman, L., Casassus, S., May, J., \& Nyman, L.-{\AA}.\ 2000, \aap, 358, 521 
\bibitem[Bronfman et al.(1988b)]{bron88} Bronfman, L., Cohen, R.~S., Alvarez, H., May, J., \& Thaddeus, P.\ 1988b, \apj, 324, 248   
\bibitem[Bronfman et al.(2008)]{bron08} Bronfman, L., Garay, G., Merello, M., et al. 2008, R.\ 2008, \apj, 672, 391   
\bibitem[Bronfman et al.(1996)]{bron96} Bronfman, L., Nyman, L.-A., \& May, J.\ 1996, \aaps, 115, 81 
\bibitem[Burton(1971)]{burton71} Burton, W.~B.\ 1971, \aap, 10, 76 
\bibitem[Burton(1988)]{burton88} Burton, W.~B.\ 1988, Galactic and ExtraGalactic Radio Astronomy (2nd ed., eds. G.L Verschuur \& K. Kellerman; New York:Springer)
\bibitem[Busfield et al.(2006)]{busfield2006} Busfield, A.~L.,Purcell, C.~R., Hoare, M.~G., et al. 2006, \mnras, 366, 1096 
\bibitem[Casoli et al.(1984)]{casoli84} Casoli, F., Combes, F., \& Gerin, M.\ 1984, \aap, 133, 99 
\bibitem[Caswell(2003)]{130} Caswell, J.~L.\ 2003, \mnras, 341, 551 
\bibitem[Caswell \& Haynes(1975)]{133} Caswell, J.~L., \& Haynes, R.~F.\ 1975, \mnras, 173, 649 
\bibitem[Caswell \& Haynes(1987)]{caswell87} Caswell, J.~L., \& Haynes, R.~F.\ 1987, \aap, 171, 261 
\bibitem[Caswell et al.(1975)]{1} Caswell, J.~L., Murray, J.~D., Roger, R.~S., Cole, D.~J., \& Cooke, D.~J.\ 1975, \aap, 45, 239 
\bibitem[Caswell \& Reynolds(2001)]{129} Caswell, J.~L., \& Reynolds, J.~E.\ 2001, \mnras, 325, 1346 
\bibitem[Caswell \& Robinson(1974)]{46} Caswell, J.~L., \& Robinson, B.~J.\ 1974, AuJPh, 27, 597 
\bibitem[Churchwell et al.(1974)]{68} Churchwell, E., Mezger, P.~G., \& Huchtmeier, W.\ 1974, \aap, 32, 283 
\bibitem[Cohen \& Davies(1976)]{cohen1976} Cohen, R.~J., \& Davies, R.~D.\ 1976, \mnras, 175, 1 
\bibitem[Combes(1991)]{combes91} Combes, F.\ 1991, \araa, 29, 195
\bibitem[Dame(1983)]{dame83} Dame, T.~M.\ 1983, Ph.D.~Thesis, Columbia University, New York.
\bibitem[Dame et al.(1986)]{dame86} Dame, T.~M., Elmegreen, B.~G., Cohen, R.~S., \& Thaddeus, P.\ 1986, \apj, 305, 892 
\bibitem[Dame \& Thaddeus(2008)]{dame2008} Dame, T.~M., \& Thaddeus, P.\ 2008, \apjl, 683, L143 
\bibitem[Dame \& Thaddeus(2011)]{dame2011} Dame, T.~M., \& Thaddeus, P.\ 2011, \apjl, 734, L24 
\bibitem[Dickman(1978)]{dickman78} Dickman, R.~L.\ 1978, \apjs, 37, 407 
\bibitem[Elmegreen(1993)]{elmegreen93b} Elmegreen, B.~G.\ 1993, Star Formation, Galaxies and the Interstellar Medium (J. Franco, F. Ferrini, \& G. Tenorio-Tagle ed.; Cambridge:Cambridge Univ. Press)
\bibitem[Elmegreen(1994)]{elmegreen94} Elmegreen, B.~G.\ 1994, \apj, 433, 39 
\bibitem[Englmaier \& Gerhard(1999)]{englmaier99} Englmaier, P., \& Gerhard, O.\ 1999, \mnras, 304, 512 
\bibitem[Evans(1999)]{evans99} Evans, N.~J., II 1999, \araa, 37, 311 
\bibitem[Fa{\'u}ndez et al.(2004)]{faundez2004} Fa{\'u}ndez, S., Bronfman, L., Garay, G., et al.\ 2004, \aap, 426, 97 
\bibitem[Fish et al.(2003)]{fish03} Fish, V.~L., Reid, M.~J., Wilner, D.~J., \& Churchwell, E.\ 2003, \apj, 587, 701 
\bibitem[Frerking et al.(1982)]{frerking82} Frerking, M.~A., Langer, W.~D., \& Wilson, R.~W.\ 1982, \apj, 262, 590 
\bibitem[Gardner \& Whiteoak(1984)]{81} Gardner, F.~F., \& Whiteoak, J.~B.\ 1984, \mnras, 210, 23 
\bibitem[Georgelin \& Georgelin(1976)]{gyg76} Georgelin, Y.~M., \& Georgelin, Y.~P.\ 1976, \aap, 49, 57 
\bibitem[Glover \& Mac Low(2011)]{glover2011} Glover, S.~C.~O., \& Mac Low, M.-M.\ 2011, \mnras, 412, 337 
\bibitem[Goss et al.(1970)]{93} Goss, W.~M., Manchester, R.~N., \& Robinson, B.~J.\ 1970, AuJPh, 23, 559 
\bibitem[Goss et al.(1972)]{106} Goss, W.~M., Radhakrishnan, V., Brooks, J.~W., \& Murray, J.~D.\ 1972, \apjs, 24, 123 
\bibitem[Grabelsky et al.(1988)]{grabelsky88} Grabelsky, D.~A., Cohen, R.~S., Bronfman, L., \& Thaddeus, P.\ 1988, \apj, 331, 181  
\bibitem[Grabelsky et al.(1987)]{grabelsky87} Grabelsky, D.~A., Cohen, R.~S., Bronfman, L., Thaddeus, P., \& May, J.\ 1987, \apj, 315, 122  
\bibitem[Green et al.(2011)]{green2011} Green, J.~A., Caswell, J.~L., McClure-Griffiths, N.~M., et al.\ 2011, \apj, 733, 27 
\bibitem[Green \& McClure-Griffiths(2011)]{green_mcclure2011} Green, J.~A., \& McClure-Griffiths, N.~M.\ 2011, \mnras, 417, 2500 
\bibitem[Hunter et al.(1997)]{hunter97} Hunter, S.~D., Bertsch, D.~L., Catelli, J.~R., et al.\ 1997, \apj, 481, 205 
\bibitem[Jones \& Dickey(2012)]{jones2012} Jones, C., \& Dickey, J.~M.\ 2012, \apj, 753, 62 
\bibitem[Kennicutt(1998)]{kennicutt98a} Kennicutt, R.~C., Jr.\ 1998, \araa, 36, 189 
\bibitem[Kerr \& Knapp(1970)]{105} Kerr, F.~J., \& Knapp, G.~R.\ 1970, AuJPA, 18, 9 
\bibitem[Kritsuk(2007)]{kritsuk2007} Kritsuk, A.\ 2007, KITP Conference: Star Formation, Then and Now: http://online.kitp.ucsb.edu/online/stars\_c07/, article \#14
\bibitem[Krumholz(2006)]{krum05} Krumholz, M.~R.\ 2006, New Horizons in Astronomy: Frank N.~Bash Symposium, 352, 31 
\bibitem[Krumholz \& McKee(2005)]{krum05b} Krumholz, M.~R., \& McKee, C.~F.\ 2005, \apj, 630, 250 
\bibitem[Larson(1981)]{larson81} Larson, R.~B.\ 1981, \mnras, 194, 809 
\bibitem[Liszt(1995)]{liszt95} Liszt, H.~S.\ 1995, \apj, 442, 163
\bibitem[Liszt et al.(1984)]{liszt84} Liszt, H.~S., Burton, W.~B., \& Xiang, D.-L.\ 1984, \aap, 140, 303 
\bibitem[Lockman(1979)]{5} Lockman, F.~J.\ 1979, \apj, 232, 761 
\bibitem[Luna et al.(2006)]{luna06} Luna, A., Bronfman, L., Carrasco, L., \& May, J.\ 2006, \apj, 641, 938 
\bibitem[Mac Low \& Klessen(2004)]{maclow03} Mac Low, M.-M., \& Klessen, R.~S.\ 2004, RvMP, 76, 125 
\bibitem[May et al.(1997)]{may97} May, J., Alvarez, H., \& Bronfman, L.\ 1997, \aap, 327, 325 
\bibitem[McClure-Griffiths et al.(2001)]{30} McClure-Griffiths, N.~M., Green, A.~J., Dickey, J.~M., et al. 2001, \apj, 551, 394 
\bibitem[McGee et al.(1967)]{94} McGee, R.~X., Gardner, F.~F., \& Robinson, B.~J.\ 1967, AuJPh, 20, 407 
\bibitem[McKee \& Ostriker(2007)]{mckee07} McKee, C.~F., \& Ostriker, E.~C.\ 2007, \araa, 45, 565 
\bibitem[Merello et al.(2013)]{merello2013} Merello, M., Bronfman, L., Garay, G., et al.\ 2013, \apj, 774, 38 
\bibitem[Myers et al.(1986)]{myers86} Myers, P.~C., Dame, T.~M., Thaddeus, P., et al. 1986, \apj, 301, 398  
\bibitem[Nyman et al.(1987)]{nyman87} Nyman, L.-A., Thaddeus, P., Bronfman, L., \& Cohen, R.~S.\ 1987, \apj, 314, 374 
\bibitem[Phillips et al.(1998)]{139} Phillips, C.~J., Norris, R.~P., Ellingsen, S.~P., \& McCulloch, P.~M.\ 1998, \mnras, 300, 1131 
\bibitem[Radhakrishnan et al.(1972)]{82} Radhakrishnan, V., Goss, W.~M., Murray, J.~D., \& Brooks, J.~W.\ 1972, \apjs, 24, 49 
\bibitem[Reid et al.(2009a)]{reid2009paralaxes} Reid, M.~J., Menten, K.~M., Brunthaler, A., et al.\ 2009a, \apj, 693, 397 
\bibitem[Reid et al.(2009b)]{reid2009} Reid, M.~J., Menten, K.~M., Zheng, X.~W., et al.\ 2009b, \apj, 700, 137
\bibitem[Robinson et al.(1971)]{98} Robinson, B.~J., Caswell, J.~L., \& Goss, W.~M.\ 1971, ApL, 9, 5
\bibitem[Rodgers et al.(1960)]{153} Rodgers, A.~W., Campbell, C.~T., \& Whiteoak, J.~B.\ 1960, \mnras, 121, 103 
\bibitem[Roman-Duval et al.(2009)]{roman2009} Roman-Duval, J., Jackson, J.~M., Heyer, M., et al.\ 2009, \apj, 699, 1153 
\bibitem[Romero-G{\'o}mez et al.(2011a)]{romero2011a} Romero-G{\'o}mez, M., Athanassoula, E., Antoja, T., \& Figueras, F.\ 2011a, \mnras, 418, 1176 
\bibitem[Romero-G{\'o}mez et al.(2011b)]{romero2011b} Romero-Gomez, M., Athanassoula, E., Antoja, T., et al.\ 2011b, RevMexAA Conference Series, 40, 92 
\bibitem[Rosolowsky(2005)]{roso05} Rosolowsky, E.\ 2005, \pasp, 117, 1403 
\bibitem[Rougoor \& Oort(1960)]{rougoor60} Rougoor, G.~W., \& Oort, J.~H.\ 1960, PNAS, 46, 1 
\bibitem[Russeil(2003)]{russeil03} Russeil, D.\ 2003, \aap, 397, 133 
\bibitem[Sanna et al.(2013)]{sanna2013} Sanna, A., Reid, M.~J., Menten, K., \& BeSSeL Survey Team 2013, American Astronomical Society Meeting Abstracts, 222, \#211.02 
\bibitem[Sanders et al.(1985)]{sanders85} Sanders, D.~B., Scoville, N.~Z., \& Solomon, P.~M.\ 1985, \apj, 289, 373 
\bibitem[Schuller et al.(2010)]{schuller2010} Schuller, F., Menten, K.~M., Wyrowski, F., et al.\ 2010, Highlights of Astronomy, 15, 780 
\bibitem[Scoville et al.(1987)]{scoville87} Scoville, N.~Z., Yun, M.~S., Sanders, D.~B., Clemens, D.~P., \& Waller, W.~H.\ 1987, \apjs, 63, 821  
\bibitem[Shaver \& Goss(1970)]{14} Shaver, P.~A., \& Goss, W.~M.\ 1970, AuJPA, 14, 133 
\bibitem[Shaver et al.(1983)]{83} Shaver, P.~A., McGee, R.~X., Newton, L.~M., Danks, A.~C., \& Pottasch, S.~R.\ 1983, \mnras, 204, 53
\bibitem[Shaver et al.(1982)]{124} Shaver, P.~A., Radhakrishnan, V., Anantharamaiah, K.~R., Retallack, D.~S., Wamsteker, W., \& Danks, A.~C.\ 1982, \aap, 106, 105 
\bibitem[Shaver et al.(1981)]{11} Shaver, P.~A., Retallack, D.~S., Wamsteker, W., \& Danks, A.~C.\ 1981, \aap, 102, 225
\bibitem[Shetty et al.(2011)]{shetty10} Shetty, R., Glover, S.~C., Dullemond, C.~P., \& Klessen, R.~S.\ 2011, \mnras, 412, 1686 
\bibitem[Solomon \& Rivolo(1989)]{solomon89} Solomon, P.~M., \& Rivolo, A.~R.\ 1989, \apj, 339, 919 
\bibitem[Solomon et al.(1987)]{solomon87} Solomon, P.~M., Rivolo, A.~R., Barrett, J., \& Yahil, A.\ 1987, \apj, 319, 730 
\bibitem[Tan(2005)]{tan05} Tan, J.~C.\ 2005, ArXiv Astrophysics e-prints, arXiv:astro-ph/0504256
\bibitem[Tan et al.(2013)]{tan2012} Tan, J.~C., Shaske, S.~N., \& Van Loo, S.\ 2013, IAU Symposium, 292, 19 
\bibitem[Tanti et al.(2012)]{tanti2012} Tanti, K.~K., Roy, J., \& Duorah, K.\ 2012, Advances in Astronomy and Space Physics, 2, \#.1.39 
\bibitem[Turner(1970)]{121} Turner, B.~E.\ 1970, ApL, 6, 99 
\bibitem[Turner(1979)]{44} Turner, B.~E.\ 1979, \aaps, 37, 1 
\bibitem[Ungerechts et al.(2000)]{ungerechts2000} Ungerechts, H., Umbanhowar, P., \& Thaddeus, P.\ 2000, \apj, 537, 221
\bibitem[Urquhart et al.(2012)]{ur2011} Urquhart, J.~S., Hoare, M.~G., Lumsden, S.~L., et al.\ 2012, \mnras, 420, 1656 
\bibitem[Walsh et al.(1997)]{20} Walsh, A.~J., Hyland, A.~R., Robinson, G., \& Burton, M.~G.\ 1997, \mnras, 291, 261 
\bibitem[Walsh et al.(2002)]{137} Walsh, A.~J., Lee, J.-K., \& Burton, M.~G.\ 2002, \mnras, 329, 475 
\bibitem[Whiteoak \& Gardner(1970)]{92} Whiteoak, J.~B., \& Gardner, F.~F.\ 1970, ApL, 5, 5 
\bibitem[Whiteoak \& Gardner(1974)]{80} Whiteoak, J.~B., \& Gardner, F.~F.\ 1974, \aap, 37, 389 
\bibitem[Williams \& McKee(1997)]{williams97} Williams, J.~P., \& McKee, C.~F.\ 1997, \apj, 476, 166 
\bibitem[Wilson et al.(1970)]{25} Wilson, T.~L., Mezger, P.~G., Gardner, F.~F., \& Milne, D.~K.\ 1970, \aap, 6, 364 
\bibitem[Wood \& Churchwell(1989)]{wood89b} Wood, D.~O.~S., \& Churchwell, E.\ 1989, \apj, 340, 265 
\bibitem[Zinnecker \& Yorke(2007)]{zinnecker07} Zinnecker, H., \& Yorke, H.~W.\ 2007, \araa, 45, 481 
\end{thebibliography}
\end{document}